\documentclass[longauth]{aa_gaia} 
\pdfoutput=1
\usepackage{graphicx}
\usepackage{txfonts}
\usepackage{pdflscape}

\usepackage{natbib,twoopt}
\usepackage{amsmath} 
\usepackage[breaklinks=true]{hyperref} 
\bibpunct{(}{)}{;}{a}{}{,} 
\makeatletter
\newcommandtwoopt{\citeads}[3][][]{\href{http://adsabs.harvard.edu/abs/#3}%
{\def\hyper@linkstart##1##2{}%
\let\hyper@linkend\@empty\citealp[#1][#2]{#3}}}
\newcommandtwoopt{\citepads}[3][][]{\href{http://adsabs.harvard.edu/abs/#3}%
{\def\hyper@linkstart##1##2{}%
\let\hyper@linkend\@empty\citep[#1][#2]{#3}}}
\newcommandtwoopt{\citetads}[3][][]{\href{http://adsabs.harvard.edu/abs/#3}%
{\def\hyper@linkstart##1##2{}%
\let\hyper@linkend\@empty\citet[#1][#2]{#3}}}
\newcommandtwoopt{\citeyearads}[3][][]%
{\href{http://adsabs.harvard.edu/abs/#3}
{\def\hyper@linkstart##1##2{}%
\let\hyper@linkend\@empty\citeyear[#1][#2]{#3}}}
\makeatother

\usepackage{color}
\definecolor{mygreen}{RGB}{0,128,0}

\hypersetup{colorlinks=true,linkcolor=blue,citecolor=blue,urlcolor=blue}

\renewcommand{\degr}{\ensuremath{^\circ}}

\newcommand{\muasyr}{\ensuremath{\,\mu\text{as\,yr}^{-1}}}
\newcommand{\masyr}{\ensuremath{\,\text{mas\,yr}^{-1}}}
\newcommand{\mssquared}{\ensuremath{\,\mathrm{m\,s^{-2}}}}

\newcommand{\kpc}{\,\text{kpc}}
\newcommand{\Myr}{\,\text{Myr}}

\newcommand{\gaia}{\textit{Gaia}}

\newcommand\gdrtwo{\gaia~DR2}
\newcommand\gdrthree{\gaia~DR3}
\newcommand{\gcrf}{\textit{Gaia}-CRF}
\newcommand{\gcrftwo}{\textit{Gaia}-CRF2}
\newcommand{\gcrfthree}{\textit{Gaia}-CRF3}

\newcommand{\gdr}[1]{{\gaia~DR#1}}
\newcommand{\gedr}[1]{{\gaia~EDR#1}}

\providecommand{\kms}{{\,\mathrm{km\,s^{-1}}}}
\providecommand{\kmsMyr}{\kms\Myr^{-1}}

\usepackage{wasysym}

\newcommand{\orcit}[1]{\protect\href{https://orcid.org/#1}{\protect\includegraphics[width=8pt]{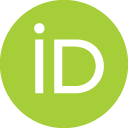}}}

\begin{document}

   \title{\textit{Gaia} Early Data Release 3}
   \subtitle{Acceleration of the solar system from \textit{Gaia} astrometry}

\author{
{\it Gaia} Collaboration
\and S.A.      ~Klioner                       \orcit{0000-0003-4682-7831}\inst{\ref{inst:0001}}
\and F.        ~Mignard                       \inst{\ref{inst:0002}}
\and L.        ~Lindegren                     \orcit{0000-0002-5443-3026}\inst{\ref{inst:0003}}
\and U.        ~Bastian                       \orcit{0000-0002-8667-1715}\inst{\ref{inst:0004}}
\and P.J.      ~McMillan                      \orcit{0000-0002-8861-2620}\inst{\ref{inst:0003}}
\and J.        ~Hern\'{a}ndez                 \inst{\ref{inst:0006}}
\and D.        ~Hobbs                         \orcit{0000-0002-2696-1366}\inst{\ref{inst:0003}}
\and M.        ~Ramos-Lerate                  \inst{\ref{inst:0008}}
\and M.        ~Biermann                      \inst{\ref{inst:0004}}
\and A.        ~Bombrun                       \inst{\ref{inst:0010}}
\and A.        ~de Torres                     \inst{\ref{inst:0010}}
\and E.        ~Gerlach                       \orcit{0000-0002-9533-2168}\inst{\ref{inst:0001}}
\and R.        ~Geyer                         \orcit{0000-0001-6967-8707}\inst{\ref{inst:0001}}
\and T.        ~Hilger                        \orcit{0000-0003-1646-0063}\inst{\ref{inst:0001}}
\and U.        ~Lammers                       \orcit{0000-0001-8309-3801}\inst{\ref{inst:0006}}
\and H.        ~Steidelm\"{ u}ller            \inst{\ref{inst:0001}}
\and C.A.      ~Stephenson                    \inst{\ref{inst:0017}}
\and A.G.A.    ~Brown                         \orcit{0000-0002-7419-9679}\inst{\ref{inst:0018}}
\and A.        ~Vallenari                     \orcit{0000-0003-0014-519X}\inst{\ref{inst:0019}}
\and T.        ~Prusti                        \orcit{0000-0003-3120-7867}\inst{\ref{inst:0020}}
\and J.H.J.    ~de Bruijne                    \orcit{0000-0001-6459-8599}\inst{\ref{inst:0020}}
\and C.        ~Babusiaux                     \orcit{0000-0002-7631-348X}\inst{\ref{inst:0022},\ref{inst:0023}}
\and O.L.      ~Creevey                       \orcit{0000-0003-1853-6631}\inst{\ref{inst:0002}}
\and D.W.      ~Evans                         \orcit{0000-0002-6685-5998}\inst{\ref{inst:0025}}
\and L.        ~Eyer                          \orcit{0000-0002-0182-8040}\inst{\ref{inst:0026}}
\and A.        ~Hutton                        \inst{\ref{inst:0027}}
\and F.        ~Jansen                        \inst{\ref{inst:0020}}
\and C.        ~Jordi                         \orcit{0000-0001-5495-9602}\inst{\ref{inst:0029}}
\and X.        ~Luri                          \orcit{0000-0001-5428-9397}\inst{\ref{inst:0029}}
\and C.        ~Panem                         \inst{\ref{inst:0031}}
\and D.        ~Pourbaix                      \orcit{0000-0002-3020-1837}\inst{\ref{inst:0032},\ref{inst:0033}}
\and S.        ~Randich                       \orcit{0000-0003-2438-0899}\inst{\ref{inst:0034}}
\and P.        ~Sartoretti                    \inst{\ref{inst:0023}}
\and C.        ~Soubiran                      \orcit{0000-0003-3304-8134}\inst{\ref{inst:0036}}
\and N.A.      ~Walton                        \orcit{0000-0003-3983-8778}\inst{\ref{inst:0025}}
\and F.        ~Arenou                        \orcit{0000-0003-2837-3899}\inst{\ref{inst:0023}}
\and C.A.L.    ~Bailer-Jones                  \inst{\ref{inst:0039}}
\and M.        ~Cropper                       \orcit{0000-0003-4571-9468}\inst{\ref{inst:0040}}
\and R.        ~Drimmel                       \orcit{0000-0002-1777-5502}\inst{\ref{inst:0041}}
\and D.        ~Katz                          \orcit{0000-0001-7986-3164}\inst{\ref{inst:0023}}
\and M.G.      ~Lattanzi                      \orcit{0000-0003-0429-7748}\inst{\ref{inst:0041},\ref{inst:0044}}
\and F.        ~van Leeuwen                   \inst{\ref{inst:0025}}
\and J.        ~Bakker                        \inst{\ref{inst:0006}}
\and J.        ~Casta\~{n}eda                 \orcit{0000-0001-7820-946X}\inst{\ref{inst:0047}}
\and F.        ~De Angeli                     \inst{\ref{inst:0025}}
\and C.        ~Ducourant                     \orcit{0000-0003-4843-8979}\inst{\ref{inst:0036}}
\and C.        ~Fabricius                     \orcit{0000-0003-2639-1372}\inst{\ref{inst:0029}}
\and M.        ~Fouesneau                     \orcit{0000-0001-9256-5516}\inst{\ref{inst:0039}}
\and Y.        ~Fr\'{e}mat                    \orcit{0000-0002-4645-6017}\inst{\ref{inst:0052}}
\and R.        ~Guerra                        \orcit{0000-0002-9850-8982}\inst{\ref{inst:0006}}
\and A.        ~Guerrier                      \inst{\ref{inst:0031}}
\and J.        ~Guiraud                       \inst{\ref{inst:0031}}
\and A.        ~Jean-Antoine Piccolo          \inst{\ref{inst:0031}}
\and E.        ~Masana                        \orcit{0000-0002-4819-329X}\inst{\ref{inst:0029}}
\and R.        ~Messineo                      \inst{\ref{inst:0058}}
\and N.        ~Mowlavi                       \inst{\ref{inst:0026}}
\and C.        ~Nicolas                       \inst{\ref{inst:0031}}
\and K.        ~Nienartowicz                  \orcit{0000-0001-5415-0547}\inst{\ref{inst:0061},\ref{inst:0062}}
\and F.        ~Pailler                       \inst{\ref{inst:0031}}
\and P.        ~Panuzzo                       \orcit{0000-0002-0016-8271}\inst{\ref{inst:0023}}
\and F.        ~Riclet                        \inst{\ref{inst:0031}}
\and W.        ~Roux                          \inst{\ref{inst:0031}}
\and G.M.      ~Seabroke                      \inst{\ref{inst:0040}}
\and R.        ~Sordo                         \orcit{0000-0003-4979-0659}\inst{\ref{inst:0019}}
\and P.        ~Tanga                         \orcit{0000-0002-2718-997X}\inst{\ref{inst:0002}}
\and F.        ~Th\'{e}venin                  \inst{\ref{inst:0002}}
\and G.        ~Gracia-Abril                  \inst{\ref{inst:0071},\ref{inst:0004}}
\and J.        ~Portell                       \orcit{0000-0002-8886-8925}\inst{\ref{inst:0029}}
\and D.        ~Teyssier                      \orcit{0000-0002-6261-5292}\inst{\ref{inst:0017}}
\and M.        ~Altmann                       \orcit{0000-0002-0530-0913}\inst{\ref{inst:0004},\ref{inst:0076}}
\and R.        ~Andrae                        \inst{\ref{inst:0039}}
\and I.        ~Bellas-Velidis                \inst{\ref{inst:0078}}
\and K.        ~Benson                        \inst{\ref{inst:0040}}
\and J.        ~Berthier                      \orcit{0000-0003-1846-6485}\inst{\ref{inst:0080}}
\and R.        ~Blomme                        \orcit{0000-0002-2526-346X}\inst{\ref{inst:0052}}
\and E.        ~Brugaletta                    \orcit{0000-0003-2598-6737}\inst{\ref{inst:0082}}
\and P.W.      ~Burgess                       \inst{\ref{inst:0025}}
\and G.        ~Busso                         \orcit{0000-0003-0937-9849}\inst{\ref{inst:0025}}
\and B.        ~Carry                         \orcit{0000-0001-5242-3089}\inst{\ref{inst:0002}}
\and A.        ~Cellino                       \orcit{0000-0002-6645-334X}\inst{\ref{inst:0041}}
\and N.        ~Cheek                         \inst{\ref{inst:0087}}
\and G.        ~Clementini                    \orcit{0000-0001-9206-9723}\inst{\ref{inst:0088}}
\and Y.        ~Damerdji                      \inst{\ref{inst:0089},\ref{inst:0090}}
\and M.        ~Davidson                      \inst{\ref{inst:0091}}
\and L.        ~Delchambre                    \inst{\ref{inst:0089}}
\and A.        ~Dell'Oro                      \orcit{0000-0003-1561-9685}\inst{\ref{inst:0034}}
\and J.        ~Fern\'{a}ndez-Hern\'{a}ndez   \inst{\ref{inst:0094}}
\and L.        ~Galluccio                     \orcit{0000-0002-8541-0476}\inst{\ref{inst:0002}}
\and P.        ~Garc\'{i}a-Lario              \inst{\ref{inst:0006}}
\and M.        ~Garcia-Reinaldos              \inst{\ref{inst:0006}}
\and J.        ~Gonz\'{a}lez-N\'{u}\~{n}ez    \orcit{0000-0001-5311-5555}\inst{\ref{inst:0087},\ref{inst:0099}}
\and E.        ~Gosset                        \inst{\ref{inst:0089},\ref{inst:0033}}
\and R.        ~Haigron                       \inst{\ref{inst:0023}}
\and J.-L.     ~Halbwachs                     \orcit{0000-0003-2968-6395}\inst{\ref{inst:0103}}
\and N.C.      ~Hambly                        \orcit{0000-0002-9901-9064}\inst{\ref{inst:0091}}
\and D.L.      ~Harrison                      \orcit{0000-0001-8687-6588}\inst{\ref{inst:0025},\ref{inst:0106}}
\and D.        ~Hatzidimitriou                \orcit{0000-0002-5415-0464}\inst{\ref{inst:0107}}
\and U.        ~Heiter                        \orcit{0000-0001-6825-1066}\inst{\ref{inst:0108}}
\and D.        ~Hestroffer                    \orcit{0000-0003-0472-9459}\inst{\ref{inst:0080}}
\and S.T.      ~Hodgkin                       \inst{\ref{inst:0025}}
\and B.        ~Holl                          \orcit{0000-0001-6220-3266}\inst{\ref{inst:0026},\ref{inst:0061}}
\and K.        ~Jan{\ss}en                    \inst{\ref{inst:0113}}
\and G.        ~Jevardat de Fombelle          \inst{\ref{inst:0026}}
\and S.        ~Jordan                        \orcit{0000-0001-6316-6831}\inst{\ref{inst:0004}}
\and A.        ~Krone-Martins                 \orcit{0000-0002-2308-6623}\inst{\ref{inst:0116},\ref{inst:0117}}
\and A.C.      ~Lanzafame                     \orcit{0000-0002-2697-3607}\inst{\ref{inst:0082},\ref{inst:0119}}
\and W.        ~L\"{ o}ffler                  \inst{\ref{inst:0004}}
\and A.        ~Lorca                         \inst{\ref{inst:0027}}
\and M.        ~Manteiga                      \orcit{0000-0002-7711-5581}\inst{\ref{inst:0122}}
\and O.        ~Marchal                       \inst{\ref{inst:0103}}
\and P.M.      ~Marrese                       \inst{\ref{inst:0124},\ref{inst:0125}}
\and A.        ~Moitinho                      \orcit{0000-0003-0822-5995}\inst{\ref{inst:0116}}
\and A.        ~Mora                          \inst{\ref{inst:0027}}
\and K.        ~Muinonen                      \orcit{0000-0001-8058-2642}\inst{\ref{inst:0128},\ref{inst:0129}}
\and P.        ~Osborne                       \inst{\ref{inst:0025}}
\and E.        ~Pancino                       \orcit{0000-0003-0788-5879}\inst{\ref{inst:0034},\ref{inst:0125}}
\and T.        ~Pauwels                       \inst{\ref{inst:0052}}
\and A.        ~Recio-Blanco                  \inst{\ref{inst:0002}}
\and P.J.      ~Richards                      \inst{\ref{inst:0135}} %
\and M.        ~Riello                        \orcit{0000-0002-3134-0935}\inst{\ref{inst:0025}}
\and L.        ~Rimoldini                     \orcit{0000-0002-0306-585X}\inst{\ref{inst:0061}}
\and A.C.      ~Robin                         \orcit{0000-0001-8654-9499}\inst{\ref{inst:0138}}
\and T.        ~Roegiers                      \inst{\ref{inst:0139}}
\and J.        ~Rybizki                       \orcit{0000-0002-0993-6089}\inst{\ref{inst:0039}}
\and L.M.      ~Sarro                         \orcit{0000-0002-5622-5191}\inst{\ref{inst:0141}}
\and C.        ~Siopis                        \inst{\ref{inst:0032}}
\and M.        ~Smith                         \inst{\ref{inst:0040}}
\and A.        ~Sozzetti                      \orcit{0000-0002-7504-365X}\inst{\ref{inst:0041}}
\and A.        ~Ulla                          \inst{\ref{inst:0145}}
\and E.        ~Utrilla                       \inst{\ref{inst:0027}}
\and M.        ~van Leeuwen                   \inst{\ref{inst:0025}}
\and W.        ~van Reeven                    \inst{\ref{inst:0027}}
\and U.        ~Abbas                         \orcit{0000-0002-5076-766X}\inst{\ref{inst:0041}}
\and A.        ~Abreu Aramburu                \inst{\ref{inst:0094}}
\and S.        ~Accart                        \inst{\ref{inst:0151}}
\and C.        ~Aerts                         \orcit{0000-0003-1822-7126}\inst{\ref{inst:0152},\ref{inst:0153},\ref{inst:0039}}
\and J.J.      ~Aguado                        \inst{\ref{inst:0141}}
\and M.        ~Ajaj                          \inst{\ref{inst:0023}}
\and G.        ~Altavilla                     \orcit{0000-0002-9934-1352}\inst{\ref{inst:0124},\ref{inst:0125}}
\and M.A.      ~\'{A}lvarez                   \orcit{0000-0002-6786-2620}\inst{\ref{inst:0159}}
\and J.        ~\'{A}lvarez Cid-Fuentes       \orcit{0000-0001-7153-4649}\inst{\ref{inst:0160}}
\and J.        ~Alves                         \orcit{0000-0002-4355-0921}\inst{\ref{inst:0161}}
\and R.I.      ~Anderson                      \orcit{0000-0001-8089-4419}\inst{\ref{inst:0162}}
\and E.        ~Anglada Varela                \orcit{0000-0001-7563-0689}\inst{\ref{inst:0094}}
\and T.        ~Antoja                        \orcit{0000-0003-2595-5148}\inst{\ref{inst:0029}}
\and M.        ~Audard                        \orcit{0000-0003-4721-034X}\inst{\ref{inst:0061}}
\and D.        ~Baines                        \orcit{0000-0002-6923-3756}\inst{\ref{inst:0017}}
\and S.G.      ~Baker                         \orcit{0000-0002-6436-1257}\inst{\ref{inst:0040}}
\and L.        ~Balaguer-N\'{u}\~{n}ez        \orcit{0000-0001-9789-7069}\inst{\ref{inst:0029}}
\and E.        ~Balbinot                      \orcit{0000-0002-1322-3153}\inst{\ref{inst:0169}}
\and Z.        ~Balog                         \orcit{0000-0003-1748-2926}\inst{\ref{inst:0004},\ref{inst:0039}}
\and C.        ~Barache                       \inst{\ref{inst:0076}}
\and D.        ~Barbato                       \inst{\ref{inst:0026},\ref{inst:0041}}
\and M.        ~Barros                        \orcit{0000-0002-9728-9618}\inst{\ref{inst:0116}}
\and M.A.      ~Barstow                       \orcit{0000-0002-7116-3259}\inst{\ref{inst:0176}}
\and S.        ~Bartolom\'{e}                 \orcit{0000-0002-6290-6030}\inst{\ref{inst:0029}}
\and J.-L.     ~Bassilana                     \inst{\ref{inst:0151}}
\and N.        ~Bauchet                       \inst{\ref{inst:0080}}
\and A.        ~Baudesson-Stella              \inst{\ref{inst:0151}}
\and U.        ~Becciani                      \orcit{0000-0002-4389-8688}\inst{\ref{inst:0082}}
\and M.        ~Bellazzini                    \orcit{0000-0001-8200-810X}\inst{\ref{inst:0088}}
\and M.        ~Bernet                        \inst{\ref{inst:0029}}
\and S.        ~Bertone                       \orcit{0000-0001-9885-8440}\inst{\ref{inst:0184},\ref{inst:0185},\ref{inst:0041}}
\and L.        ~Bianchi                       \inst{\ref{inst:0187}}
\and S.        ~Blanco-Cuaresma               \orcit{0000-0002-1584-0171}\inst{\ref{inst:0188}}
\and T.        ~Boch                          \orcit{0000-0001-5818-2781}\inst{\ref{inst:0103}}
\and D.        ~Bossini                       \orcit{0000-0002-9480-8400}\inst{\ref{inst:0190}}
\and S.        ~Bouquillon                    \inst{\ref{inst:0076}}
\and L.        ~Bramante                      \inst{\ref{inst:0058}}
\and E.        ~Breedt                        \orcit{0000-0001-6180-3438}\inst{\ref{inst:0025}}
\and A.        ~Bressan                       \orcit{0000-0002-7922-8440}\inst{\ref{inst:0195}}
\and N.        ~Brouillet                     \inst{\ref{inst:0036}}
\and B.        ~Bucciarelli                   \orcit{0000-0002-5303-0268}\inst{\ref{inst:0041}}
\and A.        ~Burlacu                       \inst{\ref{inst:0198}}
\and D.        ~Busonero                      \orcit{0000-0002-3903-7076}\inst{\ref{inst:0041}}
\and A.G.      ~Butkevich                     \inst{\ref{inst:0041}}
\and R.        ~Buzzi                         \orcit{0000-0001-9389-5701}\inst{\ref{inst:0041}}
\and E.        ~Caffau                        \orcit{0000-0001-6011-6134}\inst{\ref{inst:0023}}
\and R.        ~Cancelliere                   \orcit{0000-0002-9120-3799}\inst{\ref{inst:0203}}
\and H.        ~C\'{a}novas                   \orcit{0000-0001-7668-8022}\inst{\ref{inst:0027}}
\and T.        ~Cantat-Gaudin                 \orcit{0000-0001-8726-2588}\inst{\ref{inst:0029}}
\and R.        ~Carballo                      \inst{\ref{inst:0206}}
\and T.        ~Carlucci                      \inst{\ref{inst:0076}}
\and M.I       ~Carnerero                     \orcit{0000-0001-5843-5515}\inst{\ref{inst:0041}}
\and J.M.      ~Carrasco                      \orcit{0000-0002-3029-5853}\inst{\ref{inst:0029}}
\and L.        ~Casamiquela                   \orcit{0000-0001-5238-8674}\inst{\ref{inst:0036}}
\and M.        ~Castellani                    \orcit{0000-0002-7650-7428}\inst{\ref{inst:0124}}
\and A.        ~Castro-Ginard                 \orcit{0000-0002-9419-3725}\inst{\ref{inst:0029}}
\and P.        ~Castro Sampol                 \inst{\ref{inst:0029}}
\and L.        ~Chaoul                        \inst{\ref{inst:0031}}
\and P.        ~Charlot                       \inst{\ref{inst:0036}}
\and L.        ~Chemin                        \orcit{0000-0002-3834-7937}\inst{\ref{inst:0216}}
\and A.        ~Chiavassa                     \orcit{0000-0003-3891-7554}\inst{\ref{inst:0002}}
\and G.        ~Comoretto                     \inst{\ref{inst:0219}}
\and W.J.      ~Cooper                        \orcit{0000-0003-3501-8967}\inst{\ref{inst:0220},\ref{inst:0041}}
\and T.        ~Cornez                        \inst{\ref{inst:0151}}
\and S.        ~Cowell                        \inst{\ref{inst:0025}}
\and F.        ~Crifo                         \inst{\ref{inst:0023}}
\and M.        ~Crosta                        \orcit{0000-0003-4369-3786}\inst{\ref{inst:0041}}
\and C.        ~Crowley                       \inst{\ref{inst:0010}}
\and C.        ~Dafonte                       \orcit{0000-0003-4693-7555}\inst{\ref{inst:0159}}
\and A.        ~Dapergolas                    \inst{\ref{inst:0078}}
\and M.        ~David                         \orcit{0000-0002-4172-3112}\inst{\ref{inst:0229}}
\and P.        ~David                         \inst{\ref{inst:0080}}
\and P.        ~de Laverny                    \inst{\ref{inst:0002}}
\and F.        ~De Luise                      \orcit{0000-0002-6570-8208}\inst{\ref{inst:0232}}
\and R.        ~De March                      \orcit{0000-0003-0567-842X}\inst{\ref{inst:0058}}
\and J.        ~De Ridder                     \orcit{0000-0001-6726-2863}\inst{\ref{inst:0152}}
\and R.        ~de Souza                      \inst{\ref{inst:0235}}
\and P.        ~de Teodoro                    \inst{\ref{inst:0006}}
\and E.F.      ~del Peloso                    \inst{\ref{inst:0004}}
\and E.        ~del Pozo                      \inst{\ref{inst:0027}}
\and A.        ~Delgado                       \inst{\ref{inst:0025}}
\and H.E.      ~Delgado                       \orcit{0000-0003-1409-4282}\inst{\ref{inst:0141}}
\and J.-B.     ~Delisle                       \orcit{0000-0001-5844-9888}\inst{\ref{inst:0026}}
\and P.        ~Di Matteo                     \inst{\ref{inst:0023}}
\and S.        ~Diakite                       \inst{\ref{inst:0243}}
\and C.        ~Diener                        \inst{\ref{inst:0025}}
\and E.        ~Distefano                     \orcit{0000-0002-2448-2513}\inst{\ref{inst:0082}}
\and C.        ~Dolding                       \inst{\ref{inst:0040}}
\and D.        ~Eappachen                     \inst{\ref{inst:0247},\ref{inst:0153}}
\and H.        ~Enke                          \orcit{0000-0002-2366-8316}\inst{\ref{inst:0113}}
\and P.        ~Esquej                        \orcit{0000-0001-8195-628X}\inst{\ref{inst:0251}}
\and C.        ~Fabre                         \inst{\ref{inst:0252}}
\and M.        ~Fabrizio                      \orcit{0000-0001-5829-111X}\inst{\ref{inst:0124},\ref{inst:0125}}
\and S.        ~Faigler                       \inst{\ref{inst:0255}}
\and G.        ~Fedorets                      \inst{\ref{inst:0128},\ref{inst:0257}}
\and P.        ~Fernique                      \orcit{0000-0002-3304-2923}\inst{\ref{inst:0103},\ref{inst:0259}}
\and A.        ~Fienga                        \orcit{0000-0002-4755-7637}\inst{\ref{inst:0260},\ref{inst:0080}}
\and F.        ~Figueras                      \orcit{0000-0002-3393-0007}\inst{\ref{inst:0029}}
\and C.        ~Fouron                        \inst{\ref{inst:0198}}
\and F.        ~Fragkoudi                     \inst{\ref{inst:0264}}
\and E.        ~Fraile                        \inst{\ref{inst:0251}}
\and F.        ~Franke                        \inst{\ref{inst:0266}}
\and M.        ~Gai                           \orcit{0000-0001-9008-134X}\inst{\ref{inst:0041}}
\and D.        ~Garabato                      \orcit{0000-0002-7133-6623}\inst{\ref{inst:0159}}
\and A.        ~Garcia-Gutierrez              \inst{\ref{inst:0029}}
\and M.        ~Garc\'{i}a-Torres             \orcit{0000-0002-6867-7080}\inst{\ref{inst:0270}}
\and A.        ~Garofalo                      \orcit{0000-0002-5907-0375}\inst{\ref{inst:0088}}
\and P.        ~Gavras                        \orcit{0000-0002-4383-4836}\inst{\ref{inst:0251}}
\and P.        ~Giacobbe                      \inst{\ref{inst:0041}}
\and G.        ~Gilmore                       \orcit{0000-0003-4632-0213}\inst{\ref{inst:0025}}
\and S.        ~Girona                        \orcit{0000-0002-1975-1918}\inst{\ref{inst:0160}}
\and G.        ~Giuffrida                     \inst{\ref{inst:0124}}
\and A.        ~Gomez                         \orcit{0000-0002-3796-3690}\inst{\ref{inst:0159}}
\and I.        ~Gonzalez-Santamaria           \orcit{0000-0002-8537-9384}\inst{\ref{inst:0159}}
\and J.J.      ~Gonz\'{a}lez-Vidal            \inst{\ref{inst:0029}}
\and M.        ~Granvik                       \orcit{0000-0002-5624-1888}\inst{\ref{inst:0128},\ref{inst:0281}}
\and R.        ~Guti\'{e}rrez-S\'{a}nchez     \inst{\ref{inst:0017}}
\and L.P.      ~Guy                           \orcit{0000-0003-0800-8755}\inst{\ref{inst:0061},\ref{inst:0219}}
\and M.        ~Hauser                        \inst{\ref{inst:0039},\ref{inst:0286}}
\and M.        ~Haywood                       \orcit{0000-0003-0434-0400}\inst{\ref{inst:0023}}
\and A.        ~Helmi                         \orcit{0000-0003-3937-7641}\inst{\ref{inst:0169}}
\and S.L.      ~Hidalgo                       \orcit{0000-0002-0002-9298}\inst{\ref{inst:0289},\ref{inst:0290}}
\and N.        ~H\l{}adczuk                   \inst{\ref{inst:0006}}
\and G.        ~Holland                       \inst{\ref{inst:0025}}
\and H.E.      ~Huckle                        \inst{\ref{inst:0040}}
\and G.        ~Jasniewicz                    \inst{\ref{inst:0294}}
\and P.G.      ~Jonker                        \orcit{0000-0001-5679-0695}\inst{\ref{inst:0153},\ref{inst:0247}}
\and J.        ~Juaristi Campillo             \inst{\ref{inst:0004}}
\and F.        ~Julbe                         \inst{\ref{inst:0029}}
\and L.        ~Karbevska                     \inst{\ref{inst:0026}}
\and P.        ~Kervella                      \orcit{0000-0003-0626-1749}\inst{\ref{inst:0300}}
\and S.        ~Khanna                        \orcit{0000-0002-2604-4277}\inst{\ref{inst:0169}}
\and A.        ~Kochoska                      \orcit{0000-0002-9739-8371}\inst{\ref{inst:0302}}
\and G.        ~Kordopatis                    \orcit{0000-0002-9035-3920}\inst{\ref{inst:0002}}
\and A.J.      ~Korn                          \orcit{0000-0002-3881-6756}\inst{\ref{inst:0108}}
\and Z.        ~Kostrzewa-Rutkowska           \inst{\ref{inst:0018},\ref{inst:0247}}
\and K.        ~Kruszy\'{n}ska                \orcit{0000-0002-2729-5369}\inst{\ref{inst:0308}}
\and S.        ~Lambert                       \orcit{0000-0001-6759-5502}\inst{\ref{inst:0076}}
\and A.F.      ~Lanza                         \orcit{0000-0001-5928-7251}\inst{\ref{inst:0082}}
\and Y.        ~Lasne                         \inst{\ref{inst:0151}}
\and J.-F.     ~Le Campion                    \inst{\ref{inst:0312}}
\and Y.        ~Le Fustec                     \inst{\ref{inst:0198}}
\and Y.        ~Lebreton                      \orcit{0000-0002-4834-2144}\inst{\ref{inst:0300},\ref{inst:0315}}
\and T.        ~Lebzelter                     \orcit{0000-0002-0702-7551}\inst{\ref{inst:0161}}
\and S.        ~Leccia                        \orcit{0000-0001-5685-6930}\inst{\ref{inst:0317}}
\and N.        ~Leclerc                       \inst{\ref{inst:0023}}
\and I.        ~Lecoeur-Taibi                 \orcit{0000-0003-0029-8575}\inst{\ref{inst:0061}}
\and S.        ~Liao                          \inst{\ref{inst:0041}}
\and E.        ~Licata                        \orcit{0000-0002-5203-0135}\inst{\ref{inst:0041}}
\and H.E.P.    ~Lindstr{\o}m                  \inst{\ref{inst:0041},\ref{inst:0323}}
\and T.A.      ~Lister                        \orcit{0000-0002-3818-7769}\inst{\ref{inst:0324}}
\and E.        ~Livanou                       \inst{\ref{inst:0107}}
\and A.        ~Lobel                         \inst{\ref{inst:0052}}
\and P.        ~Madrero Pardo                 \inst{\ref{inst:0029}}
\and S.        ~Managau                       \inst{\ref{inst:0151}}
\and R.G.      ~Mann                          \orcit{0000-0002-0194-325X}\inst{\ref{inst:0091}}
\and J.M.      ~Marchant                      \inst{\ref{inst:0330}}
\and M.        ~Marconi                       \orcit{0000-0002-1330-2927}\inst{\ref{inst:0317}}
\and M.M.S.    ~Marcos Santos                 \inst{\ref{inst:0087}}
\and S.        ~Marinoni                      \orcit{0000-0001-7990-6849}\inst{\ref{inst:0124},\ref{inst:0125}}
\and F.        ~Marocco                       \orcit{0000-0001-7519-1700}\inst{\ref{inst:0335},\ref{inst:0336}}
\and D.J.      ~Marshall                      \inst{\ref{inst:0337}}
\and L.        ~Martin Polo                   \inst{\ref{inst:0087}}
\and J.M.      ~Mart\'{i}n-Fleitas            \orcit{0000-0002-8594-569X}\inst{\ref{inst:0027}}
\and A.        ~Masip                         \inst{\ref{inst:0029}}
\and D.        ~Massari                       \orcit{0000-0001-8892-4301}\inst{\ref{inst:0088}}
\and A.        ~Mastrobuono-Battisti          \orcit{0000-0002-2386-9142}\inst{\ref{inst:0003}}
\and T.        ~Mazeh                         \orcit{0000-0002-3569-3391}\inst{\ref{inst:0255}}
\and S.        ~Messina                       \orcit{0000-0002-2851-2468}\inst{\ref{inst:0082}}
\and D.        ~Michalik                      \orcit{0000-0002-7618-6556}\inst{\ref{inst:0020}}
\and N.R.      ~Millar                        \inst{\ref{inst:0025}}
\and A.        ~Mints                         \orcit{0000-0002-8440-1455}\inst{\ref{inst:0113}}
\and D.        ~Molina                        \orcit{0000-0003-4814-0275}\inst{\ref{inst:0029}}
\and R.        ~Molinaro                      \orcit{0000-0003-3055-6002}\inst{\ref{inst:0317}}
\and L.        ~Moln\'{a}r                    \orcit{0000-0002-8159-1599}\inst{\ref{inst:0350},\ref{inst:0351},\ref{inst:0352}}
\and P.        ~Montegriffo                   \inst{\ref{inst:0088}}
\and R.        ~Mor                           \orcit{0000-0002-8179-6527}\inst{\ref{inst:0029}}
\and R.        ~Morbidelli                    \orcit{0000-0001-7627-4946}\inst{\ref{inst:0041}}
\and T.        ~Morel                         \inst{\ref{inst:0089}}
\and D.        ~Morris                        \inst{\ref{inst:0091}}
\and A.F.      ~Mulone                        \inst{\ref{inst:0058}}
\and D.        ~Munoz                         \inst{\ref{inst:0151}}
\and T.        ~Muraveva                      \orcit{0000-0002-0969-1915}\inst{\ref{inst:0088}}
\and C.P.      ~Murphy                        \inst{\ref{inst:0006}}
\and I.        ~Musella                       \orcit{0000-0001-5909-6615}\inst{\ref{inst:0317}}
\and L.        ~Noval                         \inst{\ref{inst:0151}}
\and C.        ~Ord\'{e}novic                 \inst{\ref{inst:0002}}
\and G.        ~Orr\`{u}                      \inst{\ref{inst:0058}}
\and J.        ~Osinde                        \inst{\ref{inst:0251}}
\and C.        ~Pagani                        \inst{\ref{inst:0176}}
\and I.        ~Pagano                        \orcit{0000-0001-9573-4928}\inst{\ref{inst:0082}}
\and L.        ~Palaversa                     \inst{\ref{inst:0369},\ref{inst:0025}}
\and P.A.      ~Palicio                       \orcit{0000-0002-7432-8709}\inst{\ref{inst:0002}}
\and A.        ~Panahi                        \orcit{0000-0001-5850-4373}\inst{\ref{inst:0255}}
\and M.        ~Pawlak                        \orcit{0000-0002-5632-9433}\inst{\ref{inst:0373},\ref{inst:0308}}
\and X.        ~Pe\~{n}alosa Esteller         \inst{\ref{inst:0029}}
\and A.        ~Penttil\"{ a}                 \orcit{0000-0001-7403-1721}\inst{\ref{inst:0128}}
\and A.M.      ~Piersimoni                    \orcit{0000-0002-8019-3708}\inst{\ref{inst:0232}}
\and F.-X.     ~Pineau                        \orcit{0000-0002-2335-4499}\inst{\ref{inst:0103}}
\and E.        ~Plachy                        \orcit{0000-0002-5481-3352}\inst{\ref{inst:0350},\ref{inst:0351},\ref{inst:0352}}
\and G.        ~Plum                          \inst{\ref{inst:0023}}
\and E.        ~Poggio                        \orcit{0000-0003-3793-8505}\inst{\ref{inst:0041}}
\and E.        ~Poretti                       \orcit{0000-0003-1200-0473}\inst{\ref{inst:0384}}
\and E.        ~Poujoulet                     \inst{\ref{inst:0385}}
\and A.        ~Pr\v{s}a                      \orcit{0000-0002-1913-0281}\inst{\ref{inst:0302}}
\and L.        ~Pulone                        \orcit{0000-0002-5285-998X}\inst{\ref{inst:0124}}
\and E.        ~Racero                        \inst{\ref{inst:0087},\ref{inst:0389}}
\and S.        ~Ragaini                       \inst{\ref{inst:0088}}
\and M.        ~Rainer                        \orcit{0000-0002-8786-2572}\inst{\ref{inst:0034}}
\and C.M.      ~Raiteri                       \orcit{0000-0003-1784-2784}\inst{\ref{inst:0041}}
\and N.        ~Rambaux                       \inst{\ref{inst:0080}}
\and P.        ~Ramos                         \orcit{0000-0002-5080-7027}\inst{\ref{inst:0029}}
\and P.        ~Re Fiorentin                  \orcit{0000-0002-4995-0475}\inst{\ref{inst:0041}}
\and S.        ~Regibo                        \inst{\ref{inst:0152}}
\and C.        ~Reyl\'{e}                     \inst{\ref{inst:0138}}
\and V.        ~Ripepi                        \orcit{0000-0003-1801-426X}\inst{\ref{inst:0317}}
\and A.        ~Riva                          \orcit{0000-0002-6928-8589}\inst{\ref{inst:0041}}
\and G.        ~Rixon                         \inst{\ref{inst:0025}}
\and N.        ~Robichon                      \orcit{0000-0003-4545-7517}\inst{\ref{inst:0023}}
\and C.        ~Robin                         \inst{\ref{inst:0151}}
\and M.        ~Roelens                       \orcit{0000-0003-0876-4673}\inst{\ref{inst:0026}}
\and L.        ~Rohrbasser                    \inst{\ref{inst:0061}}
\and M.        ~Romero-G\'{o}mez              \orcit{0000-0003-3936-1025}\inst{\ref{inst:0029}}
\and N.        ~Rowell                        \inst{\ref{inst:0091}}
\and F.        ~Royer                         \orcit{0000-0002-9374-8645}\inst{\ref{inst:0023}}
\and K.A.      ~Rybicki                       \orcit{0000-0002-9326-9329}\inst{\ref{inst:0308}}
\and G.        ~Sadowski                      \inst{\ref{inst:0032}}
\and A.        ~Sagrist\`{a} Sell\'{e}s       \orcit{0000-0001-6191-2028}\inst{\ref{inst:0004}}
\and J.        ~Sahlmann                      \orcit{0000-0001-9525-3673}\inst{\ref{inst:0251}}
\and J.        ~Salgado                       \orcit{0000-0002-3680-4364}\inst{\ref{inst:0017}}
\and E.        ~Salguero                      \inst{\ref{inst:0094}}
\and N.        ~Samaras                       \orcit{0000-0001-8375-6652}\inst{\ref{inst:0052}}
\and V.        ~Sanchez Gimenez               \inst{\ref{inst:0029}}
\and N.        ~Sanna                         \inst{\ref{inst:0034}}
\and R.        ~Santove\~{n}a                 \orcit{0000-0002-9257-2131}\inst{\ref{inst:0159}}
\and M.        ~Sarasso                       \orcit{0000-0001-5121-0727}\inst{\ref{inst:0041}}
\and M.        ~Schultheis                    \orcit{0000-0002-6590-1657}\inst{\ref{inst:0002}}
\and E.        ~Sciacca                       \orcit{0000-0002-5574-2787}\inst{\ref{inst:0082}}
\and M.        ~Segol                         \inst{\ref{inst:0266}}
\and J.C.      ~Segovia                       \inst{\ref{inst:0087}}
\and D.        ~S\'{e}gransan                 \orcit{0000-0003-2355-8034}\inst{\ref{inst:0026}}
\and D.        ~Semeux                        \inst{\ref{inst:0252}}
\and H.I.      ~Siddiqui                      \orcit{0000-0003-1853-6033}\inst{\ref{inst:0425}}
\and A.        ~Siebert                       \orcit{0000-0001-8059-2840}\inst{\ref{inst:0103},\ref{inst:0259}}
\and L.        ~Siltala                       \orcit{0000-0002-6938-794X}\inst{\ref{inst:0128}}
\and E.        ~Slezak                        \inst{\ref{inst:0002}}
\and R.L.      ~Smart                         \orcit{0000-0002-4424-4766}\inst{\ref{inst:0041}}
\and E.        ~Solano                        \inst{\ref{inst:0431}}
\and F.        ~Solitro                       \inst{\ref{inst:0058}}
\and D.        ~Souami                        \orcit{0000-0003-4058-0815}\inst{\ref{inst:0300},\ref{inst:0434}}
\and J.        ~Souchay                       \inst{\ref{inst:0076}}
\and A.        ~Spagna                        \orcit{0000-0003-1732-2412}\inst{\ref{inst:0041}}
\and F.        ~Spoto                         \orcit{0000-0001-7319-5847}\inst{\ref{inst:0188}}
\and I.A.      ~Steele                        \orcit{0000-0001-8397-5759}\inst{\ref{inst:0330}}
\and M.        ~S\"{ u}veges                  \inst{\ref{inst:0061},\ref{inst:0440},\ref{inst:0039}}
\and L.        ~Szabados                      \orcit{0000-0002-2046-4131}\inst{\ref{inst:0350}}
\and E.        ~Szegedi-Elek                  \orcit{0000-0001-7807-6644}\inst{\ref{inst:0350}}
\and F.        ~Taris                         \inst{\ref{inst:0076}}
\and G.        ~Tauran                        \inst{\ref{inst:0151}}
\and M.B.      ~Taylor                        \orcit{0000-0002-4209-1479}\inst{\ref{inst:0446}}
\and R.        ~Teixeira                      \orcit{0000-0002-6806-6626}\inst{\ref{inst:0235}}
\and W.        ~Thuillot                      \inst{\ref{inst:0080}}
\and N.        ~Tonello                       \orcit{0000-0003-0550-1667}\inst{\ref{inst:0160}}
\and F.        ~Torra                         \orcit{0000-0002-8429-299X}\inst{\ref{inst:0047}}
\and J.        ~Torra$^\dagger$               \inst{\ref{inst:0029}}
\and C.        ~Turon                         \orcit{0000-0003-1236-5157}\inst{\ref{inst:0023}}
\and N.        ~Unger                         \orcit{0000-0003-3993-7127}\inst{\ref{inst:0026}}
\and M.        ~Vaillant                      \inst{\ref{inst:0151}}
\and E.        ~van Dillen                    \inst{\ref{inst:0266}}
\and O.        ~Vanel                         \inst{\ref{inst:0023}}
\and A.        ~Vecchiato                     \orcit{0000-0003-1399-5556}\inst{\ref{inst:0041}}
\and Y.        ~Viala                         \inst{\ref{inst:0023}}
\and D.        ~Vicente                       \inst{\ref{inst:0160}}
\and S.        ~Voutsinas                     \inst{\ref{inst:0091}}
\and M.        ~Weiler                        \inst{\ref{inst:0029}}
\and T.        ~Wevers                        \orcit{0000-0002-4043-9400}\inst{\ref{inst:0025}}
\and \L{}.     ~Wyrzykowski                   \orcit{0000-0002-9658-6151}\inst{\ref{inst:0308}}
\and A.        ~Yoldas                        \inst{\ref{inst:0025}}
\and P.        ~Yvard                         \inst{\ref{inst:0266}}
\and H.        ~Zhao                          \orcit{0000-0003-2645-6869}\inst{\ref{inst:0002}}
\and J.        ~Zorec                         \inst{\ref{inst:0467}}
\and S.        ~Zucker                        \orcit{0000-0003-3173-3138}\inst{\ref{inst:0468}}
\and C.        ~Zurbach                       \inst{\ref{inst:0469}}
\and T.        ~Zwitter                       \orcit{0000-0002-2325-8763}\inst{\ref{inst:0470}}
}
\institute{
     Lohrmann Observatory, Technische Universit\"{ a}t Dresden, Mommsenstra{\ss}e 13, 01062 Dresden, Germany\relax                                                                                                                                                                                               \label{inst:0001}
\and Universit\'{e} C\^{o}te d'Azur, Observatoire de la C\^{o}te d'Azur, CNRS, Laboratoire Lagrange, Bd de l'Observatoire, CS 34229, 06304 Nice Cedex 4, France\relax                                                                                                                                            \label{inst:0002}
\and Lund Observatory, Department of Astronomy and Theoretical Physics, Lund University, Box 43, 22100 Lund, Sweden\relax                                                                                                                                                                                        \label{inst:0003}
\and Astronomisches Rechen-Institut, Zentrum f\"{ u}r Astronomie der Universit\"{ a}t Heidelberg, M\"{ o}nchhofstr. 12-14, 69120 Heidelberg, Germany\relax                                                                                                                                                       \label{inst:0004}
\and European Space Agency (ESA), European Space Astronomy Centre (ESAC), Camino bajo del Castillo, s/n, Urbanizacion Villafranca del Castillo, Villanueva de la Ca\~{n}ada, 28692 Madrid, Spain\relax                                                                                                           \label{inst:0006}
\and Vitrociset Belgium for European Space Agency (ESA), Camino bajo del Castillo, s/n, Urbanizacion Villafranca del Castillo, Villanueva de la Ca\~{n}ada, 28692 Madrid, Spain\relax                                                                                                                            \label{inst:0008}
\and HE Space Operations BV for European Space Agency (ESA), Camino bajo del Castillo, s/n, Urbanizacion Villafranca del Castillo, Villanueva de la Ca\~{n}ada, 28692 Madrid, Spain\relax                                                                                                                        \label{inst:0010}
\and Telespazio Vega UK Ltd for European Space Agency (ESA), Camino bajo del Castillo, s/n, Urbanizacion Villafranca del Castillo, Villanueva de la Ca\~{n}ada, 28692 Madrid, Spain\relax                                                                                                                        \label{inst:0017}
\and Leiden Observatory, Leiden University, Niels Bohrweg 2, 2333 CA Leiden, The Netherlands\relax                                                                                                                                                                                                               \label{inst:0018}
\and INAF - Osservatorio astronomico di Padova, Vicolo Osservatorio 5, 35122 Padova, Italy\relax                                                                                                                                                                                                                 \label{inst:0019}
\and European Space Agency (ESA), European Space Research and Technology Centre (ESTEC), Keplerlaan 1, 2201AZ, Noordwijk, The Netherlands\relax                                                                                                                                                                  \label{inst:0020}
\and Univ. Grenoble Alpes, CNRS, IPAG, 38000 Grenoble, France\relax                                                                                                                                                                                                                                              \label{inst:0022}
\and GEPI, Observatoire de Paris, Universit\'{e} PSL, CNRS, 5 Place Jules Janssen, 92190 Meudon, France\relax                                                                                                                                                                                                    \label{inst:0023}
\and Institute of Astronomy, University of Cambridge, Madingley Road, Cambridge CB3 0HA, United Kingdom\relax                                                                                                                                                                                                    \label{inst:0025}
\and Department of Astronomy, University of Geneva, Chemin des Maillettes 51, 1290 Versoix, Switzerland\relax                                                                                                                                                                                                    \label{inst:0026}
\and Aurora Technology for European Space Agency (ESA), Camino bajo del Castillo, s/n, Urbanizacion Villafranca del Castillo, Villanueva de la Ca\~{n}ada, 28692 Madrid, Spain\relax                                                                                                                             \label{inst:0027}
\and Institut de Ci\`{e}ncies del Cosmos (ICCUB), Universitat  de  Barcelona  (IEEC-UB), Mart\'{i} i  Franqu\`{e}s  1, 08028 Barcelona, Spain\relax                                                                                                                                                              \label{inst:0029}
\and CNES Centre Spatial de Toulouse, 18 avenue Edouard Belin, 31401 Toulouse Cedex 9, France\relax                                                                                                                                                                                                              \label{inst:0031}
\and Institut d'Astronomie et d'Astrophysique, Universit\'{e} Libre de Bruxelles CP 226, Boulevard du Triomphe, 1050 Brussels, Belgium\relax                                                                                                                                                                     \label{inst:0032}
\and F.R.S.-FNRS, Rue d'Egmont 5, 1000 Brussels, Belgium\relax                                                                                                                                                                                                                                                   \label{inst:0033}
\and INAF - Osservatorio Astrofisico di Arcetri, Largo Enrico Fermi 5, 50125 Firenze, Italy\relax                                                                                                                                                                                                                \label{inst:0034}
\and Laboratoire d'astrophysique de Bordeaux, Univ. Bordeaux, CNRS, B18N, all{\'e}e Geoffroy Saint-Hilaire, 33615 Pessac, France\relax                                                                                                                                                                           \label{inst:0036}
\and Max Planck Institute for Astronomy, K\"{ o}nigstuhl 17, 69117 Heidelberg, Germany\relax                                                                                                                                                                                                                     \label{inst:0039}
\and Mullard Space Science Laboratory, University College London, Holmbury St Mary, Dorking, Surrey RH5 6NT, United Kingdom\relax                                                                                                                                                                                \label{inst:0040}
\and INAF - Osservatorio Astrofisico di Torino, via Osservatorio 20, 10025 Pino Torinese (TO), Italy\relax                                                                                                                                                                                                       \label{inst:0041}
\and University of Turin, Department of Physics, Via Pietro Giuria 1, 10125 Torino, Italy\relax                                                                                                                                                                                                                  \label{inst:0044}
\and DAPCOM for Institut de Ci\`{e}ncies del Cosmos (ICCUB), Universitat  de  Barcelona  (IEEC-UB), Mart\'{i} i  Franqu\`{e}s  1, 08028 Barcelona, Spain\relax                                                                                                                                                   \label{inst:0047}
\vfill\break\and Royal Observatory of Belgium, Ringlaan 3, 1180 Brussels, Belgium\relax                                                                                                                                                                                                                          \label{inst:0052}
\and ALTEC S.p.a, Corso Marche, 79,10146 Torino, Italy\relax                                                                                                                                                                                                                                                     \label{inst:0058}
\and Department of Astronomy, University of Geneva, Chemin d'Ecogia 16, 1290 Versoix, Switzerland\relax                                                                                                                                                                                                          \label{inst:0061}
\and Sednai S\`{a}rl, Geneva, Switzerland\relax                                                                                                                                                                                                                                                                  \label{inst:0062}
\and Gaia DPAC Project Office, ESAC, Camino bajo del Castillo, s/n, Urbanizacion Villafranca del Castillo, Villanueva de la Ca\~{n}ada, 28692 Madrid, Spain\relax                                                                                                                                                \label{inst:0071}
\and SYRTE, Observatoire de Paris, Universit\'{e} PSL, CNRS,  Sorbonne Universit\'{e}, LNE, 61 avenue de l’Observatoire 75014 Paris, France\relax                                                                                                                                                              \label{inst:0076}
\and National Observatory of Athens, I. Metaxa and Vas. Pavlou, Palaia Penteli, 15236 Athens, Greece\relax                                                                                                                                                                                                       \label{inst:0078}
\and IMCCE, Observatoire de Paris, Universit\'{e} PSL, CNRS, Sorbonne Universit{\'e}, Univ. Lille, 77 av. Denfert-Rochereau, 75014 Paris, France\relax                                                                                                                                                           \label{inst:0080}
\and INAF - Osservatorio Astrofisico di Catania, via S. Sofia 78, 95123 Catania, Italy\relax                                                                                                                                                                                                                     \label{inst:0082}
\and Serco Gesti\'{o}n de Negocios for European Space Agency (ESA), Camino bajo del Castillo, s/n, Urbanizacion Villafranca del Castillo, Villanueva de la Ca\~{n}ada, 28692 Madrid, Spain\relax                                                                                                                 \label{inst:0087}
\and INAF - Osservatorio di Astrofisica e Scienza dello Spazio di Bologna, via Piero Gobetti 93/3, 40129 Bologna, Italy\relax                                                                                                                                                                                    \label{inst:0088}
\and Institut d'Astrophysique et de G\'{e}ophysique, Universit\'{e} de Li\`{e}ge, 19c, All\'{e}e du 6 Ao\^{u}t, B-4000 Li\`{e}ge, Belgium\relax                                                                                                                                                                  \label{inst:0089}
\and CRAAG - Centre de Recherche en Astronomie, Astrophysique et G\'{e}ophysique, Route de l'Observatoire Bp 63 Bouzareah 16340 Algiers, Algeria\relax                                                                                                                                                           \label{inst:0090}
\and Institute for Astronomy, University of Edinburgh, Royal Observatory, Blackford Hill, Edinburgh EH9 3HJ, United Kingdom\relax                                                                                                                                                                                \label{inst:0091}
\and ATG Europe for European Space Agency (ESA), Camino bajo del Castillo, s/n, Urbanizacion Villafranca del Castillo, Villanueva de la Ca\~{n}ada, 28692 Madrid, Spain\relax                                                                                                                                    \label{inst:0094}
\and ETSE Telecomunicaci\'{o}n, Universidade de Vigo, Campus Lagoas-Marcosende, 36310 Vigo, Galicia, Spain\relax                                                                                                                                                                                                 \label{inst:0099}
\and Universit\'{e} de Strasbourg, CNRS, Observatoire astronomique de Strasbourg, UMR 7550,  11 rue de l'Universit\'{e}, 67000 Strasbourg, France\relax                                                                                                                                                          \label{inst:0103}
\and Kavli Institute for Cosmology Cambridge, Institute of Astronomy, Madingley Road, Cambridge, CB3 0HA\relax                                                                                                                                                                                                   \label{inst:0106}
\and Department of Astrophysics, Astronomy and Mechanics, National and Kapodistrian University of Athens, Panepistimiopolis, Zografos, 15783 Athens, Greece\relax                                                                                                                                                \label{inst:0107}
\and Observational Astrophysics, Division of Astronomy and Space Physics, Department of Physics and Astronomy, Uppsala University, Box 516, 751 20 Uppsala, Sweden\relax                                                                                                                                         \label{inst:0108}
\and Leibniz Institute for Astrophysics Potsdam (AIP), An der Sternwarte 16, 14482 Potsdam, Germany\relax                                                                                                                                                                                                        \label{inst:0113}
\and CENTRA, Faculdade de Ci\^{e}ncias, Universidade de Lisboa, Edif. C8, Campo Grande, 1749-016 Lisboa, Portugal\relax                                                                                                                                                                                          \label{inst:0116}
\and Department of Informatics, Donald Bren School of Information and Computer Sciences, University of California, 5019 Donald Bren Hall, 92697-3440 CA Irvine, United States\relax                                                                                                                              \label{inst:0117}
\and Dipartimento di Fisica e Astronomia ""Ettore Majorana"", Universit\`{a} di Catania, Via S. Sofia 64, 95123 Catania, Italy\relax                                                                                                                                                                             \label{inst:0119}
\and CITIC, Department of Nautical Sciences and Marine Engineering, University of A Coru\~{n}a, Campus de Elvi\~{n}a S/N, 15071, A Coru\~{n}a, Spain\relax                                                                                                                                                       \label{inst:0122}
\vfill\break\and INAF - Osservatorio Astronomico di Roma, Via Frascati 33, 00078 Monte Porzio Catone (Roma), Italy\relax                                                                                                                                                                                                     \label{inst:0124}
\and Space Science Data Center - ASI, Via del Politecnico SNC, 00133 Roma, Italy\relax                                                                                                                                                                                                               \label{inst:0125}
\and Department of Physics, University of Helsinki, P.O. Box 64, 00014 Helsinki, Finland\relax                                                                                                                                                                                                                   \label{inst:0128}
\and Finnish Geospatial Research Institute FGI, Geodeetinrinne 2, 02430 Masala, Finland\relax                                                                                                                                                                                                                    \label{inst:0129}
\and STFC, Rutherford Appleton Laboratory, Harwell, Didcot, OX11 0QX, United Kingdom\relax                                                                                                                                                                                                                       \label{inst:0135}
\and Institut UTINAM CNRS UMR6213, Universit\'{e} Bourgogne Franche-Comt\'{e}, OSU THETA Franche-Comt\'{e} Bourgogne, Observatoire de Besan\c{c}on, BP1615, 25010 Besan\c{c}on Cedex, France\relax                                                                                                               \label{inst:0138}
\and HE Space Operations BV for European Space Agency (ESA), Keplerlaan 1, 2201AZ, Noordwijk, The Netherlands\relax                                                                                                                                                                                              \label{inst:0139}
\and Dpto. de Inteligencia Artificial, UNED, c/ Juan del Rosal 16, 28040 Madrid, Spain\relax                                                                                                                                                                                                                     \label{inst:0141}
\and Applied Physics Department, Universidade de Vigo, 36310 Vigo, Spain\relax                                                                                                                                                                                                                                   \label{inst:0145}
\and Thales Services for CNES Centre Spatial de Toulouse, 18 avenue Edouard Belin, 31401 Toulouse Cedex 9, France\relax                                                                                                                                                                                          \label{inst:0151}
\and Instituut voor Sterrenkunde, KU Leuven, Celestijnenlaan 200D, 3001 Leuven, Belgium\relax                                                                                                                                                                                                                    \label{inst:0152}
\and Department of Astrophysics/IMAPP, Radboud University, P.O.Box 9010, 6500 GL Nijmegen, The Netherlands\relax                                                                                                                                                                                                 \label{inst:0153}
\and CITIC - Department of Computer Science and Information Technologies, University of A Coru\~{n}a, Campus de Elvi\~{n}a S/N, 15071, A Coru\~{n}a, Spain\relax                                                                                                                                                 \label{inst:0159}
\and Barcelona Supercomputing Center (BSC) - Centro Nacional de Supercomputaci\'{o}n, c/ Jordi Girona 29, Ed. Nexus II, 08034 Barcelona, Spain\relax                                                                                                                                                             \label{inst:0160}
\and University of Vienna, Department of Astrophysics, T\"{ u}rkenschanzstra{\ss}e 17, A1180 Vienna, Austria\relax                                                                                                                                                                                               \label{inst:0161}
\and European Southern Observatory, Karl-Schwarzschild-Str. 2, 85748 Garching, Germany\relax                                                                                                                                                                                                                     \label{inst:0162}
\and Kapteyn Astronomical Institute, University of Groningen, Landleven 12, 9747 AD Groningen, The Netherlands\relax                                                                                                                                                                                             \label{inst:0169}
\and School of Physics and Astronomy, University of Leicester, University Road, Leicester LE1 7RH, United Kingdom\relax                                                                                                                                                                                          \label{inst:0176}
\and Center for Research and Exploration in Space Science and Technology, University of Maryland Baltimore County, 1000 Hilltop Circle, Baltimore MD, USA\relax                                                                                                                                                  \label{inst:0184}
\and GSFC - Goddard Space Flight Center, Code 698, 8800 Greenbelt Rd, 20771 MD Greenbelt, United States\relax                                                                                                                                                                                                    \label{inst:0185}
\and EURIX S.r.l., Corso Vittorio Emanuele II 61, 10128, Torino, Italy\relax                                                                                                                                                                                                                                     \label{inst:0187}
\and Harvard-Smithsonian Center for Astrophysics, 60 Garden St., MS 15, Cambridge, MA 02138, USA\relax                                                                                                                                                                                                           \label{inst:0188}
\and CAUP - Centro de Astrofisica da Universidade do Porto, Rua das Estrelas, Porto, Portugal\relax                                                                                                                                                                                                              \label{inst:0190}
\and SISSA - Scuola Internazionale Superiore di Studi Avanzati, via Bonomea 265, 34136 Trieste, Italy\relax                                                                                                                                                                                                      \label{inst:0195}
\and Telespazio for CNES Centre Spatial de Toulouse, 18 avenue Edouard Belin, 31401 Toulouse Cedex 9, France\relax                                                                                                                                                                                               \label{inst:0198}
\and University of Turin, Department of Computer Sciences, Corso Svizzera 185, 10149 Torino, Italy\relax                                                                                                                                                                                                         \label{inst:0203}
\and Dpto. de Matem\'{a}tica Aplicada y Ciencias de la Computaci\'{o}n, Univ. de Cantabria, ETS Ingenieros de Caminos, Canales y Puertos, Avda. de los Castros s/n, 39005 Santander, Spain\relax                                                                                                                 \label{inst:0206}
\and Centro de Astronom\'{i}a - CITEVA, Universidad de Antofagasta, Avenida Angamos 601, Antofagasta 1270300, Chile\relax                                                                                                                                                                                        \label{inst:0216}
\and Vera C Rubin Observatory,  950 N. Cherry Avenue, Tucson, AZ 85719, USA\relax                                                                                                                                                                                                                                \label{inst:0219}
\vfill\break\and Centre for Astrophysics Research, University of Hertfordshire, College Lane, AL10 9AB, Hatfield, United Kingdom\relax                                                                                                                                                                                       \label{inst:0220}
\and University of Antwerp, Onderzoeksgroep Toegepaste Wiskunde, Middelheimlaan 1, 2020 Antwerp, Belgium\relax                                                                                                                                                                                                   \label{inst:0229}
\and INAF - Osservatorio Astronomico d'Abruzzo, Via Mentore Maggini, 64100 Teramo, Italy\relax                                                                                                                                                                                                                   \label{inst:0232}
\and Instituto de Astronomia, Geof\`{i}sica e Ci\^{e}ncias Atmosf\'{e}ricas, Universidade de S\~{a}o Paulo, Rua do Mat\~{a}o, 1226, Cidade Universitaria, 05508-900 S\~{a}o Paulo, SP, Brazil\relax                                                                                                              \label{inst:0235}
\and M\'{e}socentre de calcul de Franche-Comt\'{e}, Universit\'{e} de Franche-Comt\'{e}, 16 route de Gray, 25030 Besan\c{c}on Cedex, France\relax                                                                                                                                                                \label{inst:0243}
\and SRON, Netherlands Institute for Space Research, Sorbonnelaan 2, 3584CA, Utrecht, The Netherlands\relax                                                                                                                                                                                                      \label{inst:0247}
\and RHEA for European Space Agency (ESA), Camino bajo del Castillo, s/n, Urbanizacion Villafranca del Castillo, Villanueva de la Ca\~{n}ada, 28692 Madrid, Spain\relax                                                                                                                                          \label{inst:0251}
\and ATOS for CNES Centre Spatial de Toulouse, 18 avenue Edouard Belin, 31401 Toulouse Cedex 9, France\relax                                                                                                                                                                                                     \label{inst:0252}
\and School of Physics and Astronomy, Tel Aviv University, Tel Aviv 6997801, Israel\relax                                                                                                                                                                                                                        \label{inst:0255} %
\and Astrophysics Research Centre, School of Mathematics and Physics, Queen's University Belfast, Belfast BT7 1NN, UK\relax                                                                                                                                                                                      \label{inst:0257}
\and Centre de Donn\'{e}es Astronomique de Strasbourg, Strasbourg, France\relax                                                                                                                                                                                                                                  \label{inst:0259}
\and Universit\'{e} C\^{o}te d'Azur, Observatoire de la C\^{o}te d'Azur, CNRS, Laboratoire G\'{e}oazur, Bd de l'Observatoire, CS 34229, 06304 Nice Cedex 4, France\relax                                                                                                                                         \label{inst:0260}
\and Max-Planck-Institut f\"{ u}r Astrophysik, Karl-Schwarzschild-Stra{\ss}e 1, 85748 Garching, Germany\relax                                                                                                                                                                                                    \label{inst:0264}
\and APAVE SUDEUROPE SAS for CNES Centre Spatial de Toulouse, 18 avenue Edouard Belin, 31401 Toulouse Cedex 9, France\relax                                                                                                                                                                                      \label{inst:0266}
\and \'{A}rea de Lenguajes y Sistemas Inform\'{a}ticos, Universidad Pablo de Olavide, Ctra. de Utrera, km 1. 41013, Sevilla, Spain\relax                                                                                                                                                                         \label{inst:0270}
\and Onboard Space Systems, Lule\aa{} University of Technology, Box 848, S-981 28 Kiruna, Sweden\relax                                                                                                                                                                                                           \label{inst:0281}
\and TRUMPF Photonic Components GmbH, Lise-Meitner-Stra{\ss}e 13,  89081 Ulm, Germany\relax                                                                                                                                                                                                                      \label{inst:0286}
\and IAC - Instituto de Astrofisica de Canarias, Via L\'{a}ctea s/n, 38200 La Laguna S.C., Tenerife, Spain\relax                                                                                                                                                                                                 \label{inst:0289}
\and Department of Astrophysics, University of La Laguna, Via L\'{a}ctea s/n, 38200 La Laguna S.C., Tenerife, Spain\relax                                                                                                                                                                                        \label{inst:0290}
\and Laboratoire Univers et Particules de Montpellier, CNRS Universit\'{e} Montpellier, Place Eug\`{e}ne Bataillon, CC72, 34095 Montpellier Cedex 05, France\relax                                                                                                                                               \label{inst:0294}
\and LESIA, Observatoire de Paris, Universit\'{e} PSL, CNRS, Sorbonne Universit\'{e}, Universit\'{e} de Paris, 5 Place Jules Janssen, 92190 Meudon, France\relax                                                                                                                                                 \label{inst:0300}
\and Villanova University, Department of Astrophysics and Planetary Science, 800 E Lancaster Avenue, Villanova PA 19085, USA\relax                                                                                                                                                                               \label{inst:0302}
\and Astronomical Observatory, University of Warsaw,  Al. Ujazdowskie 4, 00-478 Warszawa, Poland\relax                                                                                                                                                                                                           \label{inst:0308}
\and Laboratoire d'astrophysique de Bordeaux, Univ. Bordeaux, CNRS, B18N, all\'{e}e Geoffroy Saint-Hilaire, 33615 Pessac, France\relax                                                                                                                                                                           \label{inst:0312}
\and Universit\'{e} Rennes, CNRS, IPR (Institut de Physique de Rennes) - UMR 6251, 35000 Rennes, France\relax                                                                                                                                                                                                    \label{inst:0315}
\and INAF - Osservatorio Astronomico di Capodimonte, Via Moiariello 16, 80131, Napoli, Italy\relax                                                                                                                                                                                                               \label{inst:0317}
\vfill\break\and Niels Bohr Institute, University of Copenhagen, Juliane Maries Vej 30, 2100 Copenhagen {\O}, Denmark\relax                                                                                                                                                                                                  \label{inst:0323}
\and Las Cumbres Observatory, 6740 Cortona Drive Suite 102, Goleta, CA 93117, USA\relax                                                                                                                                                                                                                          \label{inst:0324}
\and Astrophysics Research Institute, Liverpool John Moores University, 146 Brownlow Hill, Liverpool L3 5RF, United Kingdom\relax                                                                                                                                                                                \label{inst:0330}
\and IPAC, Mail Code 100-22, California Institute of Technology, 1200 E. California Blvd., Pasadena, CA 91125, USA\relax                                                                                                                                                                                         \label{inst:0335}
\and Jet Propulsion Laboratory, California Institute of Technology, 4800 Oak Grove Drive, M/S 169-327, Pasadena, CA 91109, USA\relax                                                                                                                                                                             \label{inst:0336}
\and IRAP, Universit\'{e} de Toulouse, CNRS, UPS, CNES, 9 Av. colonel Roche, BP 44346, 31028 Toulouse Cedex 4, France\relax                                                                                                                                                                                      \label{inst:0337}
\and Konkoly Observatory, Research Centre for Astronomy and Earth Sciences, MTA Centre of Excellence, Konkoly Thege Mikl\'{o}s \'{u}t 15-17, 1121 Budapest, Hungary\relax                                                                                                                                        \label{inst:0350}
\and MTA CSFK Lend\"{ u}let Near-Field Cosmology Research Group, Konkoly Observatory, CSFK, Konkoly Thege Mikl\'os \'ut 15-17, 1121 Budapest, Hungary\relax                                                                                                                                                                                                                                            \label{inst:0351}
\and ELTE E\"{ o}tv\"{ o}s Lor\'{a}nd University, Institute of Physics, 1117, P\'{a}zm\'{a}ny P\'{e}ter s\'{e}t\'{a}ny 1A, Budapest, Hungary\relax                                                                                                                                                               \label{inst:0352}
\and Ru{\dj}er Bo\v{s}kovi\'{c} Institute, Bijeni\v{c}ka cesta 54, 10000 Zagreb, Croatia\relax                                                                                                                                                                                                                   \label{inst:0369}
\and Institute of Theoretical Physics, Faculty of Mathematics and Physics, Charles University in Prague, Czech Republic\relax                                                                                                                                                                                    \label{inst:0373}
\and INAF - Osservatorio Astronomico di Brera, via E. Bianchi 46, 23807 Merate (LC), Italy\relax                                                                                                                                                                                                                 \label{inst:0384}
\and AKKA for CNES Centre Spatial de Toulouse, 18 avenue Edouard Belin, 31401 Toulouse Cedex 9, France\relax                                                                                                                                                                                                     \label{inst:0385}
\and Departmento de F\'{i}sica de la Tierra y Astrof\'{i}sica, Universidad Complutense de Madrid, 28040 Madrid, Spain\relax                                                                                                                                                                                      \label{inst:0389}
\and Department of Astrophysical Sciences, 4 Ivy Lane, Princeton University, Princeton NJ 08544, USA\relax                                                                                                                                                                                                       \label{inst:0425}
\and Departamento de Astrof\'{i}sica, Centro de Astrobiolog\'{i}a (CSIC-INTA), ESA-ESAC. Camino Bajo del Castillo s/n. 28692 Villanueva de la Ca\~{n}ada, Madrid, Spain\relax                                                                                                                                    \label{inst:0431}
\and naXys, University of Namur, Rempart de la Vierge, 5000 Namur, Belgium\relax                                                                                                                                                                                                                                 \label{inst:0434}
\and EPFL - Ecole Polytechnique f\'{e}d\'{e}rale de Lausanne, Institute of Mathematics, Station 8 EPFL SB MATH SDS, Lausanne, Switzerland\relax                                                                                                                                                                  \label{inst:0440}
\and H H Wills Physics Laboratory, University of Bristol, Tyndall Avenue, Bristol BS8 1TL, United Kingdom\relax                                                                                                                                                                                                  \label{inst:0446}
\and Sorbonne Universit\'{e}, CNRS, UMR7095, Institut d'Astrophysique de Paris, 98bis bd. Arago, 75014 Paris, France\relax                                                                                                                                                                                       \label{inst:0467}
\and Porter School of the Environment and Earth Sciences, Tel Aviv University, Tel Aviv 6997801, Israel\relax                                                                                                                                                                                                    \label{inst:0468}
\and Laboratoire Univers et Particules de Montpellier, Universit\'{e} Montpellier, Place Eug\`{e}ne Bataillon, CC72, 34095 Montpellier Cedex 05, France\relax                                                                                                                                                    \label{inst:0469}
\and Faculty of Mathematics and Physics, University of Ljubljana, Jadranska ulica 19, 1000 Ljubljana, Slovenia\relax                                                                                                                                                                                             \label{inst:0470}
}

   \date{ }

 
\abstract
    {
    \textit{Gaia} Early Data Release 3 (\textit{Gaia} EDR3) provides
    accurate astrometry for about 1.6~million compact (QSO-like)
    extragalactic sources, 1.2~million of which have the best-quality five-parameter astrometric solutions.
    }
    {
    The proper motions of QSO-like sources
    are used to reveal a systematic pattern due to the acceleration of the solar
    system barycentre with respect to the rest frame of the Universe.
    Apart from being an important scientific result
    by itself, the acceleration measured in this way is
    a good quality indicator of the \textit{Gaia} astrometric
    solution.
    }
    {
    The effect of the acceleration is
    obtained as a part of the general expansion of the vector field of
    proper motions in Vector Spherical Harmonics (VSH). Various
    versions of the VSH fit and various subsets of the sources are
    tried and compared to get the most consistent result and a realistic estimate of its
    uncertainty. Additional tests with the \textit{Gaia}
    astrometric solution are used to get a better idea on possible
    systematic errors in the estimate.
    }
    {
    Our best estimate of the acceleration based on \textit{Gaia} EDR3 is 
    $(2.32\pm0.16)\times10^{-10}\mssquared$ (or $7.33\pm0.51 \kmsMyr$) 
    towards $\alpha=269.1\degr\pm5.4\degr$,
    $\delta=-31.6\degr\pm4.1\degr$,
    corresponding to a proper motion amplitude of 
    $5.05\pm0.35$~\muasyr. 
    This is in good agreement with the acceleration expected from current models
    of the Galactic gravitational potential. 
    We expect that future \textit{Gaia} data releases will provide estimates of
    the acceleration with uncertainties substantially below 0.1~$\mu$as\,yr$^{-1}$.
    }
  {}


   \titlerunning{{\textit{Gaia}} Early Data Release 3 -- Acceleration of the solar system} 
   \authorrunning{Gaia Collaboration et al.}

   \maketitle

%

\section{Introduction} 
\label{sec:intro}

It is well known that the velocity of an observer causes the apparent 
positions of all celestial bodies to be displaced in the direction of the 
velocity, an effect referred to as the aberration of light.
If the velocity is changing with time, that is if the observer is accelerated,
the displacements are also changing, giving the impression of a pattern of 
proper motions in the direction of the acceleration. We exploit this effect 
to detect the imprint in the \gaia\ data of the 
acceleration of the solar system with respect to the rest-frame of
remote extragalactic sources.

\subsection{Historical considerations}
\label{sec:intro:historical}
In 1833 John Pond, the Astronomer Royal at that time, sent to print
the \textsl{Catalogue of 1112 stars, reduced from observations made at
  the Royal Observatory at Greenwich} \citepads{1833RGAO...18P...1P}, the
happy conclusion of a standard and tedious observatory work, and a
catalogue much praised for its accuracy
\citepads{1852hopa.book.....G}. At the end of his short introduction he
added a note discussing \textsl{Causes of Disturbance of the proper
  Motion of Stars}, in which he considered the secular aberration resulting
from the motion of the solar system in free space, stating that,
\begin{quotation}
\textsl{So long as the motion of the Sun continues uniform
  and rectilinear, this aberration or distortion from their true places
  will be constant: it will not affect our observations; nor am I
  aware that we possess any means of determining whether it exist or
  not. If the motion of the Sun be uniformly accelerated, or uniformly retarded,
  $[\ldots]$ [t]he effects of either of these suppositions would be, to
  produce uniform motion in every star according to its 
  position, and might in time be discoverable by our observations, if
  the stars had no proper motions of their own $[\ldots]$ But it is
  needless to enter into further speculation on questions that appear
  at present not likely to lead to the least practical utility, though it
  may become a subject of interest to future ages.}
\end{quotation}
This was a simple, but clever, realisation of the consequences of aberration, really
new at that time and totally outside the technical capabilities of the
time. The idea gained more visibility through the successful textbooks
of the renowned English astronomer John Herschel, first in his
\textsl{Treatise of Astronomy} (\citeads{1833tras.book.....H}, \S612) and later
in the expanded version \textsl{Outlines of Astronomy}
(\citeads{1849oast.book.....H}, \S862), both of which went through numerous editions. In
the former he referred directly to John Pond as the original source of
this `\textsl{very ingenious idea}', whereas in the latter the reference to Pond was 
dropped and the description of the effect looks unpromising:
\begin{quotation}
\textsl{
  This displacement, however, is permanent, and therefore
  unrecognizable by any ph{\ae}nomenon, so long as the solar motion
  remains invariable ; but should it, in the course of ages, alter its
  direction and velocity, both the direction and amount of the
  displacement in question would alter with it. The change, however,
  would become mixed up with other changes in the apparent proper
  motions of the stars, and it would seem hopeless to attempt
  disentangling them.}
\end{quotation}
 John Pond in 1833 wrote that the idea came to him `\textsl{many years
   ago}' but did not hint at borrowing it from someone else. 
 For such an idea to emerge, at least three devices had to be present 
 in the tool kit of a practising astronomer: a deep
 understanding of aberration, well known since James Bradley's discovery in 1728;
 the secure proof that stars have proper motion, provided by the
 Catalogue of Tobias Mayer in 1760; and the notion of the secular motion of the Sun
 towards the apex, established by William Herschel in 1783. 
 Therefore Pond was probably the first, to our knowledge, who
 combined the aberration and the free motion of the Sun among the
 stars to draw the important observable consequence in terms of
 systematic proper motions. We have found no earlier mention, and had it
 been commonly known by astronomers much earlier we would have 
 found a mention in
 Lalande's \textsl{Astronomie} \citep{Lalande1792}, the most
 encyclopaedic treatise on the subject at the time.

References to the constant aberration due to the secular motion of
the solar system as a whole appear over the course of years in some
astronomical textbooks (e.g. \citeads{1908tsa..book.....B}), but 
not in all with the hint that only a change in the apex would make it
visible in the form of a proper motion. While the bold foresight of
these forerunners was by necessity limited by their conception of
the Milky Way and the Universe as a whole, both Pond and Herschel 
recognised that even with a curved motion of the solar
system, the effect on the stars from the change in aberration would be
very difficult to separate from other sources of 
proper motion. This would remain true today if the stars
of the Milky Way had been our only means to study the effect.

However, our current view of the hierarchical structure of the Universe
puts the issue in a different and more favourable guise. The whole
solar system is in motion within the Milky Way and there are
star-like sources, very far away from us, that do not share this motion. For
them the only source of apparent proper motion could be precisely that
resulting from the change in the secular aberration. We are happily
back to the world without proper motions contemplated by Pond, and
we show in this paper that \textit{Gaia}'s observations of extragalactic sources enable
us to discern, for the first time in the optical domain, the signature of this 
systematic proper motion.

\subsection{Recent works}

Coming to the modern era, the earliest mention we have found of the
effect on extragalactic sources is by \citetads{1983jpl..rept.8339F} in
the description of the JPL software package MASTERFIT for reducing 
Very Long Baseline Interferometry (VLBI)
observations. There is a passing remark that the change in the
apparent position of the sources from the solar system motion would be
that of a proper motion of 6~{\muasyr}, nearly two orders of magnitude
smaller than the effect of source structure, but systematic. There is
no detailed modelling of the effect, but at least this was clearly
shown to be a consequence of the change in the direction of the solar system velocity vector in
the aberration factor, worthy of further consideration.
The description of the effect is given in later descriptions of MASTERFIT and also in some other publications
of the JPL VLBI group (e.g. \citeads{1996jpl..rept.8339S}; \citeads{1998RvMP...70.1393S}).

\citetads{1995IAUS..166..283E} have a contribution in IAU Symposium 166
with the title \textsl{Secular motions of the extragalactic
  radio-sources and the stability of the radio
reference frame}. This contains the first claim of seeing statistically significant
proper motions in many sources at the level of 30~{\muasyr}, about an
order of magnitude 
larger than expected. This was unfortunately
limited to an abstract, but the idea behind was to search for
the effect discussed here. Proper motions of quasars were also
investigated by \citetads{1997ApJ...485...87G} in the context of search
for low-frequency gravitational waves. The technique relied heavily on a
decomposition on VSH (Vector Spherical Harmonics), very similar to
what is reported in the core of this paper.

\citetads{1995ESASP.379...99B} rediscovered the effect in the context of the
\gaia\ mission as it was planned at the time. He describes the effect
as a variable aberration and stated clearly how it could be measured
with {\gaia} using 60 bright quasars, with the unambiguous
conclusion that `it seems quite possible that GAIA can
  significantly measure the galactocentric acceleration of the solar
  system'. This was then included as an important science objective of
{\gaia} in the mission proposal submitted to ESA in 2000 and in most
early presentations of the mission and its expected science results
(\citeads{2001A&A...369..339P}; \citeads{2002EAS.....2..327M}).  Several theoretical
discussions followed in relation to VLBI or space astrometry
(\citeads{1998RvMP...70.1393S}; \citeads{2006AJ....131.1471K}). \citetads{2003A&A...404..743K}
considered the effect on the observed motions of stars in our Galaxy, while
\citetads{2012A&A...547A..59M} showed how the systematic use of the VSH
on a large data sample like \textit{Gaia} would permit a blind search of
the acceleration without ad~hoc model fitting. They also stressed
the importance of solving simultaneously for the acceleration and the
spin to avoid signal leakage from correlations.
 
With the VLBI data gradually covering longer periods of
time, detection of the systematic patterns in the proper motions of quasars 
became a definite possibility, and in the last decade there have been several works
claiming positive detections at different levels of significance.
But even with 20 years of data, the systematic displacement of the
best-placed quasars is only $\simeq 0.1$ mas, not much larger than
the noise floor of individual VLBI positions until very
recently. So the actual detection was, and remains, challenging.

The first published solution by \citetads{1997ApJ...485...87G}, based on
323 sources, resulted in an acceleration estimate of 
$(g_x, g_y, g_z) = (1.9 \pm 6.1,~5.4\pm 6.2,~7.5\pm 5.6)~\muasyr$,
not really above the noise level.%
\footnote{Here, and in the following, the acceleration is expressed as
a proper motion through division by $c$, the speed of light; see 
Eq.~(\ref{eq:accel_components}). $(g_x, g_y, g_z)$ are the
components of the effect in the ICRS (equatorial) system.} 
Then a first detection
claim was by \citetads{2011A&A...529A..91T}, using 555 sources and 20~years
of VLBI data. From the proper motions of these sources they found
$|\vec{g}| = g = 6.4 \pm 1.5$~{\muasyr} for the amplitude of the systematic
signal, compatible with the expected magnitude and direction. Two
years later they published an improved solution from 34~years of VLBI
data, yielding $g = 6.4 \pm 1.1$~{\muasyr}
(\citeads{2013A&A...559A..95T}). A new solution by
\citetads{2018A&A...610A..36T} with a global fit of the dipole on more
than 4000 sources and 36~years of VLBI delays yielded $g = 5.2 \pm
0.2$~{\muasyr}, the best formal error so far, and a direction a few
degrees off the Galactic centre. 
\citetads{2012A&A...544A.135X} also made a direct fit of the acceleration
vector as a global parameter to the VLBI delay observations, and
found a modulus of $g = 5.82 \pm 0.32$~{\muasyr} but
with a strong component perpendicular to the Galactic plane.

The most recent review by \citetads{2019A&A...630A..93M} is a report
of the Working Group on Galactic Aberration of the International VLBI
Service (IVS). This group was established to incorporate the effect of
the galactocentric aberration into the VLBI analysis with a unique
recommended value. They make a clear distinction between the
galactocentric component that may be estimated from Galactic
kinematics, and the additional contributions due to the accelerated
motion of the Milky Way in the intergalactic space or the peculiar
acceleration of the solar system in the Galaxy. They use the term
`aberration drift' for the total effect. Clearly the observations
cannot separate the different contributions, neither in VLBI nor in
the optical domain with \textit{Gaia}. Based on their considerations,
the working group's recommendation is to use $g=5.8$~{\muasyr} for
the galactocentric component of the aberration drift. This value,
estimated directly in a global solution of the ICRF3 solution data
set, is slightly larger than the value deduced from Galactic
astronomy. This recommendation has been finally adopted in the ICRF3
catalogue, although an additional dedicated
analysis of almost 40 years of VLBI observations gave the acceleration
$g=5.83\pm0.23$~{\muasyr} towards $\alpha = 270.2\degr\pm2.3\degr$,
$\delta = -20.2\degr\pm3.6\degr$ \citep{2020A&A...ICRF3}.

To conclude this overview of related works, a totally different
approach by \citetads{2020ApJ...902L..28C} was recently put forward,
relying on highly accurate spectroscopy. With the performances
of spectrographs reached in the search for extra-solar planets, on the level of 10~cm\,s$^{-1}$, it is
conceivable to detect the variation of the line-of-sight velocity of
stars over a time baseline of at least ten years. This would be a
direct detection of the Galactic acceleration and a way to probe the
gravitational potential at $\sim$ kpc distances. Such a result would
be totally independent of the acceleration derived from the
aberration drift of the extragalactic sources and of great interest. 


Here we report on the first determination of the solar system
acceleration in the optical domain, from \textit{Gaia} observations.
The paper is organised as follows.  Section~\ref{sec:effect}
summarises the astrometric signatures of an acceleration of the solar
system barycentre with respect to the rest frame of extragalactic
sources. Theoretical expectations of the acceleration of the solar
system are presented in Sect.~\ref{sec:expectation}. The selection of
\gaia\ sources for the determination of the effect is
discussed in Sect.~\ref{sec:selection}.  Section~\ref{sec:method}
presents the method, and the analysis of the data and a discussion of
random and systematic errors are given in
Sect.~\ref{sec:analysis}. Conclusions of this study as well as the
perspectives for the future determination with \gaia\ astrometry are
presented in Sect.~\ref{sec:summary}.  In Appendix~\ref{sec:unbiased}
we discuss the general problem of estimating the length of a
vector from the estimates of its Cartesian components.

\section{The astrometric effect of an acceleration}
\label{sec:effect}



In the Introduction we described aberration as an effect changing the
`apparent position' of a source. More accurately, it should be described in terms of 
the `proper direction' to the source: this is the direction 
from which photons are seen to arrive, as measured in a
physically adequate proper reference system of the observer
(see, e.g. \citeads{2004PhRvD..69l4001K}; \citeyearads{2012aamm.book...47K}). The proper direction
which we designate with the unit vector $\vec{u}$, is what an astrometric instrument 
in space ideally measures.

The aberration of light is the displacement $\delta\vec{u}$ obtained
when comparing the proper directions to the same source, as measured
by two co-located observers moving with velocity $\vec{v}$ relative to
each other.  According to the theory of relativity (both special and
general), the proper directions as seen by the two observers are
related by a Lorentz transformation depending on the velocity
$\vec{v}$ of one observer as measured by the other.  If
$\delta\vec{u}$ is relatively large, as for the annual aberration, a
rigorous approach to the computation is needed and also used, for
example in the \textit{Gaia} data processing
\citepads{2003AJ....125.1580K}. Here we are however concerned with
small differential effects, for which first-order formulae
(equivalent to first-order classical aberration) is sufficient.
To first order in $|\vec{v}|/c$, where $c$ is the speed of
light, the aberrational effect is linear in $\vec{v}$,
%
\begin{equation}\label{eq:galaberr}
   \delta\vec{u} = \frac{\vec{v}}{c}-\frac{\vec{v}\cdot\vec{u}}{c}\,\vec{u}\, .
\end{equation}
%
Equation~(\ref{eq:galaberr}) is accurate to ${<\,}0.001~\mu$as for $|\vec{v}|< 0.02\kms$, and
to ${<\,}1\arcsec$ for $|\vec{v}|< 600\kms$ (see, however, below).

If $\vec{v}$ is changing with time, there is a corresponding
time-dependent variation of $\delta\vec{u}$, which affects all sources
on the sky in a particular systematic way.  A familiar example is the
annual aberration, where the apparent positions seen from the Earth
are compared with those of a hypothetical observer at the same
location, but at rest with respect to the solar system barycentre. The
annual variation of $\vec{v}/c$ results in the aberrational effect
that outlines a curve that is close to an ellipse with semi-major axis
about $20\arcsec$ (the curve is not exactly an ellipse since the barycentric orbit
of the Earth is not exactly Keplerian).

\begin{figure}[htbp]
\begin{center}
  \includegraphics[width=1\hsize]{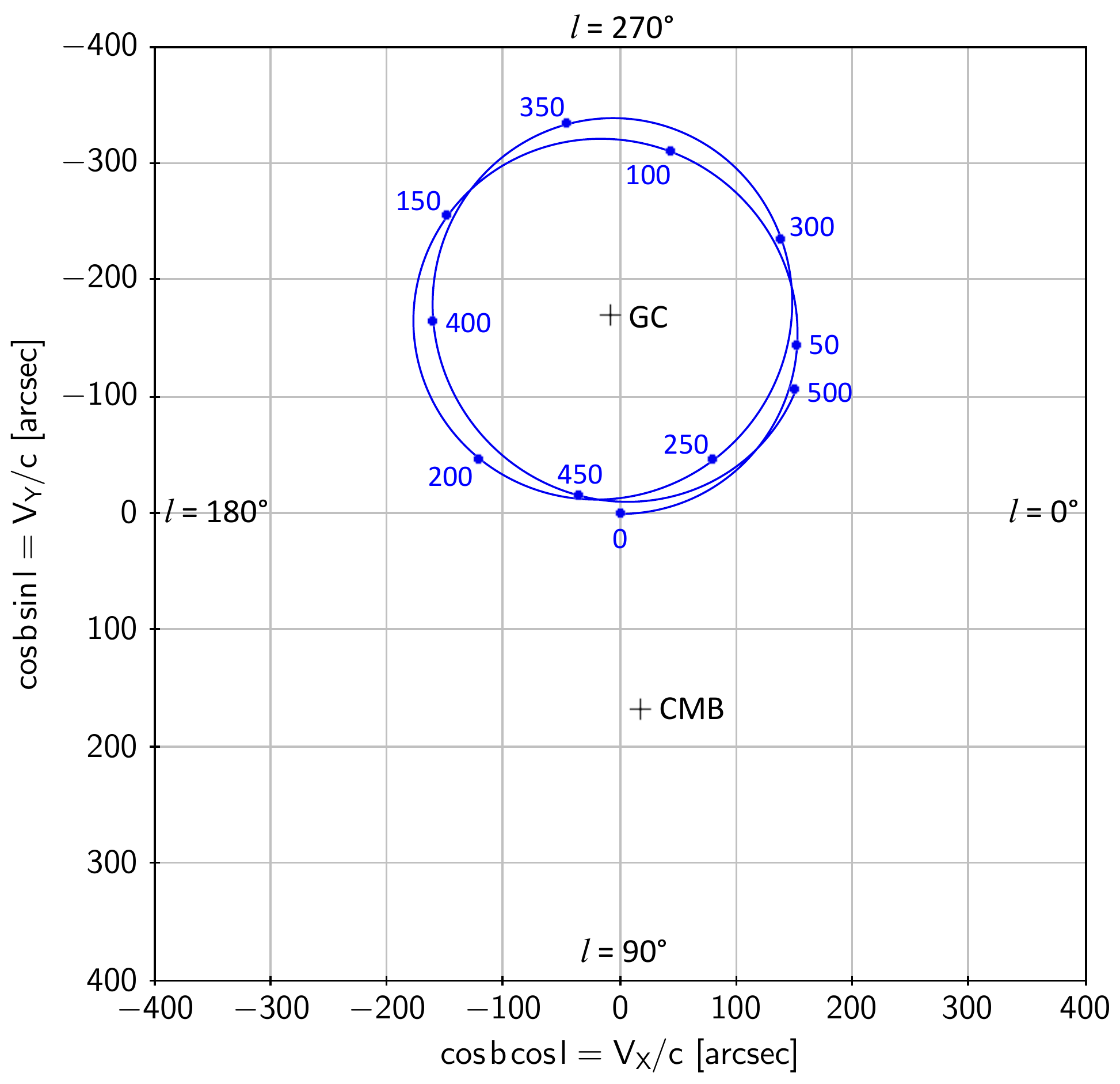}
  \caption{Galactic aberration over 500~Myr for an observer looking
    towards Galactic north. The curve shows the apparent path of a
    hypothetical quasar, currently located exactly at the north
    galactic pole, as seen from the Sun (or solar system
    barycentre). The points along the path show the apparent positions
    after 0,~50, 100,~\dots Myr due to the changing velocity of
    the Sun in its epicyclic orbit around the galactic centre. The
    point labelled GC is the position of the quasar as seen by an
    observer at rest with respect to the galactic centre. The point
    labelled CMB is the position as seen by an observer at rest with
    respect to the cosmic microwave background. The Sun's orbit was
    computed using the potential model by
    \citetads{2017MNRAS.465...76M} (see also Sect.~\ref{sec:expectation}),
    with current velocity components
    derived from the references in Sect.~\ref{sec:centripetal}. The Sun's velocity with
    respect to the CMB is taken from \citetads{2020A&A...641A...3P}.}
\label{fig:galactic-aberration}
\end{center}
\end{figure}

The motion with respect to the solar system barycentre is not the only
conceivable source of aberrational effects.  It is well known that the
whole solar system (that is, its barycentre) is in motion in the
Galaxy with a velocity of about $248\kms$
\citepads{2020ApJ...892...39R}, and that its velocity with respect to
the Cosmic Microwave Background Radiation (CMBR) is about $370\kms$
\citepads{2020A&A...641A...3P}. Therefore, if one compares the
apparent positions of the celestial sources as seen by an observer at
the barycentre of the solar system with those seen by another observer
at rest with respect to the Galaxy or the CMBR, one would see
aberrational differences up to ${\sim}171\arcsec$ or
${\sim}255\arcsec$, respectively -- effects that are so big that they
could be recognized by the naked eye (see
Fig.~\ref{fig:galactic-aberration} for an illustration of this
effect). The first of these effects is sometimes called secular
aberration. In most applications, however, there is no reason to
consider an observer that is `even more at rest' than the solar system
barycentre. The reason is that this large velocity -- for the purpose
of astrometric observations and for their accuracies -- can usually be
considered as constant; and if the velocity is constant in size and
direction, the principle of relativity imposes that the aberrational
shift cannot be detected.  In other words, without knowledge of the
`true' positions of the sources, one cannot reveal the constant
aberrational effect on their positions.

\begin{figure}
\begin{center}
  \includegraphics[width=1.0\hsize]{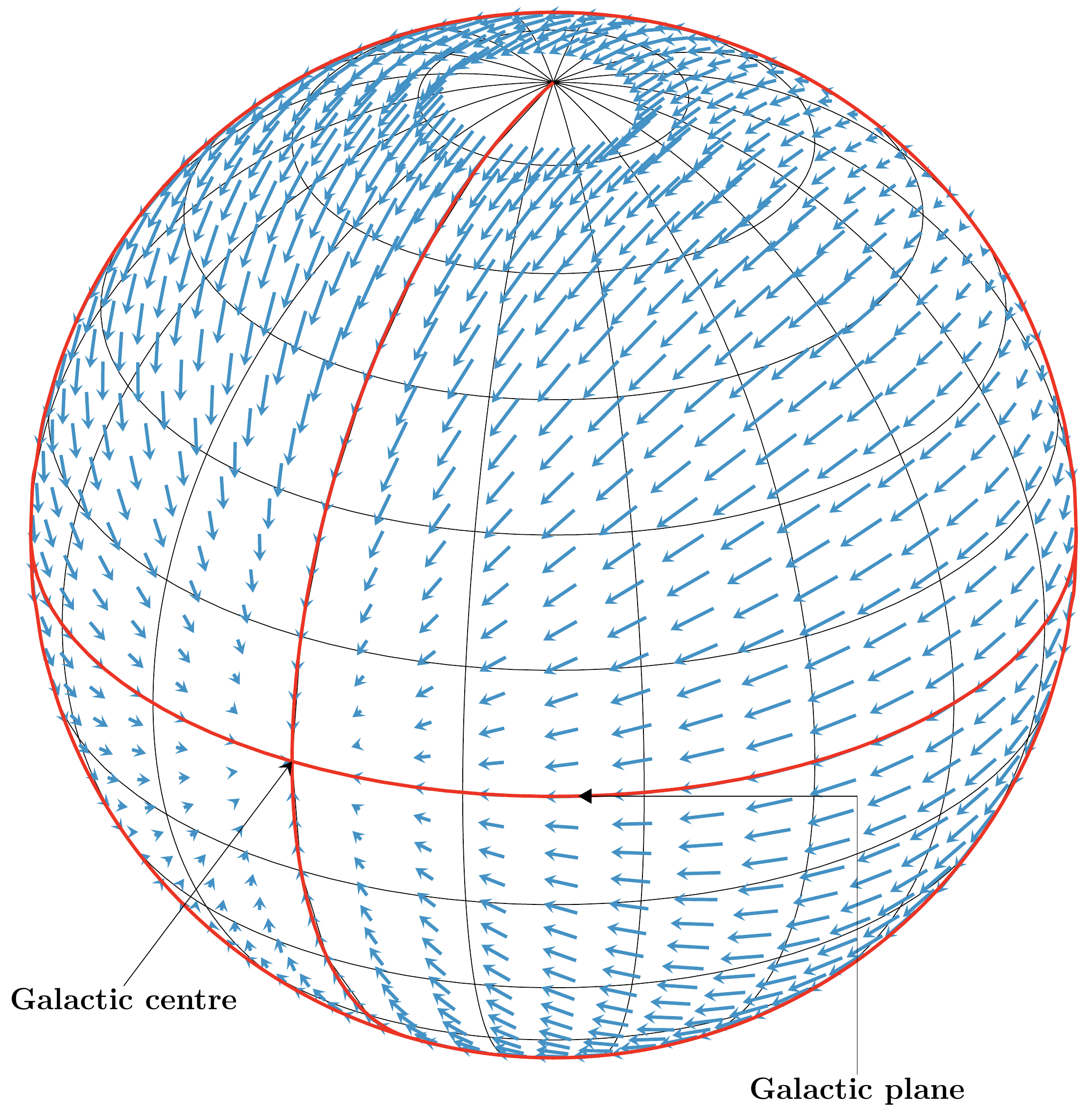}
  \caption{The proper motion field of QSO-like objects induced by the centripetal
  galactic acceleration: there is no effect in the directions of the
  galactic centre and anti-centre, and a maximum in the plane passing
  through the galactic poles with nodes at 90--270\,\degr\ in galactic
  longitudes.  The plot is in galactic coordinates with the solar
  system at the centre of the sphere, and the vector field seen from
  the exterior of the sphere. Orthographic projection with viewpoint
  at $l = 30\degr, b = 30\degr$ and an arbitrary scale for the
  vectors. See also an \href{http://www.aanda.org/XXX/olm}{online movie}.
  }
\label{fig:pm_galaccel}
\end{center}
\end{figure}

However, the velocity of the solar system is not exactly constant.
The motion of the solar system follows a curved orbit in the Galaxy, 
so its velocity vector is slowly changing with time. The secular
aberration is therefore also slowly changing with time. Considering sources that do not
participate in the galactic rotation (such as distant extragalactic sources),
we will see their apparent motions tracing out aberration `ellipses' whose period 
is the galactic `year' of 
${\sim}213$~million   
years -- they are of course not ellipses owing to the epicyclic orbit
of the solar system (see Fig.~\ref{fig:galactic-aberration}).
Over a few years, and even thousands of years, the tiny
arcs described by the sources 
cannot be distinguished from the tangent of the aberration
ellipse, and for the observer this is seen as a proper motion that can
be called additional, apparent, or spurious:
\begin{equation}\label{eq:accel}
   \frac{\text{d}(\delta \vec{u})}{\text{d}t} = \frac{\vec{a}}{c} -
   \frac{\vec{a}\cdot\vec{u}}{c}\,\vec{u}\,.
\end{equation}
Here $\vec{a}=\text{d}\vec{v}/\text{d}t$ is the acceleration of the solar
system barycentre with respect to the extragalactic sources.
For a given source, this slow drift of the observed position is
indistinguishable from its true proper motion. However, the apparent proper motion
as given by Eq.~(\ref{eq:accel}) has a global dipolar structure
with axial symmetry along the acceleration: it is maximal for sources in
the direction perpendicular to the acceleration and zero for
directions along the acceleration. This pattern is shown as a vector field in
Fig.~\ref{fig:pm_galaccel} in the case of the centripetal acceleration
directed towards the galactic centre.

Because only the change in aberration can be observed, not the aberration
itself, the underlying reference frame in Eq.~(\ref{eq:galaberr}) is irrelevant 
for the discussion. One could have considered another reference for the 
velocity, leading to a smaller or larger aberration, but the aberration drift 
would be the same and given by Eq.~(\ref{eq:accel}). Although this equation
was derived by reference to the galactic motion of the solar system, it is fully
general and tells us that any accelerated motion of the solar system
with respect to the distant sources translates into a systematic
proper-motion pattern of those sources, when the astrometric
parameters are referenced to the solar system barycentre, as it
is the case for \textit{Gaia}.  Using a rough estimate of the
centripetal acceleration of the solar system in its motion around the
galactic centre, one gets the approximate amplitude of the spurious
proper motions to be $\sim 5\,\muasyr$. A detailed discussion of
the expected acceleration is given in Sect.~\ref{sec:expectation}.

It is important to realize that the discussion in this form is
possible only when the first-order approximation given by
Eq. (\ref{eq:galaberr}) is used.  It is the linearity of
Eq. (\ref{eq:galaberr}) in $\vec{v}$ that allows one, in this
approximation, to decompose the velocity $\vec{v}$ in various parts
and simply add individual aberrational effects from those components
(e.g. annual and diurnal aberration in classical astrometry or also a
constant part and a linear variation). In the general case of a
complete relativistic description of aberration via Lorentz
transformations, the second-order aberrational effects depend also on
the velocity with respect to the underlying reference frame and can
become large. However, when the astrometric parameters are referenced
to the solar system barycentre, the underlying reference frame is at
rest with respect to the barycentre and Eq. (\ref{eq:accel}) is
correct to a fractional accuracy of about
${|\vec{v}_\text{obs}|/c}\sim10^{-4}$, where $\vec{v}_\text{obs}$ is
the barycentric velocity of the observer. While this is fully
sufficient for the present and anticipated future determinations with
\textit{Gaia}, a more sophisticated modelling is needed, if a
determination of the acceleration to better than $\sim0.01\%$ is
discussed in the future.

An alternative form of Eq.~(\ref{eq:accel}) is
\begin{equation}\label{eq:accel1}
   \vec{\mu} = \vec{g} - \left(\vec{g}\cdot\vec{u}\right)\,\vec{u}\,,
\end{equation}
where $\vec{\mu}=\text{d}(\delta\vec{u})/\text{d}t$ is the proper motion vector due to the
aberration drift and $\vec{g}=\vec{a}/c$ may be expressed
in proper motion units, for example $\mu$as~yr$^{-1}$. 
Both vectors $\vec{a}$ and $\vec{g}$ are called `acceleration' in the context of this study. 
Depending on the context, the acceleration may be given in different units, for
example \mssquared, \muasyr, or $\kmsMyr$\, (1\,\muasyr\ corresponds to 
$1.45343\kmsMyr=4.60566\times10^{-11}\mssquared$).

Equation~(\ref{eq:accel1}) can be written in component form, using
Cartesian coordinates in any suitable reference system and
the associated spherical angles.
For example, in equatorial (ICRS) reference system $(x,y,z)$ 
the associated angles are right ascension and declination $(\alpha,\delta)$.
The components of the proper motion,
$\mu_{\alpha*}\equiv\mu_\alpha\cos\delta$ and $\mu_\delta$, 
are obtained by projecting $\vec{\mu}$ on the unit vectors $\vec{e}_\alpha$ and 
$\vec{e}_\delta$ in the directions of increasing $\alpha$ and $\delta$ at the position 
of the source (see \citeads{2012A&A...547A..59M}, Fig.~1 and their Eqs.~64 and 65).
The result is
\begin{align}\label{eq:accel_components}
\begin{aligned}
\mu_{\alpha*} &= -g_x\sin\alpha + g_y\cos\alpha\,, \\
\mu_\delta &=-g_x\sin\delta\cos\alpha - g_y\sin\delta\sin\alpha +g_z\cos\delta\,,
\end{aligned}  
\end{align}
where $(g_x,\,g_y,\,g_z)$ are the corresponding components of
$\vec{g}$. A corresponding representation is valid in arbitrary coordinate
system. In this work, we will use either equatorial
(ICRS) coordinates $(x,y,z)$ or galactic coordinates $(X,Y,Z)$ and the
corresponding associated angles $(\alpha,\delta)$ and $(l,b)$,
respectively (see Sect.~\ref{sec:galactic-coordinates}).
Effects of the form in Eq.~(\ref{eq:accel_components}) are
often dubbed `glide' for the reasons explained in
Sect.~\ref{sec:method}.

\section{Theoretical expectations for the acceleration}
\label{sec:expectation}

This Section is devoted to a detailed discussion of the expected
gravitational acceleration of the solar system. We stress, however,
that the measurement of the solar system acceleration as outlined
above and further discussed in subsequent sections is absolutely
independent of the nature of the acceleration and the estimates given
here.


As briefly mentioned in Sect.~\ref{sec:effect}, the acceleration of
the solar system can, to first order, be approximated as the
centripetal acceleration towards the Galactic centre which keeps the
solar system on its not-quite circular orbit around the Galaxy. In
this section we quantify this acceleration and other likely sources of
significant acceleration. The three additional parts which we consider
are: (i) acceleration from the most significant non-axisymmetric
components of the Milky Way, specifically the Galactic bar and
spirals; (ii) the acceleration towards the Galactic plane, because the
Milky Way is a flattened system and the solar system lies slightly
above the Galactic plane; and (ii) acceleration from specific objects,
be they nearby galaxy clusters, local group galaxies, giant molecular
clouds or nearby stars.

For components of the acceleration associated with the bulk properties
of the Galaxy we describe the acceleration in galactocentric
cylindrical coordinates $(R',\,\phi',\,z')$, where $z'=0$ for the Galactic
plane, and the Sun is at $z'>0$). These are the natural model
coordinates, and we convert into an acceleration
in standard galactic coordinates $(a_X,\,a_Y,\,a_Z)$ as a final step.

\subsection{Centripetal acceleration}
\label{sec:centripetal}

The distance and proper motion of Sagittarius A* -- the super-massive
black hole at the Galactic centre -- has been measured with exquisite
precision in recent years. Since this is expected to be very close to
being at rest in the Galactic centre, the proper motion is almost
entirely a reflex of the motion of the Sun around the Galactic
centre. Its distance \citepads{2019A&A...625L..10G} is
\begin{equation*}
d_{\odot-GC}=8.178\pm0.013\; \text{(statistical)}\pm 0.022 \;\text{(systematic)}\kpc , 
\end{equation*}
and its proper motion along the Galactic plane is $-6.411\pm
0.008 \masyr$ \citepads{2020ApJ...892...39R}. The Sun is not on a circular orbit, 
so we cannot directly translate the corresponding velocity into a centripetal acceleration. To 
compensate for this, we can correct the velocity to the `local standard of rest' -- the
velocity that a circular orbit at $d_{\odot-GC}$ would have. This
correction is $12.24\pm2\kms$ \citepads{2010MNRAS.403.1829S}, in the sense
that the Sun is moving faster than a circular orbit at its
position. Considered together this gives an acceleration of
$-6.98\pm0.12\kmsMyr$ in the $R'$ direction. 
This corresponds to the centripetal acceleration of $4.80\pm0.08\muasyr$
which is compatible with the values based on measurements of Galactic rotation, discussed for example by
\citetads{2014ApJ...783..130R} and \citetads{2014jsrs.conf...44M}.

\subsection{Acceleration from non-axisymmetric components}
\label{sec:nonaxi}

The Milky Way is a barred spiral galaxy. The gravitational force from
the bar and spiral have important effects on the velocities of stars
in the Milky Way, as has been seen in numerous studies using \gdrtwo\
data (e.g.\ \citeads{2018A&A...616A..11G}). We separately consider
acceleration from the bar and the spiral. Table 1 in \citetads{2019MNRAS.490.1026H}
summarises models for the bar potential taken from the
literature. From this, assuming that the Sun lies $30\degr$ away from
the major axis of the bar \citepads{2015MNRAS.450.4050W}, most models give an
acceleration in the negative $\phi'$ direction of $0.04\kmsMyr$, with
one differing model attributed to \citetads{2017ApJ...840L...2P} which has a
$\phi'$ acceleration of $0.09\kmsMyr$. The \citetads{2017MNRAS.465.1621P} bar
model, the potential from which is illustrated in Figure 2
of \citetads{2019A&A...626A..41M}, is not included in the 
\citetads{2019MNRAS.490.1026H} table,
but is consistent with the lower value.

The recent study by \citetads{2020ApJ...900..186E} found an acceleration from the
spiral structure in the $\phi'$ direction of $0.10\kmsMyr$ in the
opposite sense to the acceleration from the bar. Statistical
uncertainties on this value are small, with systematic errors relating
to the modelling choices dominating. This spiral strength is within
the broad range considered by \citetads{2016MNRAS.461.3835M}, and we estimate the
systematic uncertainty to be of order $\pm0.05\kmsMyr$.

\subsection{Acceleration towards the Galactic plane}

The baryonic component of the Milky Way is flattened, with a stellar
disc which has an axis ratio of $\sim$1:10 and a gas disc, with
both \ion{H}{II} and H$_2$ components, which is even flatter. The Sun
is slightly above the Galactic plane, with estimates of the height
above the plane typically of the order
$z'_\odot=25\pm5\,\mathrm{pc}$ \citepads{2016ARA&A..54..529B}.

We use the Milky Way gravitational potential from \citetads{2017MNRAS.465...76M},
which has stellar discs and gas discs based on literature results, to
estimate this component of acceleration. We find an acceleration of
$0.15\pm0.03\kmsMyr$ in the negative $z'$ direction, i.e. towards the
Galactic plane. This uncertainty is found using only the uncertainty
in $d_{\odot-GC}$ and $z'_\odot$. We can estimate the systematic
uncertainty by comparison to the model from \citetads{2011MNRAS.414.2446M},
which, among other differences, has no gas discs. In this case we find
an acceleration of $0.13\pm0.02\kmsMyr$, suggesting that the
uncertainty associated with the potential is comparable to that from
the distance to the Galactic plane. For reference, if the acceleration
were directed exactly at the Galactic centre we would expect an
acceleration in the negative $z'$ direction of ${\sim}0.02\kmsMyr$ due
to the mentioned elevation of the Sun above the plane by
25\,pc, see next subsection.

Combined, this converts into an acceleration of 
$(-6.98\pm0.12,\, +0.06\pm0.05,\, -0.15\pm0.03) \kmsMyr$ 
in the $(R',\phi',z')$ directions.

\subsection{Transformation to standard galactic coordinates}
\label{sec:galactic-coordinates}

For the comparison of this model expectation with the EDR3 observations
we have to convert both into standard galactic coordinates $(X,Y,Z)$
associated with galactic longitude and latitude $(l,b)$.

The standard galactic coordinates are defined by 
the transformation between the equatorial (ICRS) and galactic coordinates given
in Sect.~1.5.3, Vol.~1 of \citet{ESA1997} using three
angles to be taken as exact quantities. In particular, the equatorial plane of the galactic
coordinates is defined by its pole at ICRS coordinates
$(\alpha=192.85948^\circ,\delta=+27.12825^\circ)$, and the origin of
galactic longitude is defined by the galactic longitude of the
ascending node of the equatorial plane of the galactic coordinates on
the ICRS equator, which is taken to be $l_\Omega =
32.93192^\circ$. This means that the point with galactic coordinates
$(l=0,b=0)$, that is the direction to the centre, is at
$(\alpha\approx266.40499^\circ, \delta\approx-28.93617^\circ)$.


The conversion of the model expectation takes into account the
above-mentioned elevation of the Sun, leading to a rotation of the $Z$
axis with respect to the $z'$ axis by $(10.5\pm 2)$\,arcmin, plus two sign
flips of the axes' directions. This leaves us with the final predicted
value of $(a_X,\,a_Y,\,a_Z) = (+6.98\pm0.12,\, -0.06\pm0.05,\,
-0.13\pm0.03)\kmsMyr$. Note that the rotation of the vertical axis is
uncertain by about 2\arcmin, due to the uncertain values of
$d_{\odot-GC}$ and $Z_\odot$. This, however, gives an uncertainty of
only 0.004$\kmsMyr$ in the predicted $a_Z$.

We should emphasize that these transformations are purely formal
ones. They should not be considered as strict in the sense that they
refer the two vectors to the true attractive center of the real
galaxy. On the one hand, they assume that the standard galactic
coordinates $(X,Y,Z)$ represent perfect knowledge of the true
orientation of the Galactic plane and the true location of the
Galactic barycentre. On the other hand, they assume that the disk is
completely flat, and that the inner part of the Galactic potential is
symmetric (apart from the effects of the bar and local spiral
structure discussed above). Both assumptions can easily be violated by
a few arcmin. This can easily be illustrated by the position of the
central black hole, Sgr~A*. It undoubtedly sits very close in the
bottom of the Galactic potential trough, by dynamical necessity. But
that bottom needs not coincide with the barycentre of the Milky Way,
nor with the precise direction of the inner galaxy's force on the
Sun. In fact, the position of Sgr~A* is off galactic longitude zero by
$-3.3$\arcmin, and off galactic latitude zero by
$-2.7$\arcmin.%
\footnote{To take the solar system as an illustrative
analogue: the bottom of the potential trough is always very close to the centre
of the Sun, but the barycentre can be off by more than one solar
radius, i.e. the attraction felt by a Kuiper belt object at, say,
30\,au can be off by more than 0.5\arcmin.} 
This latitude offset is only about a quarter of the 10.5\arcmin\ correction
derived from the Sun's altitude above the plane.

Given the present uncertainty of the measured acceleration vector by a
few degrees (see Table \ref{tab:results}), these considerations about a few arcmin are
irrelevant for the present paper. We mention them here as a matter of
principle, to be taken into account in case the measured vector would
ever attain a precision at the arcminute level.

\subsection{Specific objects}
\citetads{2016A&A...589A..71B} provide in their Table~2 an estimate of the acceleration due to
various extragalactic objects. We can use this table as an initial
guide to which objects are likely to be important, however mass
estimates of some of these objects (particularly the Large Magellanic
Cloud) have changed significantly from the values quoted there.

We note first that individual objects in the Milky Way have a
negligible effect. The acceleration from $\alpha$~Cen~AB is
${\sim}0.004\kmsMyr$, and that from any nearby giant molecular clouds is
comparable or smaller. 
In the local group, the largest effect is from
the Large Magellanic Cloud (LMC). A number of lines of evidence now
suggest that it has a mass of
$(1{-}2.5)\times10^{11}\,M_\odot$ (see \citeads{2019MNRAS.487.2685E}
and references therein), which at a distance of
$49.5\pm0.5\kpc$ \citepads{2019Natur.567..200P} gives an acceleration of
0.18 to $0.45\kmsMyr$ with components $(a_X,a_Y,a_Z)$ between 
$(+0.025,\,-0.148,\, -0.098)$ and $(+0.063,\, -0.371,\, -0.244)\kmsMyr$. 
We note therefore that the acceleration from the LMC is significantly larger
than that from either the Galactic plane or non-axisymmetric
structure.

The Small Magellanic Cloud is slightly more
distant ($62.8\pm2.5\kpc$; \citeads{2000A&A...359..601C}), and significantly less
massive. It is thought that it has been significantly tidally stripped
by the LMC (e.g.\ \citeads{2020MNRAS.495...98D}), so its mass is likely to be
substantially lower than its estimated peak mass of
${\sim}7\times10^{10}\,M_\odot$ (e.g.\ \citeads{2019MNRAS.487.5799R}), but is hard to
determine based on dynamical modelling. We follow \citetads{2020ApJ...893..121P} and
consider the range of possible masses $(0.5{-}3)\times10^{10}\,M_\odot$,
which gives an acceleration of 0.005 to $0.037\kmsMyr$.  Other local
group galaxies have a negligible effect. M31, at a distance of
$752\pm27\kpc$ \citepads{2012ApJ...745..156R}, with mass estimates in the range
$(0.7{-}2)\times10^{12}\,M_\odot$ \citepads{2013MNRAS.434.2779F} imparts an
acceleration of 0.005 to $0.016\kmsMyr$. The Sagittarius dwarf galaxy is
relatively nearby, and was once relatively massive, but has been
dramatically tidally stripped to a mass
$\lesssim4\times10^8\,M_\odot$ (\citeads{2020MNRAS.497.4162V}; \citeads{2010ApJ...714..229L}), so
provides an acceleration $\lesssim0.003\kmsMyr$.
We note that this discussion only includes the direct acceleration that these local group bodies 
apply to the Solar system. They are expected to deform the Milky Way's dark matter halo 
in a way that may also apply an acceleration  (e.g., \citeads{2020arXiv201000816G}).

We can, like \citetads{2016A&A...589A..71B}, estimate the acceleration due
to nearby galaxy clusters from their estimated masses and
distances. The Virgo cluster at a distance
$16.5\,\mathrm{Mpc}$ \citepads{2007ApJ...655..144M} and a mass
$(1.4{-}6.3)\times10^{14}\,M_\odot$ (\citeads{2012ApJS..200....4F}; \citeads{2020A&A...635A.135K})
is the most significant single influence
(0.002 to $0.010\kmsMyr$). However, we recognise that the peculiar
velocity of the Sun with respect to the Hubble flow has a component
away from the Local Void, one towards the centre of the Laniakea
supercluster, and others on larger scales that are not yet
mapped (\citeads{2008ApJ...676..184T}; \citeads{2014Natur.513...71T}), and that this is
probably reflected in the acceleration felt on the solar system
barycentre from large scale structure.

For simplicity we only add the effect of the LMC to the value given at
the end of Sect.~\ref{sec:nonaxi} to give an overall estimate of the
expected range of, adding our estimated $1\sigma$ uncertainties from
the Galactic models to our full range of possible accelerations from
the LMC to give $(a_X,\,a_Y\,,a_Z)$ as $(+6.89,\,-0.20,\,-0.20)$ to 
$(+7.17,\,-0.48,\,-0.40)\kmsMyr$.

\section{Selection of Gaia sources}
\label{sec:selection}

\subsection{QSO-like sources}
\label{sec:qso-like}

\textit{Gaia} Early Data Release 3 (EDR3; \citealt{DPACP-130})
provides high-accuracy astrometry for over 1.5~billion sources, mainly
galactic stars. However, there are good reasons to believe that a few
million sources are QSOs and other extragalactic sources
that are compact enough for \gaia\ to obtain good astrometric
solutions. These sources are hereafter referred to as `QSO-like sources'.
As explained in Sect.~\ref{sec:stars} it is only the QSO-like sources
that can be used to estimate the acceleration of the solar system.

Eventually, in later releases of \gaia\ data, we will be able
to provide astrophysical classification of the sources and thus find
all QSO-like sources based only on \gaia's own data. EDR3 may be the last
\gaia\ data release that needs to rely on external information to
identify the QSO-like sources in the main catalogue of the release. To this
end, a cross-match of the full EDR3 catalogue was performed with 17 external QSO and
AGN catalogues. The matched sources were then further filtered
to select astrometric solutions of sufficient quality in EDR3 and
to have parallaxes and proper motions compatible with zero within five
times the respective uncertainty. In this way, the contamination
of the sample by stars is reduced, even though it may also exclude some genuine
QSOs. It is important to recognise that the rejection based on significant
proper motions does not interfere with the systematic proper motions
expected from the acceleration, the latter being about two orders of
magnitude smaller than the former.  Various additional tests were
performed to avoid stellar contamination as much as possible.  As a
result, EDR3 includes $1\,614\,173$ sources that were identified as
QSO-like; these are available in the \gaia\ Archive as the table 
{\tt agn\_cross\_id}.  The full details of the selection procedure, together
with a detailed description of the resulting \gcrfthree, will be
published elsewhere \citep{DPACP-133}.

In \gaia\ EDR3 the astrometric solutions for the individual sources are
of three different types \citep{DPACP-128}:
\begin{itemize}
\item two-parameter solutions, for which only a mean position is provided;
\item five-parameter solutions, for which the position (two coordinates),
  parallax, and proper motion (two components) are provided;  
\item six-parameter solutions, for which an astrometric estimate (the 'pseudocolour') of the effective wavenumber
\footnote{The effective wavenumber $\nu_\text{eff}$ is the mean value 
of the inverse wavelength $\lambda^{-1}$, weighted by the detected photon flux 
in the \textit{Gaia} passband $G$. This quantity is extensively used to model
colour-dependent image shifts in the astrometric instrument of \textit{Gaia}.
An approximate relation between $\nu_\text{eff}$ and the colour index 
$G_\text{BP}-G_\text{RP}$ is given in \citet{DPACP-128}. The values 
$\nu_\text{eff}=1.3$, 1.6, and 1.9 roughly correspond to, respectively, 
$G_\text{BP}-G_\text{RP}= 2.4$, 0.6, and $-0.5$.\label{footnote-pseudocolor}}
is provided together with the five astrometric parameters.
\end{itemize}
Because of the astrometric filtering mentioned above, the
\gcrfthree\ sources only belong to the last two types of 
solutions: more precisely the selection comprises $1\,215\,942$
sources with five-parameter solutions and $398\,231$ sources with
six-parameter solutions. Table~\ref{tab:gaiacrf3-characteristics}
gives the main characteristics of these sources. The
\gcrfthree\ sources with six-parameter solutions are typically
fainter, redder, and have somewhat lower astrometric quality 
(as measured by the re-normalised unit weight error, RUWE) than those
with five-parameter solutions.%
\footnote{The RUWE \citep{DPACP-128} is a measure 
of the goodness-of-fit of the five- or six-parameter model to the observations
of the source. The expected value for a good fit is 1.0. A higher value
could indicate that the source is not point-like at the optical resolution of \textit{Gaia}
($\simeq 0.1\arcsec$), or has a time-variable structure.\label{fn:RUWE}}
Moreover, various studies of the astrometric quality of EDR3 
\citep[e.g.][]{DPACP-126,DPACP-128,DPACP-132}
have demonstrated
that the five-parameter solutions generally have smaller systematic errors, 
at least for $G>16$, that is for most QSO-like sources. In the following
analysis we include only the $1\,215\,942$ \gcrfthree\ sources with
five-parameter solutions.

\begin{table*}
  \caption{Characteristics of the \gcrfthree\ sources.   
\label{tab:gaiacrf3-characteristics}}
\footnotesize\setlength{\tabcolsep}{6pt}
\begin{center}
\begin{tabular}{crcccccc}
\hline\hline
\noalign{\smallskip}
\multicolumn{1}{c}{type} & \multicolumn{1}{c}{number} & \multicolumn{1}{c}{$G$} &\multicolumn{1}{c}{BP$-$RP} & \multicolumn{1}{c}{$\nu_{\rm eff}$} &\multicolumn{1}{c}{RUWE} & \multicolumn{1}{c}{$\sigma_{\mu_{\alpha*}}$} & \multicolumn{1}{c}{$\sigma_{\mu_{\delta}}$}\\
\multicolumn{1}{c}{of solution} & \multicolumn{1}{c}{of sources} & \multicolumn{1}{c}{[mag]} &\multicolumn{1}{c}{[mag]} & \multicolumn{1}{c}{[$\mu{\rm m}^{-1}$]} &\multicolumn{1}{c}{} & \multicolumn{1}{c}{[$\muasyr$]} & \multicolumn{1}{c}{[$\muasyr$] }\\
\noalign{\smallskip}\hline\noalign{\smallskip}
five-parameter & 1\,215\,942 & 19.92 & 0.64 & 1.589 & 1.013 & 457 & 423 \\ 
six-parameter &  398\,231 & 20.46 & 0.92 & -- & 1.044 & 892 & 832 \\
all         & 1\,614\,173 & 20.06 & 0.68 & -- & 1.019 & 531 & 493 \\  
\noalign{\smallskip}
\hline
\end{tabular}
\tablefoot{Columns~3--8 give median values
    of the $G$ magnitude, the BP$-$RP colour index, the effective
    wavenumber $\nu_{\rm eff}$ (see footnote~\ref{footnote-pseudocolor}; only available for the five-parameter
    solutions), the astrometric quality indicator RUWE (see footnote~\ref{fn:RUWE}),
    and the uncertainties of the equatorial proper
    motion components in $\alpha$ and $\delta$. 
   The last line (`all') is for the whole set of \gcrfthree\ sources. 
   In this study only the sources with five-parameters solutions are used.
}
\end{center}
\end{table*}

\begin{figure}[htbp]
\begin{center}
  \includegraphics[width=1\hsize]{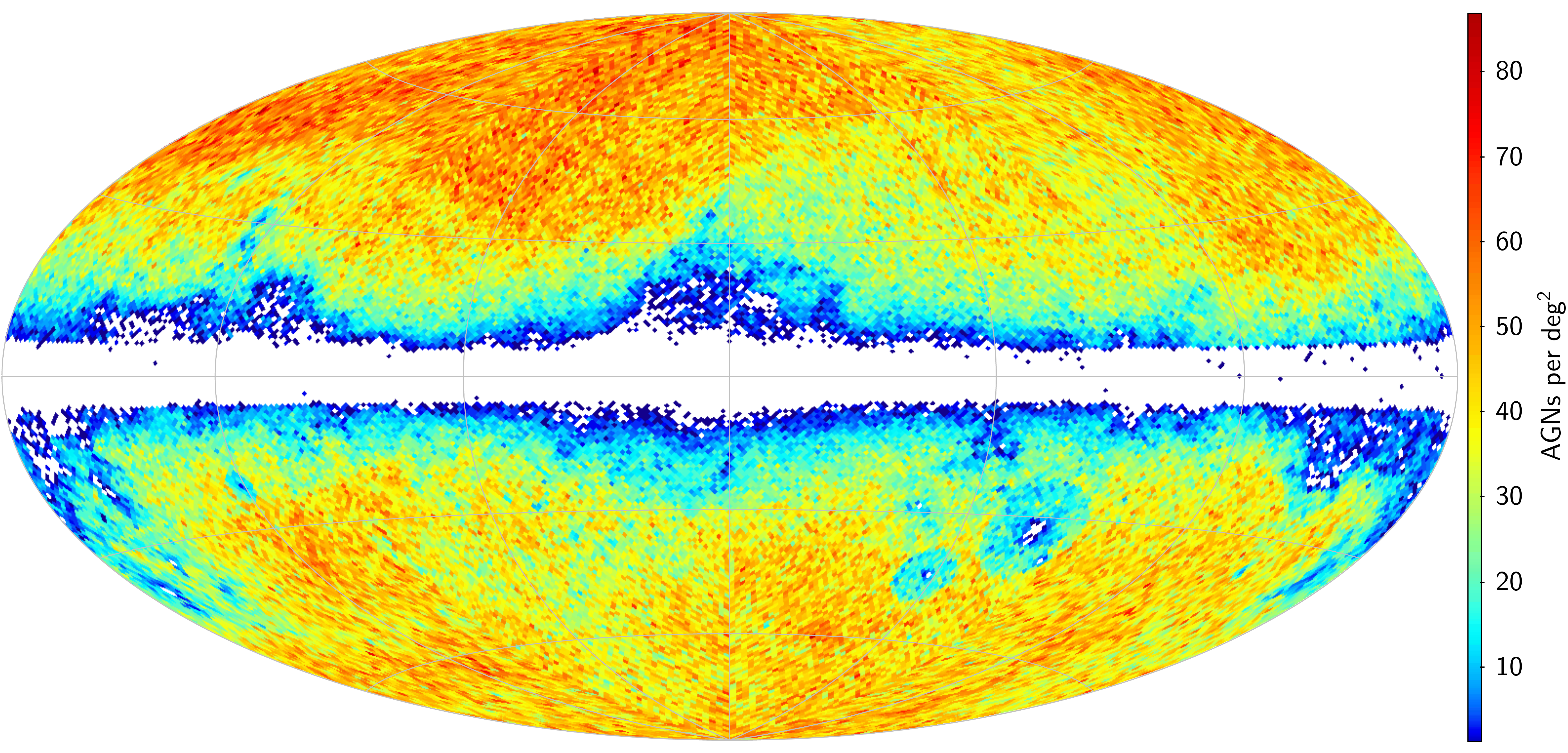}
  \caption{Distribution of the \gcrfthree\ sources with five-parameter
    solutions. The plot shows the density of sources per
    square degree computed from the source counts per pixel using
    HEALPix of level 6 (pixel size $\sim0.84$\,deg$^2$). This and
    following full-sky maps use a Hammer–Aitoff projection in galactic
    coordinates with $l = b = 0$ at the centre, north up, and $l$
    increasing from right to left.}
\label{fig:sky-distribution-5p}
\end{center}
\end{figure}

Important features of these sources are displayed in
Figs.~\ref{fig:sky-distribution-5p} and \ref{fig:histograms-5p}.  The
distribution of the sources is not homogeneous on the sky, with densities
ranging from 0 in the galactic plane to 85 sources per square
degree, and an average density of 30~deg$^{-2}$.  The distribution of
\gcrfthree\ sources primarily reflects the sky inhomogeneities of the
external QSO/AGN catalogues used to select the sources. 
In addition, to reduce the risk of source confusion in crowded areas,
the only cross-matching made in the galactic zone ($\left|\sin
b\right|<0.1$, with $b$ the galactic latitude) was with the VLBI
quasars, for which the risk of confusion is negligible thanks to their
accurate VLBI positions. One can hope that the future Releases of
\gcrf\ will substantially improve the homogeneity and remove this
selection bias (although a reduced source density at the galactic
plane may persist due to the extinction in the galactic plane).

\begin{figure}
\begin{center}
  \includegraphics[width=1.00\hsize]{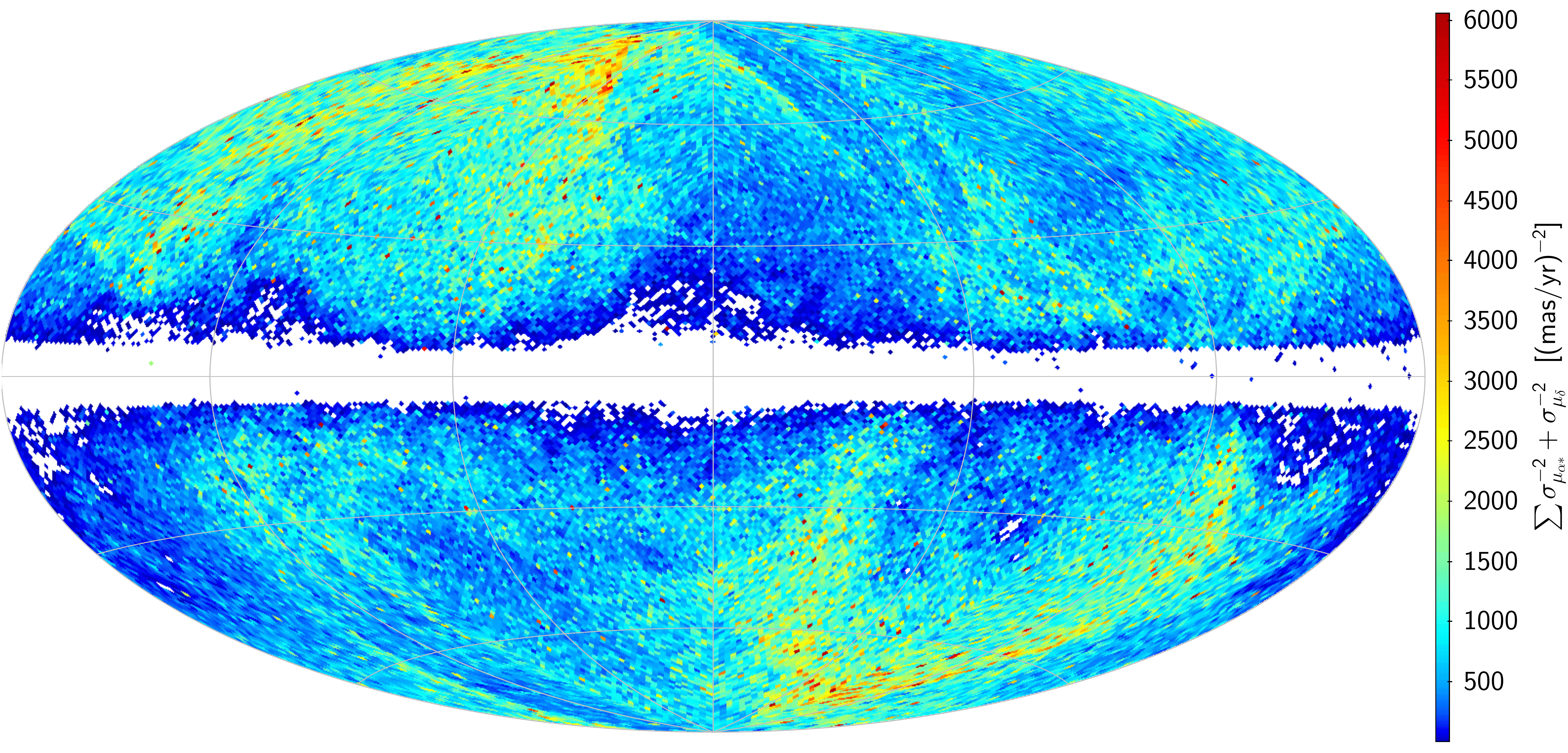}
    \caption{Distribution of the statistical weights of the proper motions of
      the \gcrfthree\ sources with five-parameter
      solutions. Statistical weight is calculated as the sum of
      $\sigma_{\mu_{\alpha*}}^{-2}+\sigma_{\mu_{\delta}}^{-2}$
      in pixels at HEALPix level~6.}    
\label{fig:sky-distribution-5p-information}
\end{center}
\end{figure}

\begin{figure}
\begin{center}
  \includegraphics[width=1.00\hsize]{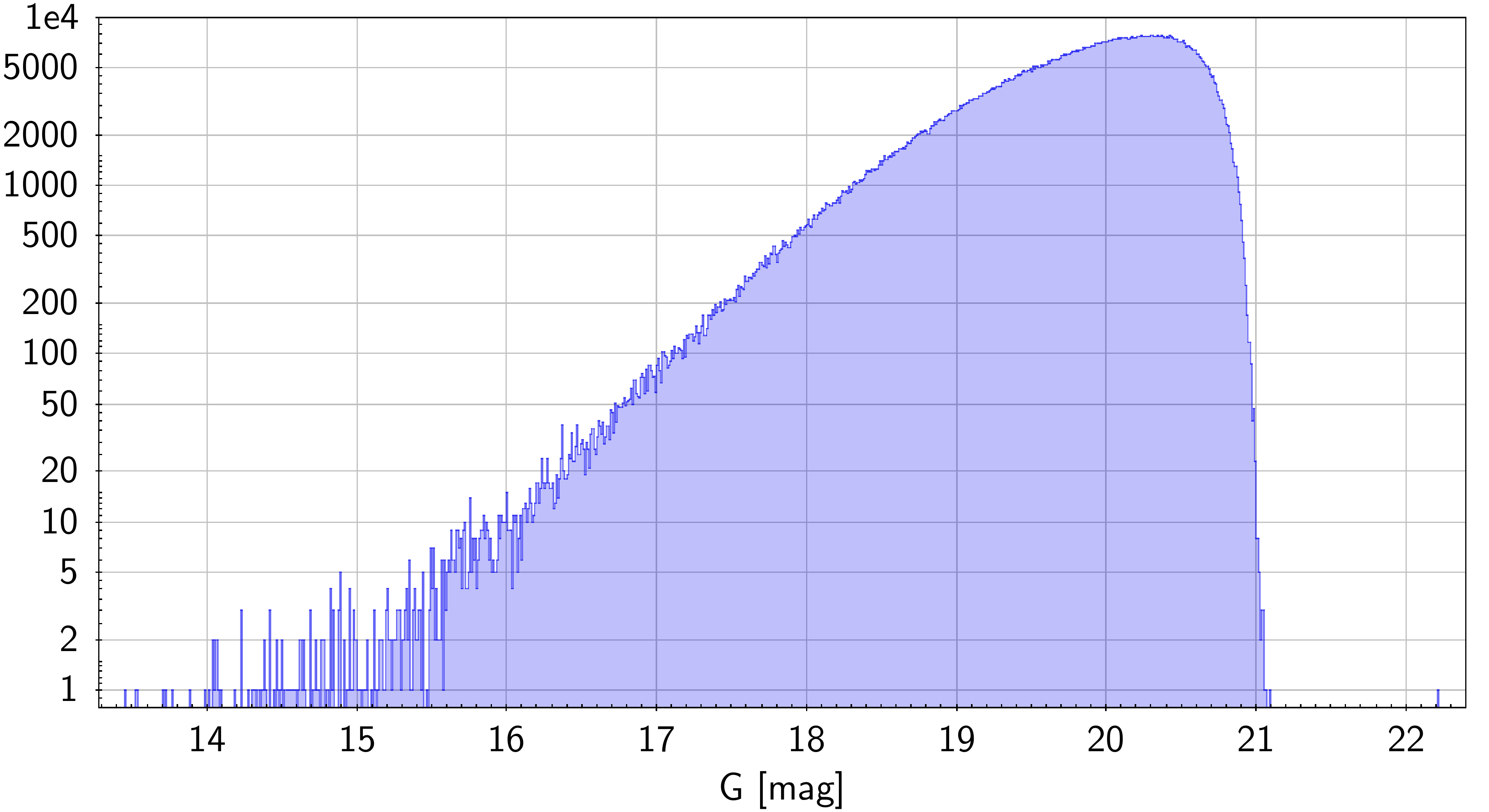}
  \includegraphics[width=1.00\hsize]{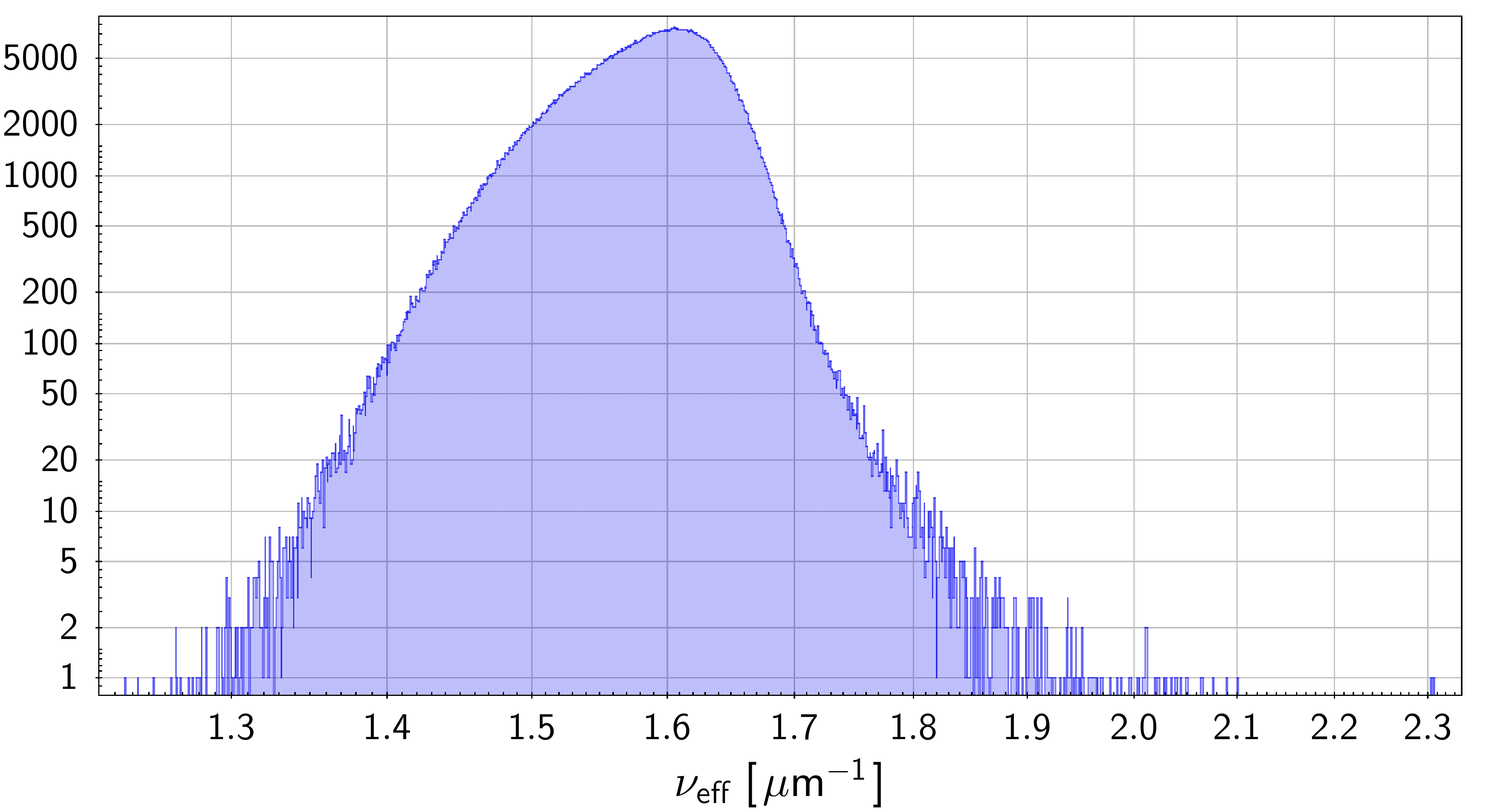}
  \includegraphics[width=1.00\hsize]{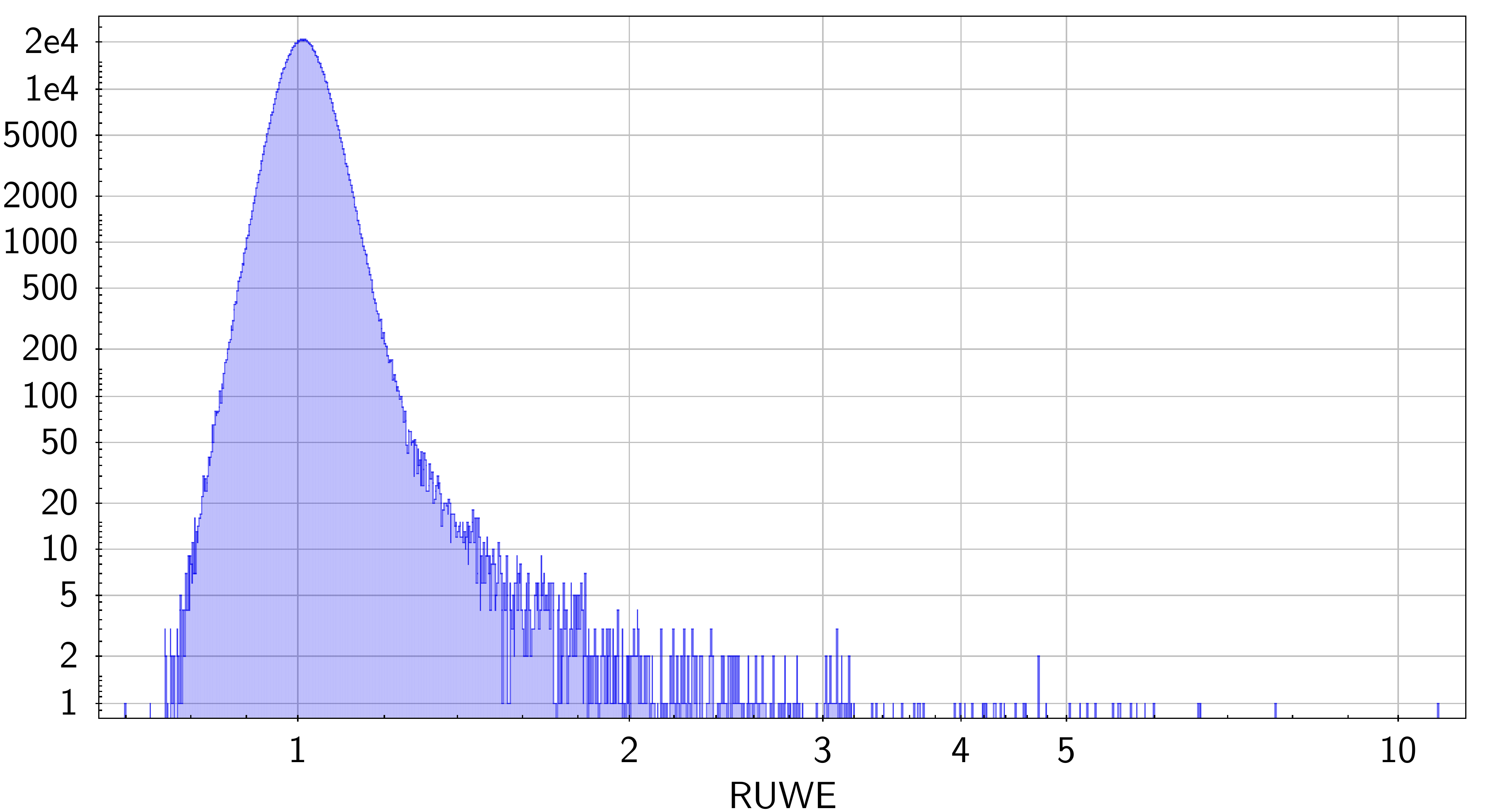}
  \caption{Histograms of some important characteristics of the \gcrfthree\ sources 
  with five-parameter solutions. From top to bottom: $G$ magnitudes,
    colours represented by the effective wavenumber $\nu_{\rm eff}$ (see footnote~\ref{footnote-pseudocolor}), 
    and the astrometric quality indicator RUWE (see footnote~\ref{fn:RUWE}).
  }
\label{fig:histograms-5p}
\end{center}
\end{figure}

As discussed below, our method for estimating
the solar system acceleration from proper motions of the
\gcrfthree\ sources involves an expansion of the vector field of
proper motions in a set of functions that are orthogonal on the sphere.
It is then advantageous if the data points 
are distributed homogeneously on the sky.  However, as shown in
Sect.~7.3 of \citepads{2012A&A...547A..59M}, what is important is not
the `kinematical homogeneity' of the sources on the sky (how many
per unit area), but the `dynamical homogeneity': the distribution of the
statistical weight of the data points over the sky (how much weight
per unit area). This distribution is shown on
Fig.~\ref{fig:sky-distribution-5p-information}.

\begin{figure}[htbp]
  \begin{center}
  \includegraphics[width=1.00\hsize]{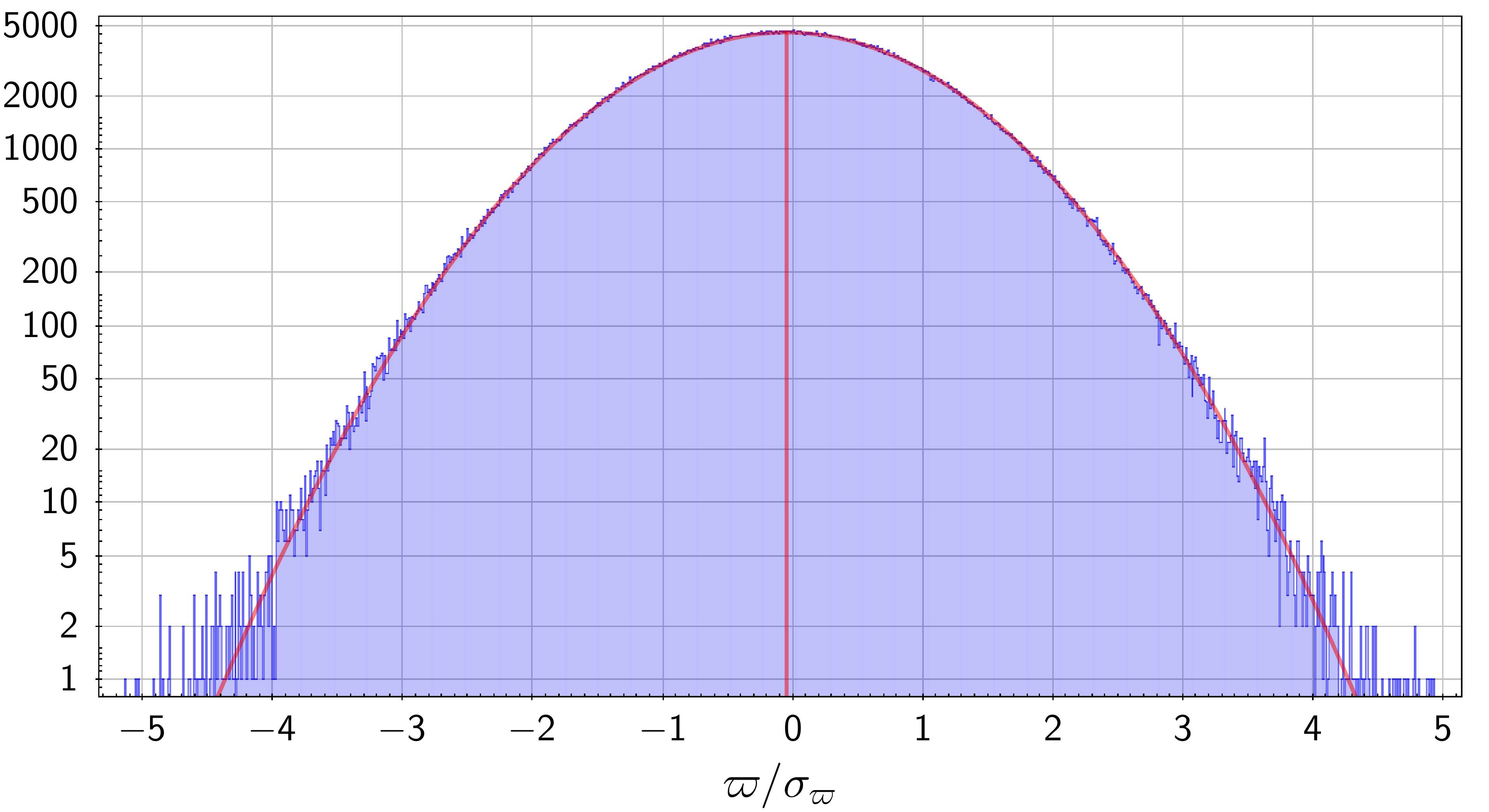}  
  \includegraphics[width=1.00\hsize]{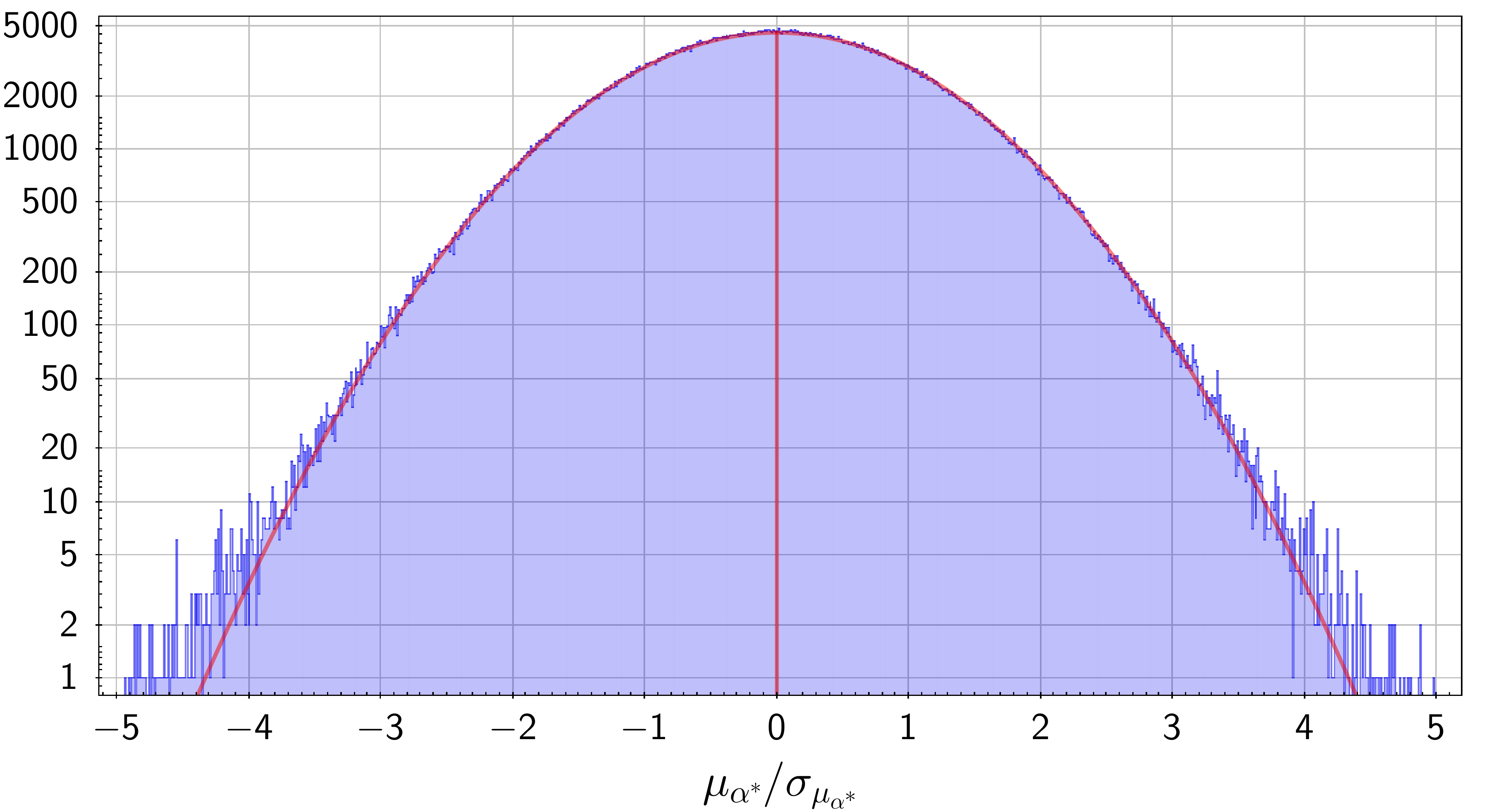}
  \includegraphics[width=1.00\hsize]{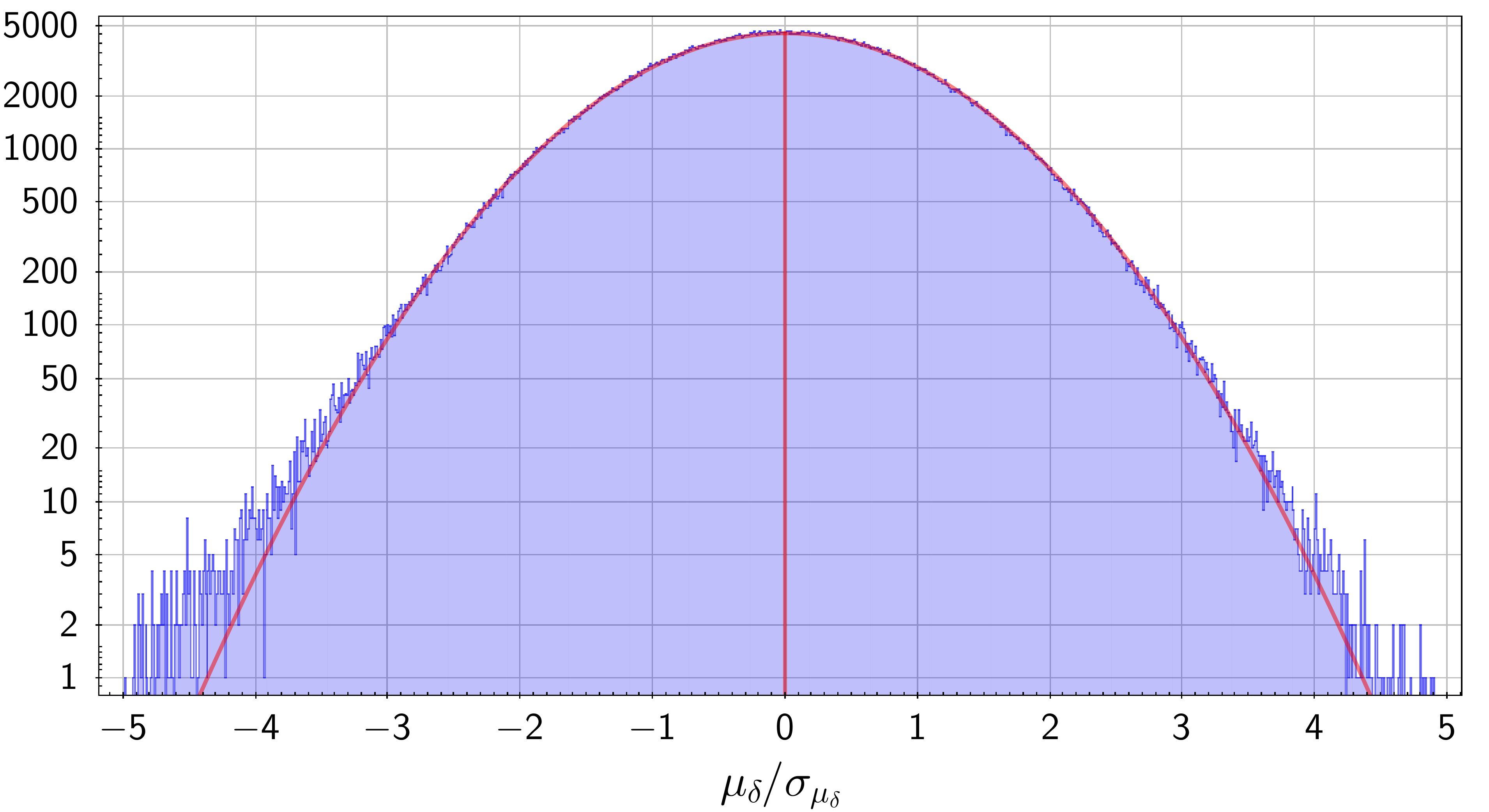}
  \caption{Distributions of the normalized parallaxes $\varpi/\sigma_\varpi$ (upper pane),
  proper motions in right ascension $\mu_{\alpha^*}/\sigma_{\mu_{\alpha*}}$
    (middle pane) and proper motions in declination $\mu_{\delta}/\sigma_{\mu_{\delta}}$ (lower pane) 
    for the \gcrfthree\ sources with five-parameter. The red lines show the corresponding best-fit Gaussian distributions. 
  }
\label{fig:pm-5p-histograms}
\end{center}
\end{figure}

For a reliable measurement of the solar system acceleration it is
important to have the cleanest possible set of QSO-like sources. A
significant stellar contamination may result in a
systematic bias in the estimated acceleration (see Sect.~\ref{sec:stars}). 
In this context the histograms of the normalised parallaxes and proper motions 
in Fig.~\ref{fig:pm-5p-histograms} are a useful diagnostic. 
For a clean sample of extragalactic
QSO-like sources one expects that the distributions of the
normalised parallaxes and proper motions are normal distributions with
(almost) zero mean and standard deviation (almost) unity. Considering
the typical uncertainties of the proper motions of over $400\muasyr$
as given in Table~\ref{tab:gaiacrf3-characteristics} it is clear that
the small effect of the solar system acceleration can be ignored in
this discussion. The best-fit normal distributions for the normalised
parallaxes and proper motions shown by red lines on
Fig. \ref{fig:pm-5p-histograms} indeed agree remarkably well with the
actual distribution of the data. The best-fit Gaussian distributions
have standard deviations of 1.052, 1.055 and 1.063, respectively for
the parallaxes ($\varpi$), proper motions in right ascension
($\mu_{\alpha*}$), and proper motions in declination ($\mu_\delta$). Small
deviations from normal distributions (note the logarithmic scale of
the histograms) can result both from statistical fluctuations
in the sample and some stellar contaminations. One can conclude that
the level of contaminations is probably very low.

\subsection{Stars of our Galaxy}
\label{sec:stars}
The acceleration of the solar system affects also the observed proper motions
of stars, albeit in a more complicated way than for the distant extragalactic sources.%
\footnote{For the proper motion of a star it is only the differential (tidal) acceleration 
between the solar system and the star that matters.}
Here it is however masked by other, much larger effects, and this
section is meant to explain why it is not useful
to look for the effect in the motions of galactic objects.

The expected size of the galactocentric acceleration term is of the
order of $5\muasyr$ (Sect.~\ref{sec:expectation}). The
galactic rotation and shear effects are of the order of 5--$10\masyr$,
i.e.\ over a thousand times bigger. In the Oort approximation they do
not contain a glide-like component, but any systematic difference 
between the solar motion and the bulk motion of some stellar
population produces a glide-like proper-motion pattern over the
whole sky. Examples of this are the solar apex motion (pointing away
from the apex direction in Hercules, $\alpha\simeq 270\degr$,
$\delta\simeq 30\degr$) and the asymmetric drift of old stars (pointing
away from the direction of rotation in Cygnus, $\alpha\simeq 318\degr$,
$\delta\simeq 48\degr$). Since these two directions -- by pure chance
-- are only $\sim40\degr$ apart on the sky, the sum of their effects 
will be in the same general direction.

But both are distance dependent, i.e.\ the size of the glide strongly
depends on the stellar sample used. The asymmetric drift is, in
addition, age dependent. Both effects attain the same order of
magnitude as the Oort terms at a distance of the order of 1~kpc. That
is, like the Oort terms they are of the order of a thousand times bigger than
the acceleration glide. Because of this huge difference in size, and the
strong dependence on the stellar sample, it is in practice impossible to
separate the tiny acceleration effect from the kinematic patterns.

Some post-Oort terms in the global galactic kinematics (e.g.\ a
non-zero second derivative of the rotation curve) can produce a big
glide component, too. And, more importantly, any asymmetries of the
galactic kinematics at the level of 0.1\% can create glides in more or less
random directions and with sizes far above the acceleration
term. Examples are halo streams in the solar vicinity, the tip of the
long galactic bar, the motion of the disk stars through a spiral wave
crest, and so on.

For all these reasons it is quite obvious that there is no hope to
discern an effect of $5\muasyr$ amongst chaotic structures of the
order of $10\masyr$ in stellar kinematics. 
In other words, we cannot use galactic objects 
to determine the glide due to the acceleration of the solar system. 

As a side remark we mention that there is a very big
($\simeq 6\masyr$) direct effect of the galactocentric acceleration in the
proper-motion pattern of stars on the galactic scale: it is not a
glide but the global rotation which is represented by the minima in
the well-known textbook double wave of the proper motions $\mu_{l*}$ in
galactic longitude $l$ as function of $l$. But this is of no relevance
in connection with the present study.

\section{Method} 
\label{sec:method}

One can think of a number of ways to estimate the acceleration from a
set of observed proper motions. For example, one could directly
estimate the components of the acceleration vector by a least-squares fit
to the proper motion components using Eq.~(\ref{eq:accel_components}). 
However, if there are other large-scale patterns present in the proper motions,
such as from a global rotation, these other effects could bias the acceleration 
estimate, because they are in general not orthogonal to the acceleration effect
for the actual weight distribution on the sky 
(Fig.~\ref{fig:sky-distribution-5p-information}).

We prefer to use a more general and more flexible mathematical
approach with Vector Spherical Harmonics (VSH). For a given set of
sources, the use of VSH allows us to mitigate the biases produced by various
large-scale patterns, thus bringing
a reasonable control over the systematic errors. The theory of VSH
expansions of arbitrary vector fields on the sphere and its
applications to the analysis of astrometric data were discussed in
detail by \citetads{2012A&A...547A..59M}. We use the
notations and definitions given in that work. In particular, to the vector 
field of proper motions
$\vec{\mu}(\alpha,\delta)=\mu_{\alpha*}\,\vec{e}_\alpha+\mu_\delta\,\vec{e}_\delta$
(where $\vec{e}_\alpha$ and $\vec{e}_\delta$ are unit vectors in the local triad as in
Fig.~1 of \citeads{2012A&A...547A..59M}) 
we fit the following VSH representation: 
\begin{equation}\label{Vexpandreal}
\begin{split}
 \vec{\mu}(\alpha,\delta) &=  \sum_{l=1}^{l_{\rm max}}\,\Biggl(
t_{l0} \vec{T}_{l0} + s_{l0} \vec{S}_{l0}\\
&\quad
+ 2\sum_{m=1}^{l}\, \left(t^{\Re}_{lm} \vec{T}^{\Re}_{lm} -  t^{\Im}_{lm} \vec{T}^{\Im}_{lm}
+s^{\Re}_{lm} \vec{S}^{\Re}_{lm} -  s^{\Im}_{lm} \vec{S}^{\Im}_{lm}
\right)\Biggr)\,.
\end{split}
\end{equation}
\noindent
Here $\vec{T}_{lm}(\alpha,\delta)$ and $\vec{S}_{lm}(\alpha,\delta)$ are
the toroidal and spheroidal vector spherical harmonics of degree $l$
and order $m$, $t_{lm}$ and $s_{lm}$ are the corresponding
coefficients of the expansion (to be fitted to the data), and the
superscripts $\Re$ and $\Im$ denote the real and imaginary parts of
the corresponding complex quantities, respectively. In general, the VSHs
are defined as complex functions and can represent complex-valued vector
fields, but the field of proper motions is real-valued and the expansion
in Eq.~(\ref{Vexpandreal}) readily uses the symmetry properties of the
expansion, so that all quantities in Eq.~(\ref{Vexpandreal}) are real.
The definitions and various properties of $\vec{T}_{lm}(\alpha,\delta)$
and $\vec{S}_{lm}(\alpha,\delta)$, as well as an efficient algorithm for their computation,
can be found in \citetads{2012A&A...547A..59M}. 

The main goal of this work is to estimate the solar system acceleration described by
Eq.~(\ref{eq:accel_components}). As explained in \citetads{2012A&A...547A..59M},
a nice property of the VSH expansion is that the first-order harmonics with $l=1$ represent
a global rotation (the toroidal harmonics $\vec{T}_{1m}$) and an effect called `glide'
(the spheroidal harmonics $\vec{S}_{1m}$). Glide has the same
mathematical form as the effect of acceleration given by Eq. (\ref{eq:accel_components}). 
One can demonstrate (Sect.~4.2 in \citeads{2012A&A...547A..59M}) that
%
%
\begin{align}\label{VSH-to-acceleration}
\begin{aligned}
  s_{10} &= \phantom{-}\sqrt{\frac{8\pi}{3}}\,g_z\,, \\
  s_{11}^\Re &= -\sqrt{\frac{4\pi}{3}}\,g_x\,,\\
  s_{11}^\Im &= \phantom{-}\sqrt{\frac{4\pi}{3}}\,g_y\,.
  \end{aligned}
\end{align}
In principle, therefore, one could restrict the model to
$l=1$. However, as already mentioned, the higher-order VSHs help to 
handle the effects of other systematic signals. The parameter
$l_{\rm max}$ in (\ref{Vexpandreal}) is the maximal order of the VSHs
that are taken into account in the model and is an important
instrument for analysing systematic signals in the data: by calculating 
a series of solutions for increasing values of 
$l_{\rm max}$, one probes how much the lower-order terms 
(and in particular the glide terms) are affected by higher-order systematics.

With the $L^2$ norm, the VSHs $\vec{T}_{lm}(\alpha,\delta)$ and $\vec{S}_{lm}(\alpha,\delta)$
form an orthonormal set of basis functions for a vector field on a
sphere. It is also known that the infinite set of these basis functions is complete on $S^2$. 
The VSHs can therefore represent arbitrary vector fields. Just as in the case of
scalar spherical harmonics, the VSHs with increasing order $l$
represent signals of higher spatial frequency on the sphere. VSHs of
different orders and degrees are orthogonal only if one has infinite
number of data points homogeneously distributed over the sphere.  For
a finite number of points and/or an inhomogeneous distribution the
VSHs are not strictly orthogonal and have a non-zero projection onto each other. 
This means that the coefficients $t_{lm}^\Re$,
$t_{lm}^\Im$, $s_{lm}^\Re$ and $s_{lm}^\Im$ are correlated when
working with observational data. The level of correlation depends on
the distribution of the statistical weight of the data over the sphere, which is illustrated by
Fig.~\ref{fig:sky-distribution-5p-information} for the source
selection used in this study. For a given weight distribution there is a
upper limit on the $l_{\rm max}$ that can be profitably used in practical
calculations. Beyond that limit the correlations between
the parameters become too high and the fit gets useless. Numerical tests
show that for our data selection it is reasonable to have $l_{\rm max}\lesssim 10$,
for which correlations are less than about 0.6 in absolute values.

Projecting Eq.~(\ref{Vexpandreal}) on the vectors $\vec{e}_\alpha$ and $\vec{e}_\delta$ of
the local triad one gets two scalar equations for each celestial
source with proper motions $\mu_{\alpha*}$ and $\mu_\delta$. For $k$
sources this gives $2k$ observation equations for 
$2l_\text{max}(l_\text{max}+2)$ 
unknowns to be solved for using a standard least-squares
estimator. The equations should be weighted using the uncertainties
of the proper motions $\sigma_{\mu_{\alpha*}}$ and
$\sigma_{\mu_\delta}$.  It is also advantageous to take into account,
in the weight matrix of the least-squares estimator, the correlation $\rho_\mu$
between $\mu_{\alpha*}$ and $\mu_\delta$ of a source.  This
correlation comes from the \gaia\ astrometric solution and is
published in the \gaia\ catalogue for each source.  The correlations
between astrometric parameters of different sources are not exactly
known and no attempt to account for these inter-source correlations
was undertaken in this study.

It is important that the fit is robust against outliers, that is sources 
that have proper motions significantly
deviating from the model in Eq.~(\ref{Vexpandreal}). Peculiar proper
motions can be caused by time-dependent structure variation of certain
sources (some but not all such sources have been rejected by the
astrometric tests at the selection level).  Outlier elimination also
makes the estimates robust against potentially bad, systematically
biased astrometric solutions of some sources.  The outlier detection
is implemented \citep{GAIA-LL-127} as an iterative elimination of all
{\sl sources} for which a measure of the post-fit residuals of the
corresponding two equations exceed the median value of that measure
computed for all sources by some chosen factor $\kappa\ge 1$, called
clip limit. As the measure $X$ of the weighted residuals for a source
we choose the post-fit residuals $\Delta\mu_{\alpha*}$ and $\Delta\mu_\delta$
of the corresponding two equations for $\mu_{\alpha*}$ and
$\mu_\delta$ for the source, weighted by the full covariance matrix
of the proper motion components:
\begin{eqnarray}
  X^2&=&
\begin{bmatrix}\Delta\mu_{\alpha*} &  \Delta\mu_\delta\end{bmatrix}
\begin{bmatrix}\sigma_{\mu_{\alpha*}}^2 & \rho_\mu\sigma_{\mu_{\alpha*}}\sigma_{\mu_\delta}\\
\rho_\mu\sigma_{\mu_{\alpha*}}\sigma_{\mu_\delta} & \sigma_{\mu_\delta}^2\end{bmatrix}^{-1}
\begin{bmatrix}\Delta\mu_{\alpha*}\\ \Delta\mu_\delta\end{bmatrix}\nonumber\\[6pt]  
&=&\frac{1}{1-\rho_\mu^2}\left[\left(\frac{\Delta\mu_{\alpha*}}{\sigma_{\mu_{\alpha*}}}\right)^2
-2\rho_\mu\left(\frac{\Delta\mu_{\alpha*}}{\sigma_{\mu_{\alpha*}}}\right)\left(\frac{\Delta\mu_\delta}{\sigma_{\mu_\delta}}\right)
+\left(\frac{\Delta\mu_\delta}{\sigma_{\mu_\delta}}\right)^2\right]\,.
\end{eqnarray}
At each iteration the least-squares fit is computed using only the sources that
were not detected as outliers in the previous iterations; the median of $X$ is 
however always computed over the whole set of sources. Iteration stops
when the set of sources identified as outliers is stable.%
\footnote{More precisely, the procedure stops the first time the set of outliers is
the same as in an earlier iteration (not necessarily the previous one).}
Identification of a whole source as an outlier and not just a single 
component of its proper motion (for example, accepting $\mu_{\alpha*}$ and
rejecting $\mu_\delta$) makes more sense from the
physical point of view and also makes the procedure independent
of the coordinate system. 

It is worth recording here that the angular covariance function $V_\mu(\theta)$, 
defined by Eq.~(17) of \citetads{2018A&A...616A...2L}, also contains information
on the glide, albeit only on its magnitude $|\vec{g}|$, not the direction. 
$V_\mu(\theta)$ quantifies the covariance of the proper motion vectors $\vec{\mu}$ 
as a function of the angular separation $\theta$ on the sky. 
Figure~14 of \citet{DPACP-128} shows this function for \gedr{3}, computed
using the same sample of QSO-like sources with five-parameter solutions as used
in the present study (but without weighting the data according to their uncertainties). 
Analogous to the case of scalar fields on a sphere (see Sect.~5.5 of \citeads{DPACP-128}), 
$V_\mu(\theta)$ is related to the VSH expansion of the vector field $\vec{\mu}(\alpha,\delta)$. 
In particular, the glide vector $\vec{g}$ gives a contribution of the form
\begin{equation}\label{V-glide}
V^{\rm glide}_\mu(\theta)=|\vec{g}|^2\,{1\over 6}\,\left(\cos^2\theta+1\right) \, .
\end{equation}
Using this expression and the $V_\mu(\theta)$ of \gedr{3} we obtain an estimate 
of $|\vec{g}|$ in reasonable agreement with the results from the VSH fit discussed 
in the next section. However, it is obvious from the plot in \citet{DPACP-128} 
that the angular covariance function contains other large-scale 
components that could bias this estimate as they are not included in the fit.
This reinforces the argument made earlier in this section, namely that the 
estimation of the glide components from the proper motion data should not be 
done in isolation, but simultaneously with the estimation of other large-scale 
patterns. This is exactly what is achieved by means of the VSH expansion.

\section{Analysis}
\label{sec:analysis}

\begin{figure}[htbp]
\begin{center}
  \includegraphics[width=1\hsize]{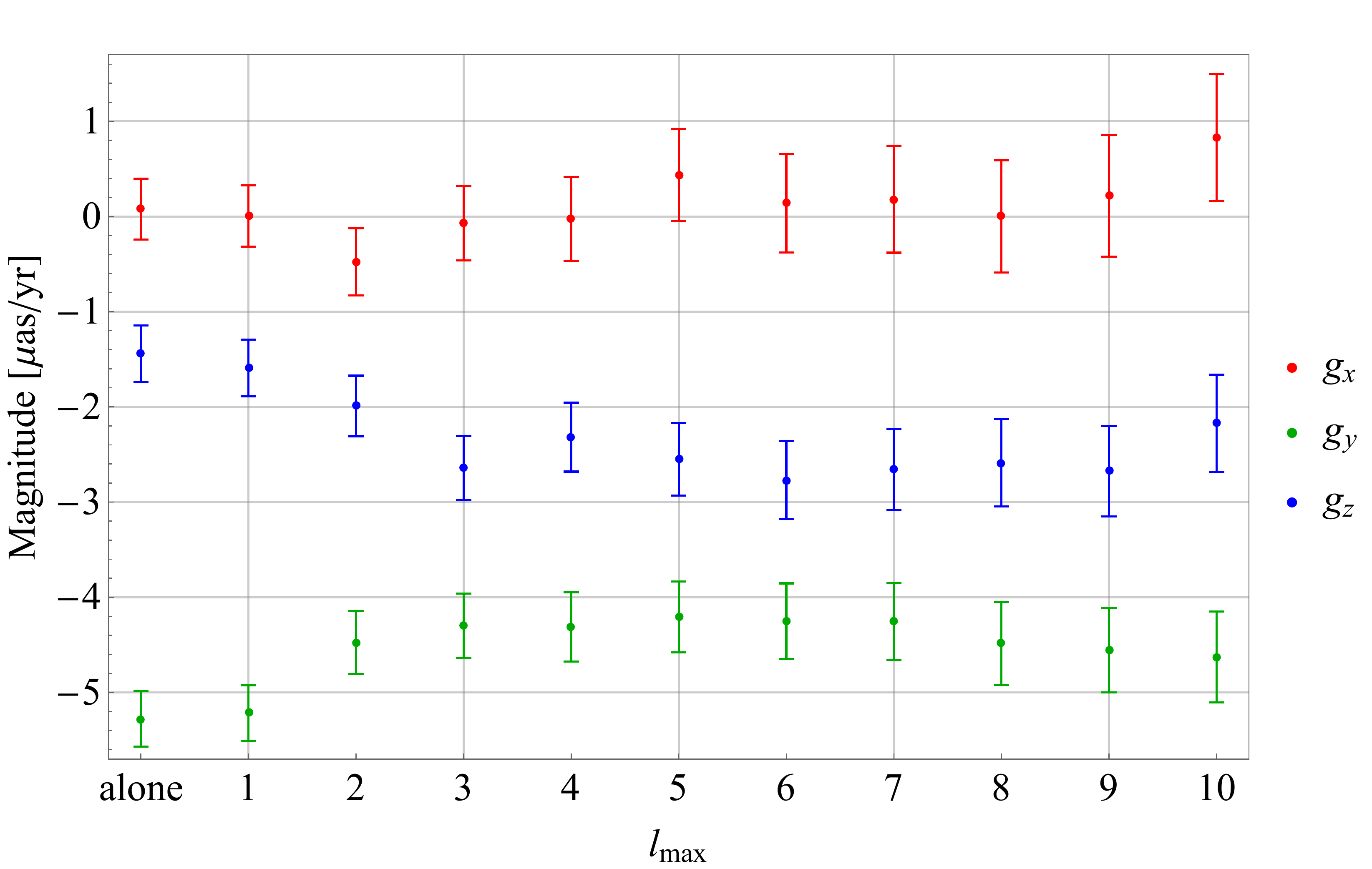}
  \includegraphics[width=1\hsize]{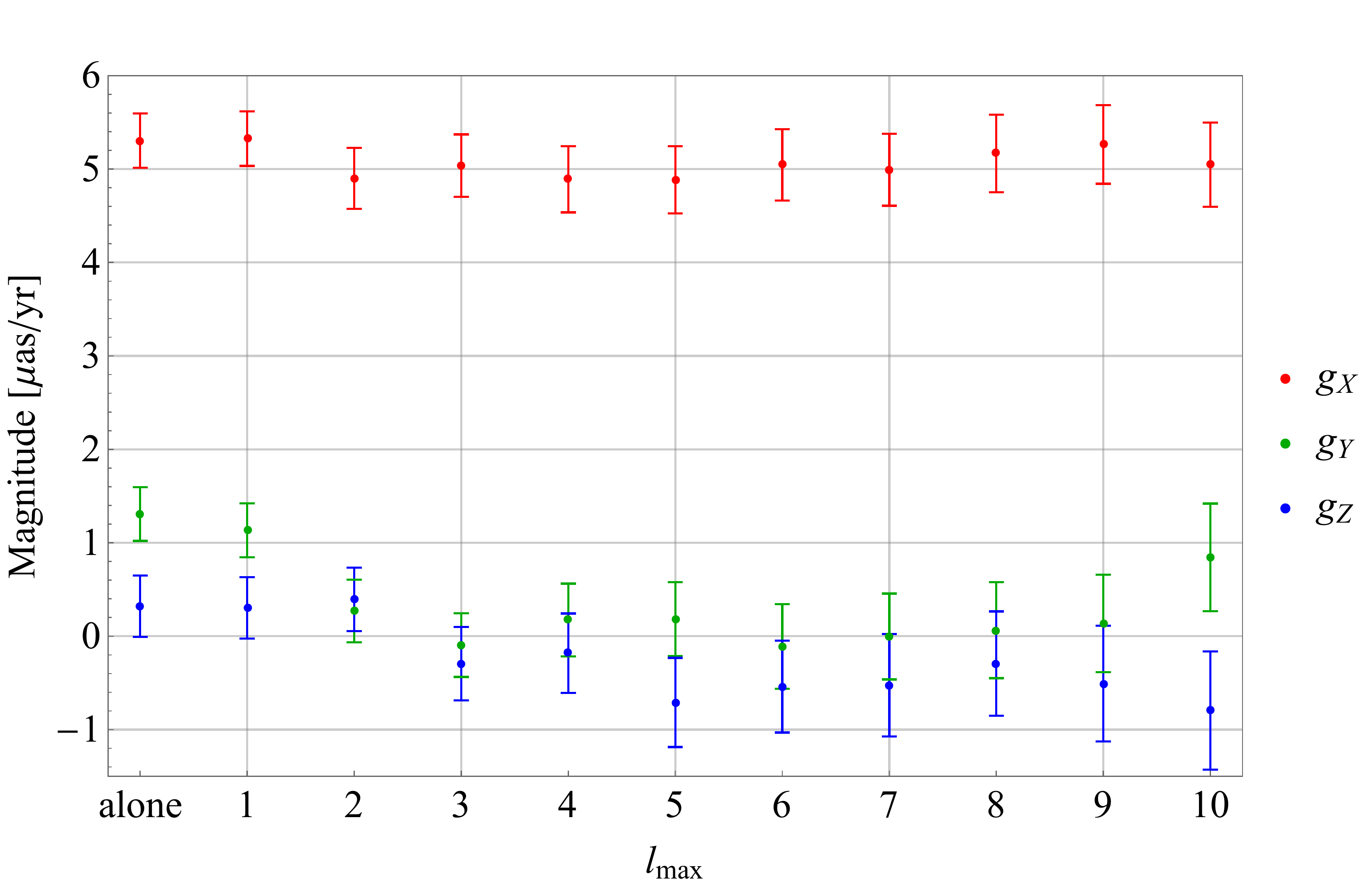}
  \caption{Equatorial (upper pane) and galactic (lower pane) components of the solar system
    acceleration for fits with different maximal
VSH order $l_{\rm max}$ (`alone' means that the three glide components
were fitted with no other VSH terms). The error bars represent $\pm 1\sigma$
uncertainties.}
\label{fig:acceleration-lmax10}
\end{center}
\end{figure}

The results for the three components of the glide vector are shown in
Fig.~\ref{fig:acceleration-lmax10}.  They have been obtained by fitting the VSH
expansion in Eq.~(\ref{Vexpandreal}) for different $l_{\rm max}$ to the proper motions of the
1\,215\,942 \gcrfthree\ sources with five-parameter solutions.  The corresponding spheroidal VSH parameters with
$l=1$ were transformed into the Cartesian components of the glide
using Eq.~(\ref{VSH-to-acceleration}).  Figure~\ref{fig:acceleration-lmax10}
displays both the equatorial components $(g_x,\,g_y,\,g_z)$ 
and the galactic components $(g_X,\,g_Y,\,g_Z)$ of the glide vector.
The equatorial components were derived
directly using the equatorial proper motions published in the
\gaia\ Archive. The galactic components can be derived either by
transforming the equatorial components of the glide and their covariance matrix to
galactic coordinates, or from a direct VSH fits using the proper motions
and covariances in galactic coordinates. We have verified that the two
procedures give strictly identical results.

One can see that starting from $l_{\rm max}=3$ the estimates are
stable and generally deviate from each other by less than the corresponding
uncertainties. 
The deviation of the results for $l_{\rm max}<3$ from those of higher $l_{\rm max}$
shows that the higher-order systematics in the data need to be
taken into account, although their effect on the glide is relatively mild.
We conclude that it is reasonable
to use the results for $l_{\rm max}=3$ as the best estimates of the
acceleration components. 

The unit weight error (square root of the reduced chi-square)
of all these fits, and of all those described below, is about 1.048.
The unit weight error calculated with all VSH terms set to zero is
also 1.048 (after applying the same outlier rejection procedure as for
the fits), which merely reflects the fact that the fitted VSH terms are much
smaller than the uncertainties of the individual proper motions.
The unit weight error is routinely used to
scale up the uncertainties of the fit. However, a more robust method
of bootstrap resampling was used to estimate the uncertainties (see
below).  

To further investigate the influence of various aspects of the
data and estimation procedure, the following
tests were done.
\begin{itemize}
\item Fits including VSH components of degree up to $l_{\rm max}=40$
  were made. They show that the variations of the estimated
  acceleration components remain at the level of a fraction of the
  corresponding uncertainties, which agrees with random variations
  expected for the fits with high $l_{\rm max}$.
\item The fits in Fig.~\ref{fig:acceleration-lmax10} used the clip limit
  $\kappa=3$, which rejected about 3800 of the 1\,215\,942 sources 
  as outliers (the exact number depends on 
  $l_{\rm max}$).  Fits with different clip limits 
  $\kappa$ (including fits without outlier rejection, corresponding
  to $\kappa=\infty$) were tried, showing that the result for the
  acceleration depends on $\kappa$ only at a level of a quarter of the
  uncertainties.
\item The use of the correlations $\rho_\mu$ between the proper
  motion components for each source in the weight matrix of the fit
  influences the acceleration estimates at a level of $\sim 0.1$ of
  the uncertainties. This should be expected since the correlations
  $\rho_\mu$ for the 1\,215\,942 \gcrfthree\ sources are relatively
  small (the distribution of $\rho_\mu$ is reasonably close to normal with 
  zero mean and standard deviation 0.28).
\end{itemize}

\begin{figure}[t]
\begin{center}
  \includegraphics[width=1.0\hsize]{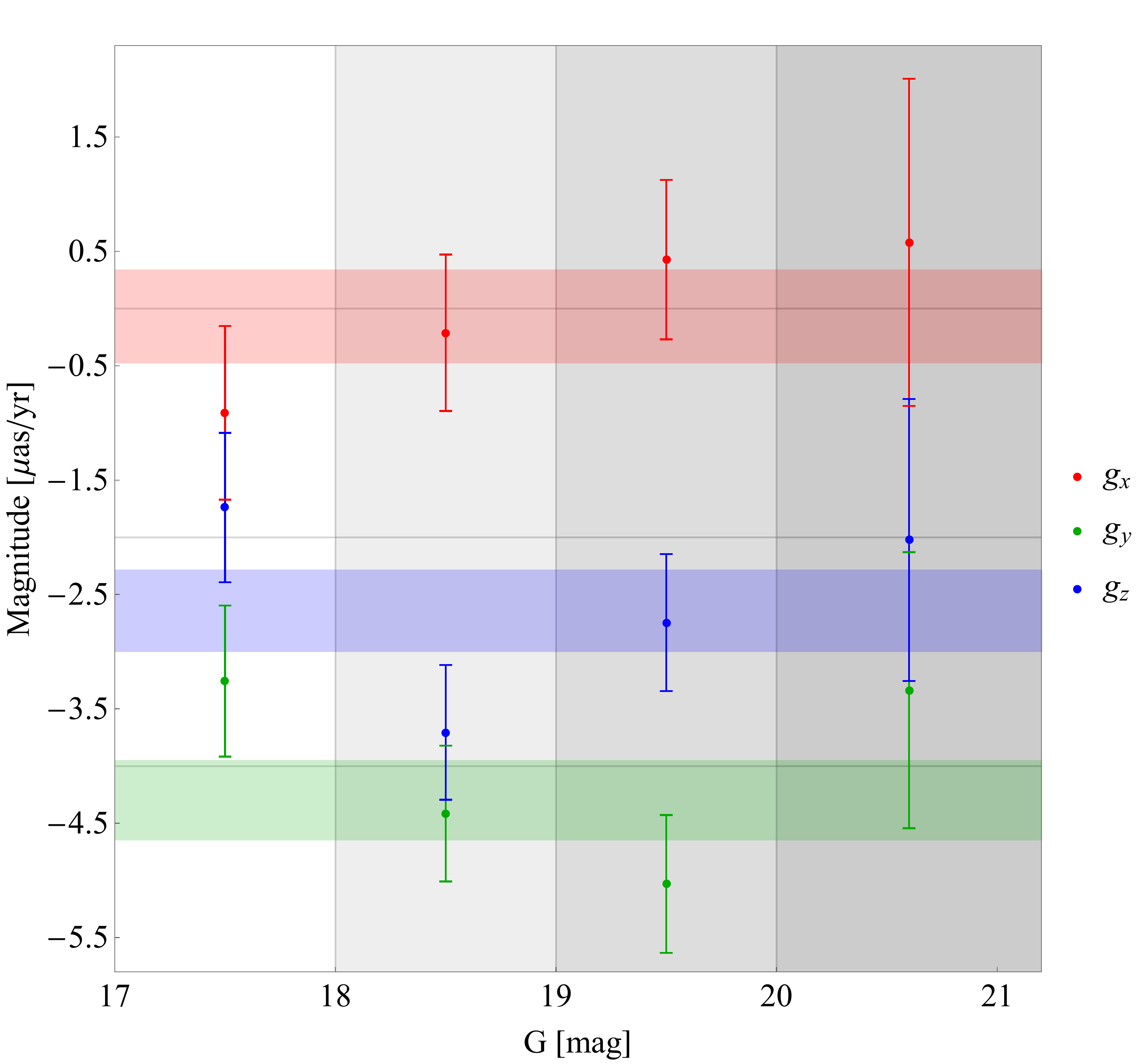}
  \caption{Equatorial components of the acceleration and their
    uncertainties for four intervals of $G$ magnitude: $G\le18$ mag
    (29\,200 sources), $18<G\le19$ mag (146\,614 sources), $19<G\le20$ mag
    (490\,161 sources), and $G>20$ mag (549\,967 sources).  The horizontal
    colour bands visualize the values and uncertainties (the height
    corresponds to twice the uncertainty) of the corresponding
    components computed from the whole data set.}
\label{fig:acceleration-gMag-selections}
\end{center}
\end{figure}

\begin{figure}[t]
\begin{center}
  \includegraphics[width=1.0\hsize]{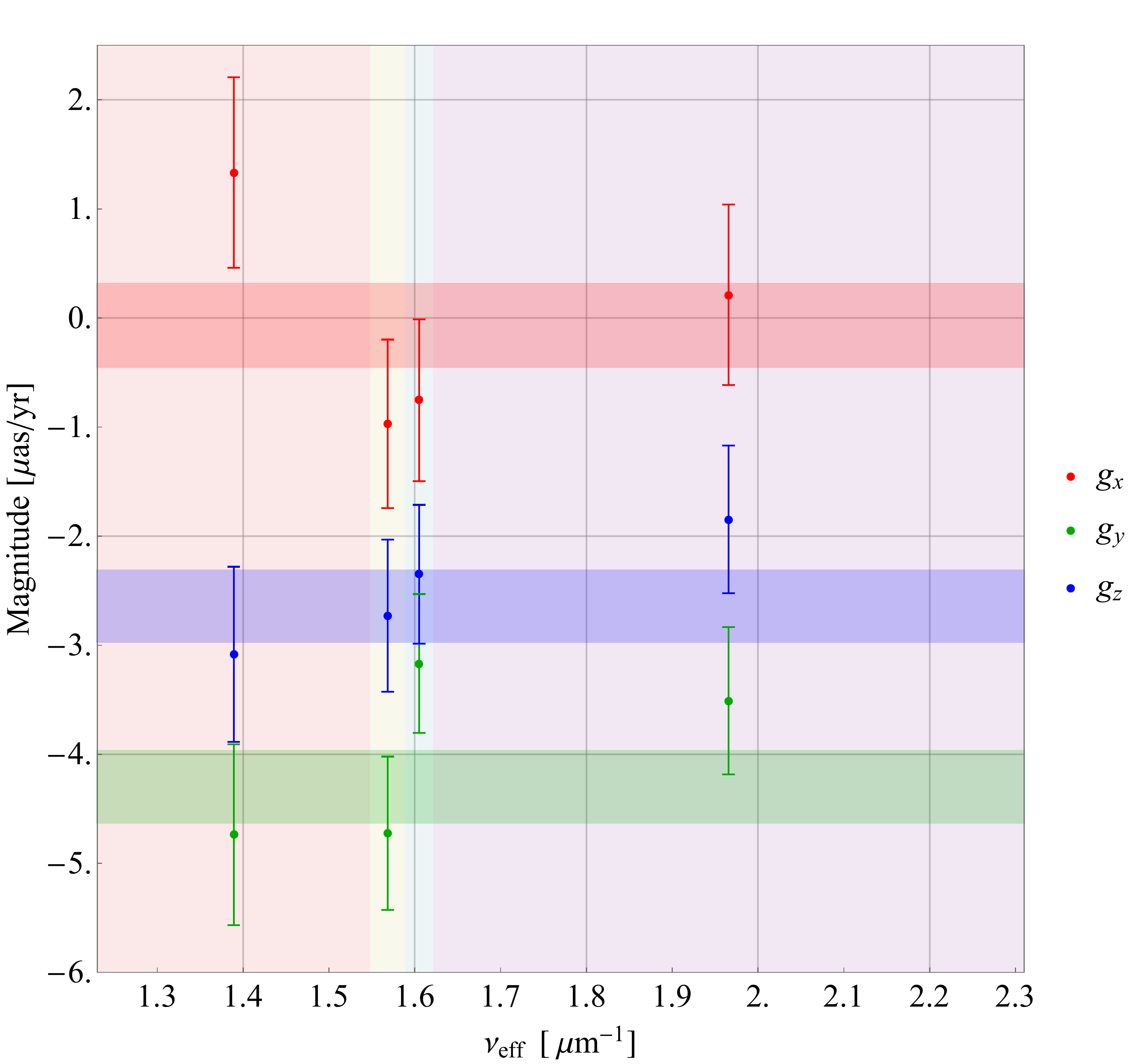}
  \caption{Equatorial components of the acceleration and their
    uncertainties for four intervals of the colour represented by the
    effective wavenumber $\nu_{\rm eff}$ used in
    \gdr{3}\ astrometry. The quartiles of the $\nu_{\rm eff}$
    distribution for the sources considered in this study are used as
    the boundaries of the $\nu_{\rm eff}$ intervals so that each
    interval contains about 304\,000 sources.  The horizontal colour
    bands visualize the values and uncertainties (the height
    corresponds to twice the uncertainty) of the corresponding
    components computed from the whole data set.}
  \label{fig:acceleration-colour-selections}
\end{center}
\end{figure}

Analysis of the \gdr{3}\ astrometry has revealed systematic errors
depending on the magnitude and colour of the sources
\citep{DPACP-128,DPACP-132}. To check how these factors 
influence the estimates, fits using $l_{\rm max}=3$ were 
made for sources split by magnitude and colour:
\begin{itemize}
\item Figure~\ref{fig:acceleration-gMag-selections} shows the acceleration 
components estimated for subsets of different mean $G$ magnitude. The
  variation of the components with $G$ is mild and the
  estimates are compatible with the estimates from the full data set
  (shown as horizontal colour bands) within their uncertainties.

\item Figure~\ref{fig:acceleration-colour-selections} is a corresponding
plot for the split by colour, as represented by the
  effective wavenumber $\nu_{\rm eff}$. Again one can
  conclude that the estimates from the data selections in colour agree
  with those from the full data set within their corresponding
  uncertainties.
\end{itemize}

It should be noted that the magnitude and colour selections are not
completely independent since the bluer QSO-like sources tend to be fainter
than the redder ones. Moreover, the magnitude and colour selections are less
homogeneous on the sky than the full set of sources (for example owing to the
Galactic extinction and reddening). However, we conclude that the
biases in the acceleration estimates, due to magnitude-
and colour-dependent effects in the \gdr{3}\ astrometry, are below the
formal uncertainties for the full sample.

Another possible cause of biases in the \textit{Gaia} data is charge transfer
inefficiency (CTI) in the CCDs (e.g.\ \citeads{2016A&A...595A...6C}). 
A detailed simulation of plausible CTI effects unaccounted for in the 
\gaia\ data processing for \gdrthree\ showed that the estimated glide
is remarkably resilient to the CTI and may be
affected only at a level below $0.1\muasyr$ -- at most a quarter of
the quoted uncertainty.


Our selection of \gaia\ sources cannot be absolutely free
from stellar contaminants.  As discussed in Sect.~\ref{sec:stars},
stars in our Galaxy have very large glide components in the vector field of their proper
motions. This means that even a small stellar contamination could bias
our estimate of the solar system acceleration.  One can hope that the
mechanism of outlier elimination used in the VSH fit in this work (see
Sect.~\ref{sec:method}) helps to eliminate at least some of the most
disturbing stellar-contamination sources. It is, however, worth to
investigate the possible biases by direct simulation. By construction,
the stellar contaminants in our list of QSO-like sources must have  
five-parameter solutions in \gdr{3}\ that satisfy the selection criteria 
discussed in Sect.~\ref{sec:qso-like} and \citep{DPACP-133}. 
It is therefore of interest to investigate the sample of sources obtained
by making exactly the same selection of \gdr{3}\ sources, but without
the cross-match to the external QSO/AGN catalogues.
There are a total of 23.6~million such sources in \gdr{3}, including
the 1.2~million (5.2\%) included in
\gcrfthree. Most of them are stars in our Galaxy, but one also sees
stars in nearby dwarf galaxies, globular clusters, and bright stars in 
other galaxies. Applying the VSH method to this sample 
gives a glide of about $360\muasyr$ in a
direction within a few degrees of $(l,b)=(270\degr,\,0\degr)$,
that is roughly opposite to the direction of motion of the Sun
in the Galaxy. This glide has obviously nothing to do
with the acceleration of the solar system (see Sect.~\ref{sec:stars})
and its precise value is irrelevant. However, it is very relevant that 
it is practically perpendicular to the glide obtained from the
QSO-like sample, for it means that a (small) stellar contamination 
will not significantly alter the magnitude of the glide $|\vec{g}|$. 
It could however bias the direction of the observed glide towards 
$(l,b)=(270\degr,\,0\degr)$, that is mainly in galactic longitude.
We do not see a clear sign of this in our estimates (the estimated 
direction is within one $\sigma$ from the Galactic centre) and we 
therefore conclude that the effect of a possible stellar contamination 
in \gcrfthree\ is negligible for the claimed estimate of the solar 
system acceleration.

\begin{figure}
\begin{center}
 \includegraphics[width=1\hsize]{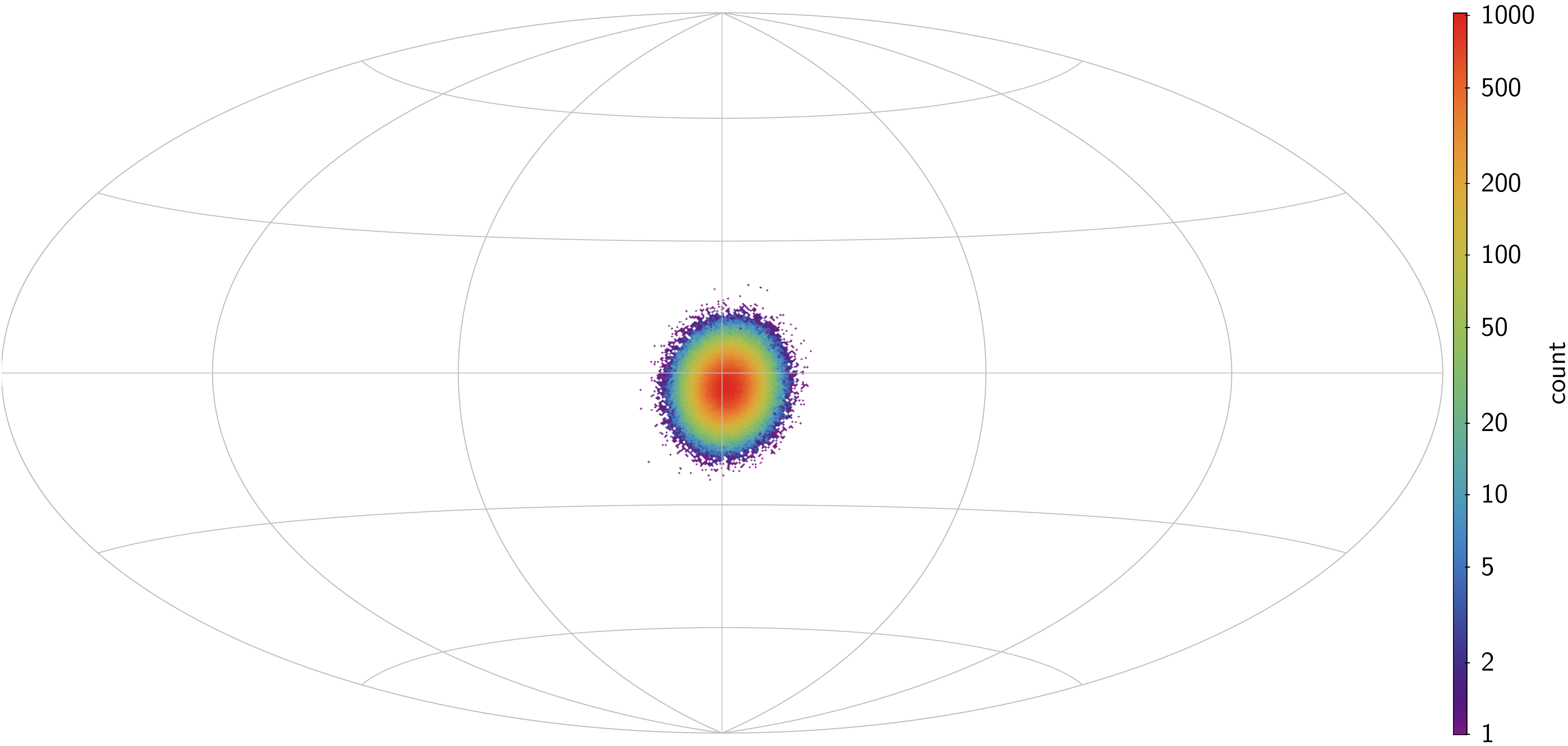}
  \caption{Visualizing of the error ellipse of the estimated direction of the 
  acceleration estimate in galactic coordinates. The plot is a density map
    of the directions from 550\,000 bootstrap resampling experiments. 
    The colour scale is logarithmic.}
\label{fig:galactic-error-ellipse}
\end{center}
\end{figure}

Finally, it should be remembered that systematic errors in the
\textit{Gaia}\ ephemeris may also bias the estimate of the solar
system acceleration.  The standard astrometric parameters in the
\textit{Gaia}\ astrometric solution are defined for a fictitious
observer located in the `solar system barycentre'. The latter is
effectively defined by the \textit{Gaia}\ ephemeris in the Barycentric
Celestial Reference Frame (BCRS; \citeads{2003AJ....126.2687S};
\citeads{2003AJ....125.1580K}) that is used in the data processing. In
particular, the \textit{Gaia}'s barycentric velocity is used to
transform the observations from the proper frame of \textit{Gaia} to
the reference frame at rest with respect to the solar system
barycentre \citepads{2004PhRvD..69l4001K}. Systematic errors in the
\textit{Gaia}\ ephemeris may result in systematic errors in the
astrometric parameters. In particular, a systematic error in the
\textit{Gaia} velocity, corresponding to a non-zero average
acceleration error over the time interval of the observations (about
33~months for \gedr{3}), will produce the same systematic error in the
measured solar system acceleration.

The barycentric ephemeris of \textit{Gaia} is obtained by combining
the geocentric orbit determination, made by the Mission Operations
Centre at ESOC (Darmstadt, Germany) using various Doppler and ranging
techniques, with a barycentric ephemeris of the Earth\footnote{The
  `geocentric' orbit of \textit{Gaia}\ is also defined in the BCRS
  and represents the difference of the BCRS coordinates of
  \textit{Gaia}\ and those of the geocentre.}. For the latter, the
INPOP10e planetary ephemerides \citepads{2016NSTIM.104.....F} was used
in \gedr{3}.  The errors in the geocentric orbit have very different
characteristics from those of the planetary ephemerides, and the two
contributions need to be considered separately. For the geocentric
part, one can rule out an acceleration bias greater than about
$2\times 10^{-13}\,\text{m}\,\text{s}^{-2}$ persisting over the
33~months, because it would produce an offset in the position of
\textit{Gaia} of the order of a km, well above the accuracy obtained
by the ranging. For the barycentric ephemeris of the Earth, we can
obtain an order-of-magnitude estimate of possible systematics by
comparing the INPOP10e ephemerides with the latest version, INPOP19a
\citepads{2019NSTIM.109.....F}, which will be used for
\gdr{4}. Averaged over 33~months, the difference in the acceleration
of the Earth between the two versions is of the order of
$10^{-12}\,\text{m}\,\text{s}^{-2}$, that is about 0.5\% of the
observed (and expected) acceleration of the solar system
barycentre. These differences in the Earth ephemeris come from the
improvements in the dynamical modelling of the solar system and the
new observational data allowing for more accurate determination of the
parameters of the solar system bodies.  One can expect that the
process of improvement will continue and involve, in particular, more
objects in the outer solar system that can potentially influence the
definition of the solar system barycentre. For example, the
hypothetical Planet Nine would have an effect that was at most
$5\times 10^{-13}\,\text{m}\,\text{s}^{-2}$
\citepads{2020A&A...640A...6F}. Taking all these aspects into account,
we conclude that plausible systematic errors in the barycentric
ephemeris of \textit{Gaia}\ are too small, by at least two orders of
magnitude, to invalidate our result. Nevertheless, special care should
be taken for this source of systematic errors when considerably more
accurate measurements of the solar system acceleration -- e.g. from a
combination of the \textit{Gaia} and \textit{GaiaNIR} data
\citepads{2016arXiv160907325H} -- will become available.

The various tests and arguments reported above strengthen our confidence in the
final results, which are summarised in Table~\ref{tab:results}. 
Both the equatorial and galactic components are given with their
uncertainties and correlations. The uncertainties were estimated by bootstrap
resampling \citep{efron1994introduction}, which in our case
increased the uncertainties from the fit (already
inflated by the unit weight error) by factors of 1.05 to 1.08.
As shown already in Fig.~\ref{fig:acceleration-lmax10}, the
direction of the measured acceleration is very close to the Galactic
centre. This is also illustrated in Fig.~\ref{fig:galactic-error-ellipse}, 
which shows the directions obtained in the bootstrap resampling.

\begin{table}
  \caption{Principal results of this work: equatorial and galactic components 
  of the estimated acceleration of the solar system, with uncertainties and correlations.
\label{tab:results}}
  \footnotesize\setlength{\tabcolsep}{4pt}
\begin{center}
\begin{tabular}{lrr}
\hline\hline
\noalign{\smallskip}
quantity & value & uncertainty \\[1pt]
\hline
\noalign{\smallskip}
\multicolumn{3}{c}{equatorial components} \\[2pt]
$g_x$ [$\muasyr$] & $-0.07$ & 0.41 \\[2pt]
$g_y$ [$\muasyr$] & $-4.30$ & 0.35 \\[2pt]
$g_z$ [$\muasyr$] & $-2.64$ & 0.36 \\[4pt]
$\alpha$ & $269.1\degr$ & $5.4\degr$ \\
$\delta$ & $-31.6\degr$ & $4.1\degr$ \\[3pt]
\multicolumn{3}{c}{correlations} \\[1pt]
$\rho_{g_x,g_y}$ & $+0.001$ &\\
$\rho_{g_x,g_z}$ & $-0.094$ &\\
$\rho_{g_y,g_z}$ & $-0.025$ &\\[3pt]
$\rho_{\alpha,\delta}$ & $-0.081$ &\\[5pt]
\hline
\noalign{\smallskip}
\multicolumn{3}{c}{galactic components} \\
$g_X$ [$\muasyr$] & $+5.04$ & 0.35 \\
$a_X$ [$\kmsMyr$] & $+7.32$ & 0.51 \\[2pt]
$g_Y$ [$\muasyr$] & $-0.10$ & 0.36 \\
$a_Y$ [$\kmsMyr$] & $-0.14$ & 0.52 \\[2pt]
$g_Z$ [$\muasyr$] & $-0.29$ & 0.41 \\
$a_Z$ [$\kmsMyr$] & $-0.43$ & 0.60 \\[3pt]
$l$ & $358.9\degr$ & $4.1\degr$ \\
$b$ & $-3.3\degr$ & $4.6\degr$ \\[3pt]
\multicolumn{3}{c}{correlations} \\[1pt]
$\rho_{g_X,g_Y}$ & $+0.036$ &\\
$\rho_{g_X,g_Z}$ & $-0.014$ &\\
$\rho_{g_Y,g_Z}$ & $-0.079$ &\\[3pt]
$\rho_{l,b}$ & $-0.078$ &\\[5pt]
\hline
\noalign{\smallskip}
$|\,\vec{g}\,|$ [$\muasyr$] & 5.05 & 0.35 \\
$|\,\vec{a}\,|$ [$\kmsMyr$] & 7.33 & 0.51 \\
\phantom{$|\,\vec{a}\,|$} [$10^{-10}\mssquared$] & 2.32 & 0.16 \\[3pt]
\hline
\end{tabular}
\end{center}
\tablefoot{All uncertainties are $\pm1\sigma$ estimates obtained using bootstrap
  resampling. The absolute values of the acceleration are computed as
  the Euclidean norm of the estimated vector, and may be biased as discussed
  in Appendix~\ref{sec:unbiased}.}
\end{table}

\section{Conclusions and prospects}
\label{sec:summary}

The exquisite quality of the \gdrthree\ astrometry together with a
careful selection of the \gcrfthree\ sources (Sect.~\ref{sec:qso-like}) 
have allowed us to detect the acceleration of the
solar system with respect to the rest-frame of the remote
extragalactic sources, with a relative precision better than 10\%.
The stability of the derived estimates was extensively checked by 
numerous experiments as discussed in Sect.~\ref{sec:analysis}. 
The consistency of the results 
support the overall claim of a significant detection. We
note that our estimate of the solar system acceleration agrees with the
theoretical expectations from galactic dynamics (Sect.~\ref{sec:expectation})
within the corresponding uncertainties.

We stress that the detection of the solar system acceleration in the
Gaia astrometry does not require any dedicated astrometric
solution. The astrometric data used in this work to detect the acceleration
and analyze its properties are those of the 
astrometric solution published in \gedr{3}. 


Although the relative accuracy obtained in the estimate 
is very satisfactory for this data release, it is at
this stage impossible to tell whether there are acceleration
contributions from other components than the motion of the solar
system in the Milky Way. As discussed in
Sect.~\ref{sec:expectation}, even this contribution is complex
and cannot be modelled with sufficient certainty to disentangle the
different contributions.

We can ask ourselves what should be expected from \gaia\ in the
future. The astrometric results in \gedr{3}\ are based only on
33~months of data, while the future \gdr{4}\ will be based on about
66~months of data and the final \gdr{5}\ may use up to 120~months of
data.  Since the effect of the acceleration is equivalent to proper
motions, the random uncertainty of its measurement improves with
observational time $T$ as $T^{-3/2}$. Therefore, we can expect that
the random errors of the acceleration estimated in \gdr{4}\ and
\gdr{5}\ could go down by factors of about $0.35$ and $0.15$,
respectively.

But random error is just one side of the story. What has made this
solution possible with \gedr{3}, while it was not possible with the
\gdr2 data, is the spectacular decrease of the systematic errors in
the astrometry. To illustrate this point, the glide determined from
the \gcrftwo\ data (Sect.~3.3 in \citeads{2018A&A...616A..14G}) was at
the level of $10\muasyr$ per component, much higher than a solution
strictly limited by random errors.  With the \gedr{3} we have a random
error on each proper motion of about $\simeq 400\muasyr$ and just over
1~million sources. So one could hope to reach $0.4\muasyr$ in the
formal uncertainty of the glide components, essentially what is now
achieved. In future releases, improvement for the solar system
acceleration will come both from the better random errors and the
reduced systematic errors, although only the random part can be
quantified with some certainty.  In the transition from \gdr2 to
\gedr{3} a major part of the gain came from the diminishing of
systematic effects.

The number of QSO-like sources that can become available in
future \gaia\ data releases is another interesting aspect.  In
general, a reliable answer is not known. Two attempts
(\citeads{2019MNRAS.489.4741S};\citeads{2019MNRAS.490.5615B}) to find QSO-like
sources in \gdr{2}\ data ended up with about 2.7~million sources each
(and even more together). Although an important part of those
catalogues did not show the level of reliability we require for \gcrfthree,
one can hope that the number of QSO-like sources with
\gaia\ astrometry will be doubled in the future compared to
\gdr{3}. Taking all these aspects into account, it is reasonable to
hope the uncertainty of the acceleration to reach the level of well
below $0.1\muasyr$ in the future \gaia\ releases.

Considering the expected accuracy, an interesting question here is if we
could think of any other effects that would give systematic patterns
in the proper motions of QSO-like sources at the level of expected
accuracy. Such effects are indeed known (a good overview of these
effects can be found e.g. in \citepads{2016A&A...589A..71B}). One such
effect is the `cosmological proper motion' \citepads{1986SvA....30..501K},
or `secular extragalactic parallax' \citepads{2020ApJ...890..146P},
caused by the motion of the solar system with respect to the rest frame of the CMB 
at a speed of $370\kms \approx 78 {\,\mathrm{au\,{yr}^{-1}}}$
towards the point with galactic
coordinates $l=264.02\degr, b=48.25\degr$ (\citeads{2020A&A...641A...3P}; 
see also Sect.~\ref{sec:effect}). This gives a reflex proper motion of 
$78\muasyr\,\times \left(1\,\text{Mpc}\,/\,d\right)\,\sin\beta$, where $d$ is the distance
to the object and $\beta$ is the angle between the object and the direction of motion \citepads{2016A&A...589A..71B}. 
The effect is analogous to the systematic proper motions of nearby stars caused by
the apex motion of the Sun (Sect.~\ref{sec:stars}), and like it decreases with the
inverse distance to the sources. At a redshift of 0.2 the systematic proper 
motion should be about $0.1\muasyr$ at right angle to the solar motion. 
However, only a few thousand QSO-like objects can be expected at such small 
redshifts, and, as discussed e.g. by \citetads{2020ApJ...890..146P}, the effect is 
muddled by the peculiar velocities of the objects and deviations of their bulk
motions from the Hubble flow due to the gravitational interactions with large-scale 
structures. It therefore remains questionable if this systematic proper motion will become 
accessible to \gaia\ in the future.

Another secular shift of the positions of extragalactic sources
comes from the light bending in the gravitational field of the
Galaxy, which depends (among other things) on the angle between the source 
and the Galactic centre. The motion of the solar system in the Galaxy results 
in a slow variation of this angle, which causes a variation of the light bending.
This will be seen as a proper motion of the extragalactic source. 
The effect is independent of the distance to the source (as long as it is far 
away from the Milky Way), but depends on its position on the sky according
to the details of the Galactic potential. The VSH technique used in this work 
seems to be very well suited for disentangling this effect from that of the
solar system acceleration.


\section{Acknowledgements\label{sec:acknowl}}

This work presents results from the European Space Agency (ESA) space mission \gaia. \gaia\ data are being processed by the \gaia\ Data Processing and Analysis Consortium (DPAC). Funding for the DPAC is provided by national institutions, in particular the institutions participating in the \gaia\ MultiLateral Agreement (MLA). The \gaia\ mission website is \url{https://www.cosmos.esa.int/gaia}. The \gaia\ archive website is \url{https://archives.esac.esa.int/gaia}.

The \gaia\ mission and data processing have financially been supported by, in alphabetical order by country:
\begin{itemize}
\item the Algerian Centre de Recherche en Astronomie, Astrophysique et G\'{e}ophysique of Bouzareah Observatory;
\item the Austrian Fonds zur F\"{o}rderung der wissenschaftlichen Forschung (FWF) Hertha Firnberg Programme through grants T359, P20046, and P23737;
\item the BELgian federal Science Policy Office (BELSPO) through various PROgramme de D\'eveloppement d'Exp\'eriences scientifiques (PRODEX) grants and the Polish Academy of Sciences - Fonds Wetenschappelijk Onderzoek through grant VS.091.16N, and the Fonds de la Recherche Scientifique (FNRS);
\item the Brazil-France exchange programmes Funda\c{c}\~{a}o de Amparo \`{a} Pesquisa do Estado de S\~{a}o Paulo (FAPESP) and Coordena\c{c}\~{a}o de Aperfeicoamento de Pessoal de N\'{\i}vel Superior (CAPES) - Comit\'{e} Fran\c{c}ais d'Evaluation de la Coop\'{e}ration Universitaire et Scientifique avec le Br\'{e}sil (COFECUB);
\item the National Science Foundation of China (NSFC) through grants 11573054 and 11703065 and the China Scholarship Council through grant 201806040200;  
\item the Tenure Track Pilot Programme of the Croatian Science Foundation and the \'{E}cole Polytechnique F\'{e}d\'{e}rale de Lausanne and the project TTP-2018-07-1171 'Mining the Variable Sky', with the funds of the Croatian-Swiss Research Programme;
\item the Czech-Republic Ministry of Education, Youth, and Sports through grant LG 15010 and INTER-EXCELLENCE grant LTAUSA18093, and the Czech Space Office through ESA PECS contract 98058;
\item the Danish Ministry of Science;
\item the Estonian Ministry of Education and Research through grant IUT40-1;
\item the European Commission’s Sixth Framework Programme through the European Leadership in Space Astrometry (\href{https://www.cosmos.esa.int/web/gaia/elsa-rtn-programme}{ELSA}) Marie Curie Research Training Network (MRTN-CT-2006-033481), through Marie Curie project PIOF-GA-2009-255267 (Space AsteroSeismology \& RR Lyrae stars, SAS-RRL), and through a Marie Curie Transfer-of-Knowledge (ToK) fellowship (MTKD-CT-2004-014188); the European Commission's Seventh Framework Programme through grant FP7-606740 (FP7-SPACE-2013-1) for the \gaia\ European Network for Improved data User Services (\href{https://gaia.ub.edu/twiki/do/view/GENIUS/}{GENIUS}) and through grant 264895 for the \gaia\ Research for European Astronomy Training (\href{https://www.cosmos.esa.int/web/gaia/great-programme}{GREAT-ITN}) network;
\item the European Research Council (ERC) through grants 320360 and 647208 and through the European Union’s Horizon 2020 research and innovation and excellent science programmes through Marie Sk{\l}odowska-Curie grant 745617 as well as grants 670519 (Mixing and Angular Momentum tranSport of massIvE stars -- MAMSIE), 687378 (Small Bodies: Near and Far), 682115 (Using the Magellanic Clouds to Understand the Interaction of Galaxies), and 695099 (A sub-percent distance scale from binaries and Cepheids -- CepBin);
\item the European Science Foundation (ESF), in the framework of the \gaia\ Research for European Astronomy Training Research Network Programme (\href{https://www.cosmos.esa.int/web/gaia/great-programme}{GREAT-ESF});
\item the European Space Agency (ESA) in the framework of the \gaia\ project, through the Plan for European Cooperating States (PECS) programme through grants for Slovenia, through contracts C98090 and 4000106398/12/NL/KML for Hungary, and through contract 4000115263/15/NL/IB for Germany;
\item the Academy of Finland and the Magnus Ehrnrooth Foundation;
\item the French Centre National d’Etudes Spatiales (CNES), the Agence Nationale de la Recherche (ANR) through grant ANR-10-IDEX-0001-02 for the 'Investissements d'avenir' programme, through grant ANR-15-CE31-0007 for project 'Modelling the Milky Way in the Gaia era' (MOD4Gaia), through grant ANR-14-CE33-0014-01 for project 'The Milky Way disc formation in the Gaia era' (ARCHEOGAL), and through grant ANR-15-CE31-0012-01 for project 'Unlocking the potential of Cepheids as primary distance calibrators' (UnlockCepheids), the Centre National de la Recherche Scientifique (CNRS) and its SNO Gaia of the Institut des Sciences de l’Univers (INSU), the 'Action F\'{e}d\'{e}ratrice Gaia' of the Observatoire de Paris, the R\'{e}gion de Franche-Comt\'{e}, and the Programme National de Gravitation, R\'{e}f\'{e}rences, Astronomie, et M\'{e}trologie (GRAM) of CNRS/INSU with the Institut National Polytechnique (INP) and the Institut National de Physique nucléaire et de Physique des Particules (IN2P3) co-funded by CNES;
\item the German Aerospace Agency (Deutsches Zentrum f\"{u}r Luft- und Raumfahrt e.V., DLR) through grants 50QG0501, 50QG0601, 50QG0602, 50QG0701, 50QG0901, 50QG1001, 50QG1101, 50QG1401, 50QG1402, 50QG1403, 50QG1404, and 50QG1904 and the Centre for Information Services and High Performance Computing (ZIH) at the Technische Universit\"{a}t (TU) Dresden for generous allocations of computer time;
\item the Hungarian Academy of Sciences through the Lend\"{u}let Programme grants LP2014-17 and LP2018-7 and through the Premium Postdoctoral Research Programme (L.~Moln\'{a}r), and the Hungarian National Research, Development, and Innovation Office (NKFIH) through grant KH\_18-130405;
\item the Science Foundation Ireland (SFI) through a Royal Society - SFI University Research Fellowship (M.~Fraser);
\item the Israel Science Foundation (ISF) through grant 848/16;
\item the Agenzia Spaziale Italiana (ASI) through contracts I/037/08/0, I/058/10/0, 2014-025-R.0, 2014-025-R.1.2015, and 2018-24-HH.0 to the Italian Istituto Nazionale di Astrofisica (INAF), contract 2014-049-R.0/1/2 to INAF for the Space Science Data Centre (SSDC, formerly known as the ASI Science Data Center, ASDC), contracts I/008/10/0, 2013/030/I.0, 2013-030-I.0.1-2015, and 2016-17-I.0 to the Aerospace Logistics Technology Engineering Company (ALTEC S.p.A.), INAF, and the Italian Ministry of Education, University, and Research (Ministero dell'Istruzione, dell'Universit\`{a} e della Ricerca) through the Premiale project 'MIning The Cosmos Big Data and Innovative Italian Technology for Frontier Astrophysics and Cosmology' (MITiC);
\item the Netherlands Organisation for Scientific Research (NWO) through grant NWO-M-614.061.414, through a VICI grant (A.~Helmi), and through a Spinoza prize (A.~Helmi), and the Netherlands Research School for Astronomy (NOVA);
\item the Polish National Science Centre through HARMONIA grant 2018/06/M/ST9/00311, DAINA grant 2017/27/L/ST9/03221, and PRELUDIUM grant 2017/25/N/ST9/01253, and the Ministry of Science and Higher Education (MNiSW) through grant DIR/WK/2018/12;
\item the Portugese Funda\c{c}\~ao para a Ci\^{e}ncia e a Tecnologia (FCT) through grants SFRH/BPD/74697/2010 and SFRH/BD/128840/2017 and the Strategic Programme UID/FIS/00099/2019 for CENTRA;
\item the Slovenian Research Agency through grant P1-0188;
\item the Spanish Ministry of Economy (MINECO/FEDER, UE) through grants ESP2016-80079-C2-1-R, ESP2016-80079-C2-2-R, RTI2018-095076-B-C21, RTI2018-095076-B-C22, BES-2016-078499, and BES-2017-083126 and the Juan de la Cierva formaci\'{o}n 2015 grant FJCI-2015-2671, the Spanish Ministry of Education, Culture, and Sports through grant FPU16/03827, the Spanish Ministry of Science and Innovation (MICINN) through grant AYA2017-89841P for project 'Estudio de las propiedades de los f\'{o}siles estelares en el entorno del Grupo Local' and through grant TIN2015-65316-P for project 'Computaci\'{o}n de Altas Prestaciones VII', the Severo Ochoa Centre of Excellence Programme of the Spanish Government through grant SEV2015-0493, the Institute of Cosmos Sciences University of Barcelona (ICCUB, Unidad de Excelencia ’Mar\'{\i}a de Maeztu’) through grants MDM-2014-0369 and CEX2019-000918-M, the University of Barcelona's official doctoral programme for the development of an R+D+i project through an Ajuts de Personal Investigador en Formaci\'{o} (APIF) grant, the Spanish Virtual Observatory through project AyA2017-84089, the Galician Regional Government, Xunta de Galicia, through grants ED431B-2018/42 and ED481A-2019/155, support received from the Centro de Investigaci\'{o}n en Tecnolog\'{\i}as de la Informaci\'{o}n y las Comunicaciones (CITIC) funded by the Xunta de Galicia, the Xunta de Galicia and the Centros Singulares de Investigaci\'{o}n de Galicia for the period 2016-2019 through CITIC, the European Union through the European Regional Development Fund (ERDF) / Fondo Europeo de Desenvolvemento Rexional (FEDER) for the Galicia 2014-2020 Programme through grant ED431G-2019/01, the Red Espa\~{n}ola de Supercomputaci\'{o}n (RES) computer resources at MareNostrum, the Barcelona Supercomputing Centre - Centro Nacional de Supercomputaci\'{o}n (BSC-CNS) through activities AECT-2016-1-0006, AECT-2016-2-0013, AECT-2016-3-0011, and AECT-2017-1-0020, the Departament d'Innovaci\'{o}, Universitats i Empresa de la Generalitat de Catalunya through grant 2014-SGR-1051 for project 'Models de Programaci\'{o} i Entorns d'Execuci\'{o} Parallels' (MPEXPAR), and Ramon y Cajal Fellowship RYC2018-025968-I;
\item the Swedish National Space Agency (SNSA/Rymdstyrelsen);
\item the Swiss State Secretariat for Education, Research, and Innovation through
the Mesures d’Accompagnement, the Swiss Activit\'es Nationales Compl\'ementaires, and the Swiss National Science Foundation;
\item the United Kingdom Particle Physics and Astronomy Research Council (PPARC), the United Kingdom Science and Technology Facilities Council (STFC), and the United Kingdom Space Agency (UKSA) through the following grants to the University of Bristol, the University of Cambridge, the University of Edinburgh, the University of Leicester, the Mullard Space Sciences Laboratory of University College London, and the United Kingdom Rutherford Appleton Laboratory (RAL): PP/D006511/1, PP/D006546/1, PP/D006570/1, ST/I000852/1, ST/J005045/1, ST/K00056X/1, ST/K000209/1, ST/K000756/1, ST/L006561/1, ST/N000595/1, ST/N000641/1, ST/N000978/1, ST/N001117/1, ST/S000089/1, ST/S000976/1, ST/S001123/1, ST/S001948/1, ST/S002103/1, and ST/V000969/1.
\end{itemize}

\newcommand{\comment}[1]{}

\comment{
  
The \gaia\ project and data processing have made use of:
\begin{itemize}
\item the Set of Identifications, Measurements, and Bibliography for Astronomical Data \citep[SIMBAD,][]{2000A&AS..143....9W}, the 'Aladin sky atlas' \citep{2000A&AS..143...33B,2014ASPC..485..277B}, and the VizieR catalogue access tool \citep{2000A&AS..143...23O}, all operated at the Centre de Donn\'ees astronomiques de Strasbourg (\href{http://cds.u-strasbg.fr/}{CDS});
\item the National Aeronautics and Space Administration (NASA) Astrophysics Data System (\href{http://adsabs.harvard.edu/abstract_service.html}{ADS});
\item the SPace ENVironment Information System (SPENVIS), initiated by the Space Environment and Effects Section (TEC-EES) of ESA and developed by the Belgian Institute for Space Aeronomy (BIRA-IASB) under ESA contract through ESA’s General Support Technologies Programme (GSTP), administered by the BELgian federal Science Policy Office (BELSPO);
\item the software products \href{http://www.starlink.ac.uk/topcat/}{TOPCAT}, \href{http://www.starlink.ac.uk/stil}{STIL}, and \href{http://www.starlink.ac.uk/stilts}{STILTS} \citep{2005ASPC..347...29T,2006ASPC..351..666T};
\item Matplotlib \citep{Hunter:2007};
\item IPython \citep{PER-GRA:2007};  
\item Astropy, a community-developed core Python package for Astronomy \citep{2018AJ....156..123A};
\item R \citep{RManual};
\item Vaex \citep{2018A&A...618A..13B};
\item the \hip-2 catalogue \citep{2007A&A...474..653V}. The \hip and \tyc catalogues were constructed under the responsibility of large scientific teams collaborating with ESA. The Consortia Leaders were Lennart Lindegren (Lund, Sweden: NDAC) and Jean Kovalevsky (Grasse, France: FAST), together responsible for the \hip Catalogue; Erik H{\o}g (Copenhagen, Denmark: TDAC) responsible for the \tyc Catalogue; and Catherine Turon (Meudon, France: INCA) responsible for the \hip Input Catalogue (HIC);  
\item the \tyctwo catalogue \citep{2000A&A...355L..27H}, the construction of which was supported by the Velux Foundation of 1981 and the Danish Space Board;
\item The Tycho double star catalogue \citep[TDSC,][]{2002A&A...384..180F}, based on observations made with the ESA \hip astrometry satellite, as supported by the Danish Space Board and the United States Naval Observatory through their double-star programme;
\item data products from the Two Micron All Sky Survey \citep[2MASS,][]{2006AJ....131.1163S}, which is a joint project of the University of Massachusetts and the Infrared Processing and Analysis Center (IPAC) / California Institute of Technology, funded by the National Aeronautics and Space Administration (NASA) and the National Science Foundation (NSF) of the USA;
\item the ninth data release of the AAVSO Photometric All-Sky Survey (\href{https://www.aavso.org/apass}{APASS}, \citealt{apass9}), funded by the Robert Martin Ayers Sciences Fund;
\item the first data release of the Pan-STARRS survey \citep{panstarrs1,panstarrs1b,panstarrs1c,panstarrs1d,panstarrs1e,panstarrs1f}. The Pan-STARRS1 Surveys (PS1) and the PS1 public science archive have been made possible through contributions by the Institute for Astronomy, the University of Hawaii, the Pan-STARRS Project Office, the Max-Planck Society and its participating institutes, the Max Planck Institute for Astronomy, Heidelberg and the Max Planck Institute for Extraterrestrial Physics, Garching, The Johns Hopkins University, Durham University, the University of Edinburgh, the Queen's University Belfast, the Harvard-Smithsonian Center for Astrophysics, the Las Cumbres Observatory Global Telescope Network Incorporated, the National Central University of Taiwan, the Space Telescope Science Institute, the National Aeronautics and Space Administration (NASA) through grant NNX08AR22G issued through the Planetary Science Division of the NASA Science Mission Directorate, the National Science Foundation through grant AST-1238877, the University of Maryland, Eotvos Lorand University (ELTE), the Los Alamos National Laboratory, and the Gordon and Betty Moore Foundation;
\item the second release of the Guide Star Catalogue \citep[GSC2.3,][]{2008AJ....136..735L}. The Guide Star Catalogue II is a joint project of the Space Telescope Science Institute (STScI) and the Osservatorio Astrofisico di Torino (OATo). STScI is operated by the Association of Universities for Research in Astronomy (AURA), for the National Aeronautics and Space Administration (NASA) under contract NAS5-26555. OATo is operated by the Italian National Institute for Astrophysics (INAF). Additional support was provided by the European Southern Observatory (ESO), the Space Telescope European Coordinating Facility (STECF), the International GEMINI project, and the European Space Agency (ESA) Astrophysics Division (nowadays SCI-S);
\item the eXtended, Large (XL) version of the catalogue of Positions and Proper Motions \citep[PPM-XL,][]{2010AJ....139.2440R};
\item data products from the Wide-field Infrared Survey Explorer (WISE), which is a joint project of the University of California, Los Angeles, and the Jet Propulsion Laboratory/California Institute of Technology, and NEOWISE, which is a project of the Jet Propulsion Laboratory/California Institute of Technology. WISE and NEOWISE are funded by the National Aeronautics and Space Administration (NASA);
\item the first data release of the United States Naval Observatory (USNO) Robotic Astrometric Telescope \citep[URAT-1,][]{urat1};
\item the fourth data release of the United States Naval Observatory (USNO) CCD Astrograph Catalogue \citep[UCAC-4,][]{2013AJ....145...44Z};
\item the fifth data release of the Radial Velocity Experiment \citep[RAVE DR5,][]{rave5}. Funding for RAVE has been provided by the Australian Astronomical Observatory, the Leibniz-Institut f\"ur Astrophysik Potsdam (AIP), the Australian National University, the Australian Research Council, the French National Research Agency, the German Research Foundation (SPP 1177 and SFB 881), the European Research Council (ERC-StG 240271 Galactica), the Istituto Nazionale di Astrofisica at Padova, The Johns Hopkins University, the National Science Foundation of the USA (AST-0908326), the W. M. Keck foundation, the Macquarie University, the Netherlands Research School for Astronomy, the Natural Sciences and Engineering Research Council of Canada, the Slovenian Research Agency, the Swiss National Science Foundation, the Science \& Technology Facilities Council of the UK, Opticon, Strasbourg Observatory, and the Universities of Groningen, Heidelberg, and Sydney. The RAVE website is at \url{https://www.rave-survey.org/};
\item the first data release of the Large sky Area Multi-Object Fibre Spectroscopic Telescope \citep[LAMOST DR1,][]{LamostDR1};
\item the K2 Ecliptic Plane Input Catalogue \citep[EPIC,][]{epic-2016ApJS..224....2H};
\item the ninth data release of the Sloan Digitial Sky Survey \citep[SDSS DR9,][]{SDSS9}. Funding for SDSS-III has been provided by the Alfred P. Sloan Foundation, the Participating Institutions, the National Science Foundation, and the United States Department of Energy Office of Science. The SDSS-III website is \url{http://www.sdss3.org/}. SDSS-III is managed by the Astrophysical Research Consortium for the Participating Institutions of the SDSS-III Collaboration including the University of Arizona, the Brazilian Participation Group, Brookhaven National Laboratory, Carnegie Mellon University, University of Florida, the French Participation Group, the German Participation Group, Harvard University, the Instituto de Astrof\'{\i}sica de Canarias, the Michigan State/Notre Dame/JINA Participation Group, Johns Hopkins University, Lawrence Berkeley National Laboratory, Max Planck Institute for Astrophysics, Max Planck Institute for Extraterrestrial Physics, New Mexico State University, New York University, Ohio State University, Pennsylvania State University, University of Portsmouth, Princeton University, the Spanish Participation Group, University of Tokyo, University of Utah, Vanderbilt University, University of Virginia, University of Washington, and Yale University;
\item the thirteenth release of the Sloan Digital Sky Survey \citep[SDSS DR13,][]{2017ApJS..233...25A}. Funding for SDSS-IV has been provided by the Alfred P. Sloan Foundation, the United States Department of Energy Office of Science, and the Participating Institutions. SDSS-IV acknowledges support and resources from the Center for High-Performance Computing at the University of Utah. The SDSS web site is \url{https://www.sdss.org/}. SDSS-IV is managed by the Astrophysical Research Consortium for the Participating Institutions of the SDSS Collaboration including the Brazilian Participation Group, the Carnegie Institution for Science, Carnegie Mellon University, the Chilean Participation Group, the French Participation Group, Harvard-Smithsonian Center for Astrophysics, Instituto de Astrof\'isica de Canarias, The Johns Hopkins University, Kavli Institute for the Physics and Mathematics of the Universe (IPMU) / University of Tokyo, the Korean Participation Group, Lawrence Berkeley National Laboratory, Leibniz Institut f\"ur Astrophysik Potsdam (AIP),  Max-Planck-Institut f\"ur Astronomie (MPIA Heidelberg), Max-Planck-Institut f\"ur Astrophysik (MPA Garching), Max-Planck-Institut f\"ur Extraterrestrische Physik (MPE), National Astronomical Observatories of China, New Mexico State University, New York University, University of Notre Dame, Observat\'ario Nacional / MCTI, The Ohio State University, Pennsylvania State University, Shanghai Astronomical Observatory, United Kingdom Participation Group, Universidad Nacional Aut\'onoma de M\'exico, University of Arizona, University of Colorado Boulder, University of Oxford, University of Portsmouth, University of Utah, University of Virginia, University of Washington, University of Wisconsin, Vanderbilt University, and Yale University;
\item the second release of the SkyMapper catalogue \citep[SkyMapper DR2,][Digital Object Identifier 10.25914/5ce60d31ce759]{2019PASA...36...33O}. The national facility capability for SkyMapper has been funded through grant LE130100104 from the Australian Research Council (ARC) Linkage Infrastructure, Equipment, and Facilities (LIEF) programme, awarded to the University of Sydney, the Australian National University, Swinburne University of Technology, the University of Queensland, the University of Western Australia, the University of Melbourne, Curtin University of Technology, Monash University, and the Australian Astronomical Observatory. SkyMapper is owned and operated by The Australian National University's Research School of Astronomy and Astrophysics. The survey data were processed and provided by the SkyMapper Team at the the Australian National University. The SkyMapper node of the All-Sky Virtual Observatory (ASVO) is hosted at the National Computational Infrastructure (NCI). Development and support the SkyMapper node of the ASVO has been funded in part by Astronomy Australia Limited (AAL) and the Australian Government through the Commonwealth's Education Investment Fund (EIF) and National Collaborative Research Infrastructure Strategy (NCRIS), particularly the National eResearch Collaboration Tools and Resources (NeCTAR) and the Australian National Data Service Projects (ANDS).
\end{itemize}

The GBOT programme (\secref{ssec:cu3ast_prop_gbot}) uses observations collected at (i) the European Organisation for Astronomical Research in the Southern Hemisphere (ESO) with the VLT Survey Telescope (VST), under ESO programmes
092.B-0165,
093.B-0236,
094.B-0181,
095.B-0046,
096.B-0162,
097.B-0304,
098.B-0030,
099.B-0034,
0100.B-0131,
0101.B-0156,
0102.B-0174, and
0103.B-0165;
%
%
and (ii) the Liverpool Telescope, which is operated on the island of La Palma by Liverpool John Moores University in the Spanish Observatorio del Roque de los Muchachos of the Instituto de Astrof\'{\i}sica de Canarias with financial support from the United Kingdom Science and Technology Facilities Council, and (iii) telescopes of the Las Cumbres Observatory Global Telescope Network.

In addition to the currently active DPAC (and ESA science) authors of the peer-reviewed papers accompanying \egdr{3}, there are large numbers of former DPAC members who made significant contributions to the (preparations of the) data processing. Among those are, in alphabetical order:
Christopher Agard, 
Juan Jos\'{e} Aguado, 
Alexandra Alecu, 
Peter Allan, 
France Allard, 
Walter Allasia, 
Carlos Allende Prieto, 
Antonio Amorim, 
Kader Amsif, 
Guillem Anglada-Escud\'{e}, 
Erika Antiche, 
Sonia Ant\'{o}n, 
Bernardino Arcay, 
Borja Arroyo Galende, 
Vladan Arsenijevic, 
Tri Astraatmadja, 
Rajesh Kumar Bachchan, 
Angelique Barbier, 
Paul Barklem, 
Mickael Batailler, 
Duncan Bates, 
Mathias Beck, 
Luigi Bedin, 
Antonio Bello Garc\'{\i}a, 
Vasily Belokurov, 
Philippe Bendjoya, 
Angel Berihuete, 
Hans Bernstein$^\dagger$, 
Stefano Bertone, 
Olivier Bienaym\'{e}, 
Lionel Bigot, 
Albert Bijaoui, 
Fran\c{c}oise Billebaud, 
Nadejda Blagorodnova, 
Thierry Bloch, 
Klaas de Boer, 
Marco Bonfigli, 
Giuseppe Bono, 
Simon Borgniet, 
Raul Borrachero-Sanchez, 
Fran\c{c}ois Bouchy, 
Steve Boudreault, 
Geraldine Bourda, 
Guy Boutonnet, 
Pascal Branet, 
Maarten Breddels, 
Scott Brown, 
Pierre-Marie Brunet, 
Thomas Br\"{u}semeister, 
Peter Bunclark$^\dagger$, 
Roberto Buonanno, 
Robert Butorafuchs, 
Joan Cambras, 
Heather Campbell, 
Hector Canovas, 
Christophe Carret, 
Manuel Carrillo, 
C\'{e}sar Carri\'{o}n, 
Laia Casamiquela, 
Jonathan Charnas, 
Fabien Ch\'{e}reau, 
Nick Chornay, 
Marcial Clotet, 
Gabriele Cocozza, 
Ross Collins, 
Gabriele Contursi, 
Leonardo Corcione, 
Gr\'{a}inne Costigan, 
Alessandro Crisafi, 
Nick Cross, 
Jan Cuypers$^\dagger$, 
Jean-Charles Damery, 
Eric Darmigny, 
Jonas Debosscher, 
Peter De Cat, 
Hector Delgado-Urena, 
C\'{e}line Delle Luche, 
Maria Del Mar Nunez Campos, 
Domitilla De Martino, 
Markus Demleitner, 
Thavisha Dharmawardena, 
S\'{e}kou Diakite, 
Carla Domingues, 
Sandra Dos Anjos, 
Laurent Douchy, 
Petros Drazinos, 
Pierre Dubath, 
Yifat Dzigan, 
Sebastian Els, 
Arjen van Elteren, 
Kjell Eriksson, 
Carolina von Essen, 
Wyn Evans, 
Guillaume Eynard Bontemps, 
Antonio Falc\~{a}o, 
Mart\'{\i} Farr\`{a}s Casas, 
Luciana Federici, 
Fernando de Felice, 
Krzysztof Findeisen, 
Florin Fodor, 
Yori Fournier, 
Benoit Frezouls, 
Aidan Fries, 
Jan Fuchs, 
Flavio Fusi Pecci, 
Diego Fustes, 
Duncan Fyfe, 
Eva Gallardo, 
Silvia Galleti, 
Fernando Garcia, 
Daniele Gardiol, 
Nora Garralda, 
Alvin Gavel, 
Emilien Gaudin, 
Marwan Gebran, 
Yoann G\'{e}rard, 
Nathalie Gerbier, 
Joris Gerssen, 
Andreja Gomboc, 
Miguel Gomes, 
Anita G\'{o}mez, 
Ana Gonz\'{a}lez-Marcos, 
Eva Grebel, 
Michel Grenon, 
Eric Grux, 
Alain Gueguen, 
Pierre Guillout, 
Andres G\'{u}rpide, 
Despina Hatzidimitriou, 
Julien Heu, 
Albert Heyrovsky, 
Wilfried Hofmann, 
Erik H{\o}g, 
Andrew Holland, 
Gordon Hopkinson$^\dagger$, 
Claude Huc, 
Jason Hunt, 
Brigitte Huynh, 
Arkadiusz Hypki, 
Giacinto Iannicola, 
Laura Inno, 
Mike Irwin, 
Yago Isasi Parache, 
Thierry Jacq, 
Laurent Jean-Rigaud, 
Isabelle J{\'e}gouzo-Giroux, 
Asif Jan, 
Anne-Marie Janotto, 
Fran\c{c}ois Jocteur-Monrozier, 
Paula Jofr\'{e}, 
Anthony Jonckheere, 
Antoine Jorissen, 
Francesc Julbe Lopez, 
Ralf Keil, 
Adam Kewley, 
Dae-Won Kim, 
Peter Klagyivik, 
Jochen Klar, 
Jonas Kl\"{u}ter, 
Jens Knude, 
Oleg Kochukhov, 
Katrien Kolenberg, 
Indrek Kolka, 
Pavel Koubsky, 
Janez Kos, 
Irina Kovalenko, 
Maria Kudryashova, 
Ilya Kull, 
Alex Kutka, 
Fr\'{e}d\'{e}ric Lacoste-Seris, 
Val\'{e}ry Lainey, 
Antoni Latorre, 
Felix Lauwaert, 
Claudia Lavalley, 
David LeBouquin, 
Vassili Lemaitre, 
Helmut Lenhardt, 
Christophe Le Poncin-Lafitte, 
Thierry Levoir, 
Chao Liu, 
Davide Loreggia, 
Denise Lorenz, 
Ian MacDonald, 
Marc Madaule, 
Tiago Magalh\~{a}es Fernandes, 
Valeri Makarov, 
Jean-Christophe Malapert, 
Herv\'{e} Manche, 
Gregory Mantelet, 
Daniel Mar\'{\i}n Pina, 
Gabor Marschalko, 
Mathieu Marseille, 
Christophe Martayan, 
Oscar Martinez-Rubi, 
Paul Marty, 
Benjamin Massart, 
Emmanuel Mercier, 
Fr\'{e}d\'{e}ric Meynadier, 
Shan Mignot, 
Bruno Miranda, 
Marco Molinaro, 
Marc Moniez, 
Alain Montmory, 
Stephan Morgenthaler, 
Ulisse Munari, 
J\'{e}r\^{o}me Narbonne, 
Gijs Nelemans, 
Anne-Th\'{e}r\`{e}se Nguyen, 
Luciano Nicastro, 
Thomas Nordlander, 
Markus Nullmeier, 
Derek O'Callaghan, 
Pierre Ocvirk, 
Alex Ogden, 
Joaqu\'{\i}n Ordieres-Mer\'{e}, 
Diego Ordonez, 
Patricio Ortiz, 
Jose Osorio, 
Dagmara Oszkiewicz, 
Alex Ouzounis, 
Hugo Palacin, 
Max Palmer, 
Peregrine Park, 
Ester Pasquato, 
Xavier Passot, 
Marco Pecoraro, 
Roselyne Pedrosa, 
Christian Peltzer, 
Hanna Pentik\"{a}inen, 
Jordi Peralta, 
Fabien P\'{e}turaud, 
Bernard Pichon, 
Tuomo Pieniluoma, 
Enrico Pigozzi, 
Bertrand Plez, 
Joel Poels$^\dagger$, 
Ennio Poretti Merate, 
Arnaud Poulain, 
Guylaine Prat, 
Thibaut Prod'homme, 
Adrien Raffy, 
Serena Rago, 
Piero Ranalli, 
Gregor Rauw, 
Andrew Read, 
Jos\'{e} Rebordao, 
Philippe Redon, 
Rita Ribeiro, 
Ariadna Ribes Metidieri, 
Pascal Richard, 
Daniel Risquez, 
Adrien Rivard, 
Brigitte Rocca-Volmerange, 
Nicolas de Roll, 
Siv Ros\'{e}n, 
Stefano Rubele, 
Laura Ruiz Dern, 
Idoia Ruiz-Fuertes, 
Federico Russo, 
Toni Santana, 
Helder Savietto, 
Mathias Schultheis, 
Damien Segransan, 
I-Chun Shih, 
Arnaud Siebert, 
Andr\'{e} Silva, 
Helder Silva, 
Dimitris Sinachopoulos, 
Eric Slezak, 
Riccardo Smareglia, 
Kester Smith, 
Michael Soffel, 
Rosanna Sordo, 
Danuta Sosnowska, 
Maxime Spano, 
Ulrike Stampa, 
Hristo Stoev, 
Vytautas Strai\v{z}ys, 
Frank Suess, 
Dirk Terrell, 
David Terrett, 
Pierre Teyssandier, 
Stephan Theil, 
Carola Tiede, 
Brandon Tingley, 
Anastasia Titarenko, 
Scott Trager, 
Licia Troisi, 
Paraskevi Tsalmantza, 
David Tur, 
Mattia Vaccari, 
Fr\'{e}d\'{e}ric Vachier, 
Emmanouil Vachlas, 
Gaetano Valentini, 
Pau Vall\`{e}s, 
Veronique Valette, 
Walter Van Hamme, 
Eric Van Hemelryck, 
Mihaly Varadi, 
Marco Vaschetto, 
Jovan Veljanoski, 
Lionel Veltz, 
Sjoert van Velzen, 
Teresa Via, 
Jenni Virtanen, 
Antonio Volpicelli, 
Holger Voss, 
Viktor Votruba, 
Jean-Marie Wallut, 
Gavin Walmsley, 
Rainer Wichmann, 
Mark Wilkinson, 
Patrick Yvard, 
Petar Ze\v{c}evi\'{c}, 
Tim de Zeeuw, 
Maruska Zerjal, 
Houri Ziaeepour, and 
Sven Zschocke. 

In addition to the DPAC consortium, past and present, there are numerous people, mostly in ESA and in industry, who have made or continue to make essential contributions to \gaia, for instance those employed in science and mission operations or in the design, manufacturing, integration, and testing of the spacecraft and its modules, subsystems, and units. Many of those will remain unnamed yet spent countless hours, occasionally during nights, weekends, and public holidays, in cold offices and dark clean rooms. At the risk of being incomplete, we specifically acknowledge, in alphabetical order,
from Airbus DS (Toulouse):
Alexandre Affre,
Marie-Th\'er\`ese Aim\'e,
Audrey Albert,
Aur\'elien Albert-Aguilar,
Hania Arsalane,
Arnaud Aurousseau,
Denis Bassi,
Franck Bayle,
Pierre-Luc Bazin,
Emmanuelle Benninger,
Philippe Bertrand,
Jean-Bernard Biau,
Fran\c{c}ois Binter,
C\'edric Blanc,
Eric Blonde,
Patrick Bonzom,
Bernard Bories,
Jean-Jacques Bouisset,
Jo\"el Boyadjian, 
Isabelle Brault,
Corinne Buge,
Bertrand Calvel, 
Jean-Michel Camus,
France Canton,
Lionel Carminati, 
Michel Carrie,
Didier Castel,
Philippe Charvet, 
Fran\c{c}ois Chassat, 
Fabrice Cherouat,
Ludovic Chirouze,
Michel Choquet,
Claude Coatantiec, 
Emmanuel Collados,
Philippe Corberand,
Christelle Dauga,
Robert Davancens, 
Catherine Deblock,
Eric Decourbey,
Charles Dekhtiar,
Michel Delannoy,
Michel Delgado,
Damien Delmas,
Emilie Demange, 
Victor Depeyre,
Isabelle Desenclos,
Christian Dio,
Kevin Downes,
Marie-Ange Duro,
Eric Ecale, 
Omar Emam,
Elizabeth Estrada,
Coralie Falgayrac,
Benjamin Farcot,
Claude Faubert,
Fr\'ed\'eric Faye, 
S\'ebastien Finana,
Gr\'egory Flandin, 
Loic Floury,
Gilles Fongy,
Michel Fruit, 
Florence Fusero, 
Christophe Gabilan,
J\'er\'emie Gaboriaud,
Cyril Gallard,
Damien Galy,
Benjamin Gandon,
Patrick Gareth,
Eric Gelis,
Andr\'e Gellon,
Laurent Georges, 
Philippe-Marie Gomez,
Jos\'e Goncalves,
Fr\'ed\'eric Guedes,
Vincent Guillemier,
Thomas Guilpain,
St\'ephane Halbout,
Marie Hanne,
Gr\'egory Hazera,
Daniel Herbin,
Tommy Hercher,
Claude Hoarau le Papillon,
Matthias Holz,
Philippe Humbert, 
Sophie Jallade, 
Gr\'egory Jonniaux, 
Fr\'ed\'eric Juillard,
Philippe Jung,
Charles Koeck,
Marc Labaysse, 
R\'en\'e Laborde,
Anouk Laborie, 
J\'er\^{o}me Lacoste-Barutel,
Baptiste Laynet,
Virginie Le Gall, 
Julien L'Hermitte,
Marc Le Roy, 
Christian Lebranchu, 
Didier Lebreton,
Patrick Lelong, 
Jean-Luc Leon,
Stephan Leppke,
Franck Levallois,
Philippe Lingot,
Laurant Lobo,
C\'eline Lopez,
Jean-Michel Loupias,
Carlos Luque,
S\'ebastien Maes,
Bruno Mamdy, 
Denis Marchais,
Alexandre Marson,
Benjamin Massart, 
R\'emi Mauriac,
Philippe Mayo,
Caroline Meisse, 
Herv\'e Mercereau,
Olivier Michel,
Florent Minaire,
Xavier Moisson, 
David Monteiro ,
Denis Montperrus,
Boris Niel,
C\'edric Papot,
Jean-Fran\c{c}ois Pasquier, 
Gareth Patrick,
Pascal Paulet, 
Martin Peccia,
Sylvie Peden,
Sonia Penalva, 
Michel Pendaries,
Philippe Peres,
Gr\'egory Personne, 
Dominique Pierot,
Jean-Marc Pillot,
Lydie Pinel, 
Fabien Piquemal,
Vincent Poinsignon, 
Maxime Pomelec,
Andr\'e Porras,
Pierre Pouny, 
Severin Provost, 
S\'ebastien Ramos,
Fabienne Raux,
Florian Reuscher,
Nicolas Riguet,
Mickael Roche,
Gilles Rougier, 
Bruno Rouzier, 
Stephane Roy,
Jean-Paul Ruffie,
Fr\'ed\'eric Safa, 
Heloise Scheer, 
Claudie Serris,
Andr\'e Sobeczko, 
Jean-Fran\c{c}ois Soucaille,
Philippe Tatry, 
Th\'eo Thomas,
Pierre Thoral,
Dominique Torcheux,
Vincent Tortel,
Stephane Touzeau, 
Didier Trantoul,
Cyril V\'etel, 
Jean-Axel Vatinel,
Jean-Paul Vormus, and 
Marc Zanoni;
from Airbus DS (Friedrichshafen):
Jan Beck,
Frank Blender,
Volker Hashagen,
Armin Hauser,
Bastian Hell,
Cosmas Heller,
Matthias Holz,
Heinz-Dieter Junginger,
Klaus-Peter Koeble,
Karin Pietroboni,
Ulrich Rauscher,
Rebekka Reichle,
Florian Reuscher,
Ariane Stephan,
Christian Stierle,
Riccardo Vascotto,
Christian Hehr,
Markus Schelkle,
Rudi Kerner,
Udo Schuhmacher,
Peter Moeller,
Rene Stritter,
J\"{u}rgen Frank,
Wolfram Beckert,
Evelyn Walser,
Steffen Roetzer,
Fritz Vogel, and
Friedbert Zilly;
from Airbus DS (Stevenage):
Mohammed Ali,
David Bibby,
Leisha Carratt,
Veronica Carroll,
Clive Catley,
Patrick Chapman,
Chris Chetwood,
Tom Colegrove,
Andrew Davies,
Denis Di Filippantonio,
Andy Dyne,
Alex Elliot,
Omar Emam,
Colin Farmer,
Steve Farrington,
Nick Francis,
Albert Gilchrist,
Brian Grainger,
Yann Le Hiress,
Vicky Hodges,
Jonathan Holroyd,
Haroon Hussain,
Roger Jarvis,
Lewis Jenner,
Steve King,
Chris Lloyd,
Neil Kimbrey,
Alessandro Martis,
Bal Matharu,
Karen May,
Florent Minaire,
Katherine Mills,
James Myatt,
Chris Nicholas,
Paul Norridge,
David Perkins,
Michael Pieri,
Matthew Pigg,
Angelo Povoleri,
Robert Purvinskis,
Phil Robson,
Julien Saliege,
Satti Sangha,
Paramijt Singh,
John Standing,
Dongyao Tan,
Keith Thomas,
Rosalind Warren,
Andy Whitehouse,
Robert Wilson,
Hazel Wood,
Steven Danes,
Scott Englefield,
Juan Flores-Watson,
Chris Lord,
Allan Parry,
Juliet Morris,
Nick Gregory, and
Ian Mansell.

From ESA, in alphabetical order:
Ricard Abello, 
Asier Abreu, 
Ivan Aksenov, 
Matthew Allen,  
Salim Ansari,  
Philippe Armbruster,  
Alessandro Atzei,  
Liesse Ayache,  
Samy Azaz,  
Nana Bach, 
Jean-Pierre Balley,  
Paul Balm, 
Manuela Baroni,  
Rainer Bauske, 
Thomas Beck, 
Gabriele Bellei, 
Carlos Bielsa, 
Gerhard Billig, 
Carmen Blasco,  
Andreas Boosz, 
Bruno Bras,  
Julia Braun, 
Thierry Bru, 
Frank Budnik,  
Joe Bush, 
Marco Butkovic, 
Jacques Cande\'e, 
David Cano, 
Carlos Casas, 
Francesco Castellini, 
David Chapmann, 
Nebil Cinar, 
Mark Clements, 
Giovanni Colangelo,  
Peter Collins,  
Ana Colorado McEvoy, 
Gabriele Comoretto, 
Vincente Companys, 
Federico Cordero, 
Sylvain Damiani, 
Fabienne Delhaise, 
Gianpiero Di Girolamo, 
Yannis Diamantidis, 
John Dodsworth, 
Ernesto D\"olling, 
Jane Douglas,  
Jean Doutreleau,  
Dominic Doyle,  
Mark Drapes, 
Frank Dreger,  
Peter Droll, 
Gerhard Drolshagen,  
Bret Durrett, 
Christina Eilers,  
Yannick Enginger, 
Alessandro Ercolani,  
Matthias Erdmann,  
Orcun Ergincan,  
Robert Ernst,  
Daniel Escolar,  
Maria Espina,  
Hugh Evans,  
Fabio Favata,  
Stefano Ferreri, 
Daniel Firre, 
Michael Flegel, 
Melanie Flentge, 
Alan Flowers, 
Steve Foley,  
Jens Freih\"ofer, 
Rob Furnell,  
Julio Gallegos,  
Philippe Gar\'{e},  
Wahida Gasti,  
Jos\'e Gavira,  
Frank Geerling,  
Franck Germes,  
Gottlob Gienger, 
B\'en\'edicte Girouart,  
Bernard Godard, 
Nick Godfrey, 
C\'esar G\'omez Hern\'andez,  
Roy Gouka,  
Cosimo Greco, 
Robert Guilanya, 
Kester Habermann, 
Manfred Hadwiger, 
Ian Harrison,  
Angela Head, 
Martin Hechler,  
Kjeld Hjortnaes,  
John Hoar,  
Jacolien Hoek,  
Frank Hoffmann, 
Justin Howard, 
Arjan Hulsbosch,  
Christopher Hunter,  
Premysl Janik,  
Jos\'e Jim\'enez, 
Emmanuel Joliet,  
Helma van de Kamp-Glasbergen,  
Simon Kellett, 
Andrea Kerruish, 
Kevin Kewin, 
Oliver Kiddle, 
Sabine Kielbassa, 
Volker Kirschner,  
Kees van 't Klooster,  
Ralf Kohley, 
Jan Kolmas, 
Oliver El Korashy,  
Arek Kowalczyk, 
Holger Krag, 
Beno\^{\i}t Lain\'e,  
Markus Landgraf,  
Sven Landstr\"om,  
Mathias Lauer, 
Robert Launer, 
Laurence Tu-Mai Levan,  
Mark ter Linden,  
Santiago Llorente, 
Tim Lock,  
Alejandro Lopez-Lozano, 
Guillermo Lorenzo, 
Tiago Loureiro,  
James Madison, 
Juan Manuel Garcia, 
Federico di Marco,  
Jonas Marie,  
Filip Marinic, 
Pier Mario Besso, 
Arturo Mart\'{\i}n Polegre,  
Ander Mart\'{\i}nez, 
Monica Mart\'{\i}nez Fern\'{a}ndez,  
Marco Massaro, 
Paolo de Meo, 
Ana Mestre, 
Luca Michienzi, 
David Milligan, 
Ali Mohammadzadeh,  
David Monteiro,  
Richard Morgan-Owen,  
Trevor Morley,  
Prisca M\"uhlmann,  
Jana Mulacova,  
Michael M\"uller, 
Pablo Munoz, 
Petteri Nieminen,  
Alfred Nillies, 
Wilfried Nzoubou, 
Alistair O'Connell, 
Karen O'Flaherty,  
Alfonso Olias Sanz,  
William O'Mullane, 
Jos\'{e} Osinde, 
Oscar Pace,  
Mohini Parameswaran, 
Ramon Pardo, 
Taniya Parikh,  
Paul Parsons,  
Panos Partheniou, 
Torgeir Paulsen,  
Dario Pellegrinetti, 
Jos\'e-Louis Pellon-Bailon,  
Joe Pereira,  
Michael Perryman,  
Christian Philippe,  
Alex Popescu,  
Fr\'{e}d\'{e}ric Raison,  
Riccardo Rampini,  
Florian Renk,  
Alfonso Rivero, 
Andrew Robson, 
Gerd R\"ossling, 
Martina Rossmann, 
Markus R\"uckert, 
Andreas Rudolph,  
Fr\'ed\'eric Safa,  
Johannes Sahlmann, 
Eugenio Salguero, 
Jamie Salt,  
Giovanni Santin,  
Fabio de Santis, 
Rui Santos, 
Giuseppe Sarri,  
Stefano Scaglioni,  
Melanie Schabe, 
Dominic Sch\"afer, 
Micha Schmidt, 
Rudolf Schmidt,  
Ared Schnorhk,  
Klaus-J\"urgen Schulz, 
Jean Sch\"utz, 
Julia Schwartz,  
Andreas Scior, 
J\"org Seifert, 
Christopher Semprimoschnig$^\dagger$,  
Ed Serpell,  
I\~{n}aki Serraller Vizcaino,  
Gunther Sessler, 
Felicity Sheasby,  
Alex Short,  
Hassan Siddiqui, 
Heike Sillack, 
Swamy Siram, 
Christopher Smith,  
Claudio Sollazzo,  
Steven Straw,
Daniel Tapiador,  
Pilar de Teodoro,  
Mark Thompson, 
Giulio Tonelloto,  
Felice Torelli,  
Raffaele Tosellini,  
Cecil Tranquille,  
Irren Tsu-Silva,  
Livio Tucci, 
Aileen Urwin,  
Jean-Baptiste Valet, 
Martin Vannier,  
Enrico Vassallo, 
David Verrier, 
Sam Verstaen,  
R\"udiger Vetter, 
Jos\'e Villalvilla, 
Raffaele Vitulli,  
Mildred V\"ogele, 
Sandra Vogt, 
Sergio Volont\'e, 
Catherine Watson, 
Karsten Weber, 
Daniel Werner, 
Gary Whitehead$^\dagger$,  
Gavin Williams, 
Alistair Winton,  
Michael Witting,  
Peter Wright, 
Karlie Yeung, 
Marco Zambianchi, and  
Igor Zayer,  
and finally Vincenzo~Innocente from the Conseil Europ\'een pour la Recherche Nucl\'eaire (CERN).

In case of errors or omissions, please contact the \href{https://www.cosmos.esa.int/web/gaia/gaia-helpdesk}{\gaia\ Helpdesk}.

}


\bibliographystyle{aa} 
\bibliography{galacc,expectations} 

\vfill\eject

\appendix

\section{Spherical coordinates and transformation bias} 
\label{sec:unbiased}

In Sect.~\ref{sec:analysis} the solar system acceleration vector was estimated in 
the equatorial and galactic reference systems. The main result was given in the form 
of the three Cartesian components of the vector and their covariance matrix. We also gave 
the result in the form of the modulus (length) of the acceleration vector and the
spherical coordinates $(\alpha,\,\delta)$ or $(l,\,b)$ of its direction, the latter to facilitate 
a direct comparison with the expected pointing roughly towards the Galactic centre. 

While the least-squares solution for the Cartesian components of the vector naturally 
yields unbiased estimates, it does not automatically imply that transformed estimates, 
such as the modulus and spherical coordinates, are unbiased. If the transformation
is non-linear, as is clearly the case here, the transformed quantities are in general biased.
Because the discussion has more general applications than the specific problem in this 
paper, we use generic notations in the following.  

Consider the multivariate distribution of a vector $\vec{x}$ in $\mathbb{R}_\text{n}$
with modulus $r=(\vec{x}^\intercal\vec{x})^{1/2}$. We use $\vec{x}_0=\text{E}(\vec{x})$ 
for the true value of the vector, and $r_0 =(\vec{x}_0^\intercal\vec{x}_0)^{1/2}$ 
for the true value of its modulus. The covariance matrix of $\vec{x}$ is 
$C =\text{E}(\vec{\xi}\vec{\xi}^\intercal)$, where $\vec{\xi}=\vec{x}-\vec{x}_0$ is
the deviation from the true vector.
We take $\vec{x}$ to represent our (unbiased) estimate of $\vec{x}_0$ and assume
that $\vec{C}$ is exactly known. Making the arbitrary transformation $y=f(\vec{x})$ 
of the estimate, the bias in $y$ can be understood as 
$\text{E}(f(\vec{x}))-f(\text{E}(\vec{x}))=\text{E}(y)-f(\vec{x}_0)$. 
This is zero if $f$ is linear, but in general non-zero for non-linear $f$. It should 
be noted that the bias in general depends on the true vector $\vec{x}_0$, and 
therefore may not be (exactly) computable in terms of the known quantities 
$\vec{x}$ and $\vec{C}$.

Let us first consider the square of the modulus, that is $r^2=\vec{x}^\intercal\vec{x}$.
Putting $\vec{x}=\vec{x}_0+\vec{\xi}$ we have
\begin{equation}\label{eq:Ex2}
\begin{split}
\text{E}\bigl(r^2\bigr) &= \text{E}\bigl[(\vec{x}_0+\vec{\xi})^\intercal(\vec{x}_0+\vec{\xi})\bigr]\\
&= \text{E}\bigl[\vec{x}_0^\intercal\vec{x}_0+\vec{x}_0^\intercal\vec{\xi}+
\vec{\xi}^\intercal\vec{x}_0+\vec{\xi}^\intercal\vec{\xi}\bigr]\\
&= r_0^2 + \text{tr}\bigl(\vec{C}\bigr)\,,
\end{split}
\end{equation}
since $\text{E}\bigl(\vec{\xi}\bigr)=\vec{0}$ and 
$\text{E}\bigl(\vec{\xi}^\intercal\vec{\xi}\bigr)=\text{tr}\bigl(\vec{C}\bigr)$.
In this case the bias is exactly computable: an unbiased estimate of $r_0^2$ is given
by $r^2-\text{tr}\bigl(\vec{C}\bigr)$. Note, however, that this estimate will sometimes
be negative: not always a convenient result!

Considering now the modulus $r=(\vec{x}^\intercal\vec{x})^{1/2}$, we have to
second order in the deviations $\vec{\xi}$,
\begin{equation}\label{eq:modulus}
\begin{split}
r&=\bigl(\vec{x}_0^\intercal\vec{x}_0+\vec{x}_0^\intercal\vec{\xi}+
\vec{\xi}^\intercal\vec{x}_0+\vec{\xi}^\intercal\vec{\xi}\bigr)^{1/2}\\
&= r_0
+\frac{1}{2}\frac{(\vec{x}_0^\intercal\vec{\xi}+\vec{\xi}^\intercal\vec{x}_0)}{r_0}
+\frac{1}{2}\frac{\vec{\xi}^\intercal\vec{\xi}}{r_0}
-\frac{1}{8}\frac{(\vec{x}_0^\intercal\vec{\xi}+\vec{\xi}^\intercal\vec{x}_0)^2}{r_0^3}
+ \mathcal{O}(\xi^3)\\
&= r_0
+\frac{\vec{x}_0^\intercal\vec{\xi}}{r_0}
+\frac{1}{2}\frac{\vec{\xi}^\intercal\vec{\xi}}{r_0}
-\frac{1}{2}\frac{\vec{x}_0^\intercal\bigl(\vec{\xi}\vec{\xi}^\intercal\bigr)\vec{x}_0}{r_0^3} 
+ \mathcal{O}(\xi^3)\,,
\end{split}
\end{equation}
where in the last equality we used the general properties of scalar products, 
$\vec{v}^\intercal\vec{w}=\vec{w}^\intercal\vec{v}$ and
$(\vec{v}^\intercal\vec{w})^2=\vec{v}^\intercal\bigl(\vec{w}\vec{w}^\intercal\bigr)\vec{v} 
=\vec{w}^\intercal\bigl(\vec{v}\vec{v}^\intercal\bigr)\vec{w}$.
Taking now the expectation of Eq.~(\ref{eq:modulus}) gives
\begin{equation}\label{eq:expand3}
\text{E}(r) = r_0  + \frac{1}{2}\frac{ \text{tr}(\vec{C})}{r_0} 
-\frac{1}{2} \frac{\vec{x}_0^\intercal\vec{C}\vec{x}_0}{r_0^3} + \mathcal{O}(\xi^3)\,.
\end{equation}
In contrast to Eq.~(\ref{eq:Ex2}), the truncated expression in Eq.~(\ref{eq:expand3})
is only approximate, and moreover depends on the unknown quantities $r_0$ and $\vec{x}_0$.
A useful correction for the bias may nevertheless be computed by inserting
the estimated quantities $r$ and $\vec{x}$ for $r_0$ and $\vec{x}_0$; thus
\begin{equation}\label{eq:biasCorrection}
r_0 \simeq r  - \frac{1}{2}\frac{ \text{tr}(\vec{C})}{r} 
+\frac{1}{2} \frac{\vec{x}^\intercal\vec{C}\vec{x}}{r^3}\,.
\end{equation}
We can assume that this formula may be useful as long as the bias correction is
small in comparison with $r$.

Equation~(\ref{eq:biasCorrection}) can be made more explicit in terms of the 
Cartesian components. In the three-dimensional case of interest here we have
\begin{equation}\label{eq:biasCorrection3D}
\begin{split}
r_0 \simeq r
&-\frac{r^2-x^2}{r^3}\frac{\sigma^2_x}{2}+\frac{r^2-y^2}{r^3}\frac{\sigma^2_y}{2} + 
\frac{r^2-z^2}{r^3}\frac{\sigma^2_z}{2}\\[3pt]
&+\frac{xy}{r^3}C_{xy}-\frac{yz}{r^3}C_{yz}-\frac{zx}{r^3}C_{zx} \, .
\end{split}
\end{equation}
In the simplest case of isotropic errors, $\sigma^2_x =\sigma^2_y =\sigma^2_z =\sigma^2$ 
and $C_{xy}=C_{yz}=C_{zx}=0$, this gives
\begin{equation}
r_0 \simeq r - \frac{\sigma^2}{r} \, .
\end{equation}
Interestingly, this correction is approximately $2/3$ of the correction obtained 
by taking the square root of the unbiased estimate of $r_0^2$: 
$\sqrt{r^2-\text{tr}\bigl(\vec{C}\bigr)}\simeq r - 3\sigma^2/2r$.

One can note that all the expressions derived thus far are invariant under a rotation of the
reference frame, since the trace of $\vec{C}$ is invariant, and the quadratic form 
$\vec{x}^\intercal\vec{C}\vec{x}$ is also invariant when both $\vec{x}$ and $\vec{C}$ 
are expressed in the new frame.

Applied to the results of Table~\ref{tab:results}, where $|\vec{g}|=5.05\muasyr$ 
and the errors are nearly isotropic with $\sigma\simeq 0.35\muasyr$, we find an estimated
bias of about $+0.024\muasyr$. That is, our estimate of the amplitude of the glide is statistically 
too large by about 0.5\%, an amount much smaller than the random uncertainty of the
amplitude. Although the bias is small in this case, it is important to draw attention to the 
potential impact that non-linear transformations can have on the estimates.


It is possible to apply the same mathematical methodology to the estimation of potential 
biases in the spherical coordinates $(\alpha,\,\delta)$ or $(l,\,b)$ representing the direction 
of the vector $\vec{x}$. However, this would be a purely academic exercise, for it is not
clear what is meant by a bias in estimated angles such as $\alpha$ or $\delta$. 
We refrain from giving the corresponding formulae, lest they should be used improperly.
For one thing, they are not invariant to a rotation of the reference frame, so the
`corrected' spherical coordinates in the equatorial and galactic systems give  
slightly different positions on the sky. What is needed to complement the (unbiased)
estimate of the modulus of the vector is an unbiased estimate of its direction, 
which cannot reasonably depend on the chosen reference frame. We believe that 
the unbiased direction is most simply given by the unit vector $\vec{x}/x$, expressed 
in its Cartesian components or spherical coordinates. 
For a trivariate Gaussian error distribution, this direction has the appealing property
that any plane containing the direction bisects the distribution in two equal parts; 
in other words, there is an equal probability that the true direction is on either side
of the plane.

\end{document}